\documentclass[amsmath,12pt,amssymb,preprint,prd,aps,nofootinbib]{revtex4-2}
\usepackage{amsfonts}
\usepackage{rotating} 
\usepackage{graphicx} 
\usepackage{epsfig}
\usepackage{multirow}
\usepackage{bm}
\usepackage{hyperref}
\usepackage{cleveref}
\usepackage[toc,page]{appendix}
\usepackage{tabularx}
\newcommand{\spinhalf}{spin$-\frac{1}{2}^+$}
\newcommand{\spinthhalf}{spin$-\frac{3}{2}^+$}

\newcommand{\diracslash}[1]{#1\llap{/\kern2pt}}

\newcommand{\be}{\begin{equation}}
	\newcommand{\ee}{\end{equation}}
\newcommand{\bea}{\begin{eqnarray}}
	\newcommand{\eea}{\end{eqnarray}}
\newcommand{\ba}[1]{\begin{array}{#1}}
	\newcommand{\ea}{\end{array}}

\newcommand{\bt}{\begin{tabular}}
	\newcommand{\et}{\end{tabular}}

\newcommand{\beas}{\begin{eqnarray*}}
	\newcommand{\eeas}{\end{eqnarray*}}

\begin{document}
	\title{Effective  masses and magnetic moments of charmed baryons in asymmetric hot strange hadronic matter}$   $
	\author{\large Suneel Dutt}
	\email{dutts@nitj.ac.in}
	\author{\large Arvind Kumar}
	\email{kumara@nitj.ac.in}
	\author{\large Harleen Dahiya}
	\email{dahiyah@nitj.ac.in}
	
	\affiliation{Department of Physics, Dr. B R Ambedkar National Institute of Technology Jalandhar,$   $
		Jalandhar -- 144008, Punjab, India}

	\def\be{\begin{equation}}
		\def\ee{\end{equation}}
	\def\bearr{\begin{eqnarray}}
		\def\eearr{\end{eqnarray}}
	\def\zbf#1{{\bf {#1}}}
	\def\bfm#1{\mbox{\boldmath $#1$}}
	\def\hf{\frac{1}{2}}
	\def\kp{\zbf k+\frac{\zbf q}{2}}
	\def\km{-\zbf k+\frac{\zbf q}{2}}
	\def\hwo{\hat\omega_1}
	\def\hwt{\hat\omega_2}

\begin{abstract}
In the present work, we have studied the   masses and magnetic moments of spin$-{\frac{1}{2}}^+$
and spin$-{\frac{3}{2}}^+$ singly and doubly charmed baryons in the strange hadronic medium at finite temperature using the chiral SU(3) quark mean field model.  The properties of baryons within the framework of chiral SU(3) mean field model  are defined in terms of constituent quark masses
and energies, which are modified through the exchange of  scalar fields $\sigma$, $\zeta$ and $\delta$ and the vector fields $\omega$, $\rho$ and $\phi$. The scalar-isovector field, $\delta$  and the vector-isovector field, $\rho$ contribute when medium has finite isospin asymmetry.
We have calculated the effective  masses of constituent quarks and charmed baryons in the nuclear and strange matter within the chiral SU(3) quark mean field model and have used these as the input in SU(4) constituent chiral quark model to compute the effective magnetic moments of these baryons. Considering the configuration mixing, the contributions of valence quarks, quark sea and orbital angular momentum of quark sea have been considered explicitly to calculate the in-medium magnetic moments.
	\end{abstract}
	
	\maketitle
\newpage	
	\section{\label{intro}Introduction}
To understand the non-perturbative aspects of the theory of strong interactions: quantum chromodynamics (QCD), the in-medium properties of hadrons play a very significant role.  Various experimental investigations  hint towards the modifications in the properties of light hadrons (composed of three lightest quark flavors $u$, $d$ and $s$) when they are subjected to finite density and temperature \cite{CrystalBall:2000tgu,Messchendorp:2002au,Hayano:2008vn,Metag:2005ai}.  Even though the theory of lattice QCD has been applied to study the properties of hadrons at finite temperature and zero baryon density, due to sign problem it is not  applicable in the situation when one aspire to explore the impact of finite baryon density  \cite{Sugano:2017yqz,Kelly:2017opi}. Some solutions
to lattice QCD, e.g., Taylor expansion have been proposed to overcome the sign problem, however they are limited to low baryon chemical potential regime only \cite{Ejiri:2009hq,Allton:2003vx}.
 Therefore, various other non-perturbative approaches such as QCD sum rules, Nambu Jona Lasino (NJL) model and its extended Polyakov NJL versions
 \cite{Forbes:2004ww,Aguirre:2024clo}, coupled channel approach$   $
 \cite{Tolos:2002ud,Tolos:2010fq}, linear and non-linear sigma models \cite{Sakai:2013nba}, quark meson model (QMC) \cite{Cobos-Martinez:2023hbp}, chiral hadronic models \cite{Mishra:2006wy,Mishra:2008dj,Mishra:2014rha,Kumar:2020gpu} etc. have been used to study the properties of hadrons
in the hot and dense medium.
The medium modifications of light hadrons is closely connected to the spontaneous and explicit breaking of chiral symmetry and its restoration at high temperature and baryon density. Experimental facilities such relativistic heavy ion collider (RHIC) and large hadron collider (LHC) probe the dense matter at high temperature and almost zero baryon density, whereas, the future facility for antiproton and ion research  (FAIR) at GSI, Germany, Japan proton accelerator research complex (J-PARC) of KEK and nuclotron-based ion collider facility (NICA) at JINR, DUBNA aim to study the properties of medium at high baryon density.

Interest has recently grown in understanding the impact of finite temperature and baryon density on the medium modifications of mesons and baryons which contain one or more heavy $c$ or $b$ quark.
The study of in-medium properties of heavy hadrons may play a crucial role in understanding their production rate
\cite{Rapp:2008tf,Andronic:2007zu}
and quarkonium suppression which
may also be significant to understand the formation of quark gluon plasma (QGP) in heavy-ion collision experiments
such as RHIC and LHC
 \cite{Matsui-Satz1986}.
The modifications in the properties such as in-medium masses and decay widths of 
open and hidden charmed mesons have been studied using different theoretical approaches \cite{Sibirtsev:1999jr,Friman:2002fs,Tsushima:2002cc,Tsushima:2002sm,Tolos:2007vh,Mishra:2008cd,Kumar:2010gb}. 
The mass modifications of heavy baryons were studied using the QMC model  \cite{Tsushima:2002cc,Tsushima:2002sm}.
Within the QCD sum rules, the studies on in-medium masses of heavy baryons were initiated initially 
in Refs. \cite{Wang:2011hta,Wang:2011yj,Wang:2012xk,Azizi:2016dmr}.
The effective masses and self energies of charmed baryon $\Lambda_c$ have been calculated using the parity projected QCD sum rules \cite{Ohtani:2017wdc}.
The QCD sum rules were applied to study the masses, residues and self-energies of heavy spin$-3/2$ charmed and bottom baryon resonances, $\Sigma_Q^*,
\Xi_{Q}^*$ and $\Omega_Q^*$  (here $Q$ is $c$ or $b$ quark) in the nuclear matter at zero temperature \cite{Azizi:2018dtb}.
Effective potential of $\Lambda_c$ baryons has also been calculated  in cold and hot nuclear medium using the heavy quark effective field theory \cite{Yasui:2018sxz}.
In Ref.  \cite{Carames:2018xek}, the  density and temperature dependence of  masses of 
$\Lambda_c, \Sigma_c$ and $\Sigma_c^*$ as well as the binding energies of $\Lambda_c N $ and $\Lambda_c \Lambda_c$
systems have been studied using the chiral constituent quark model developed in Refs.
\cite{Valcarce:2005em,Vijande:2004he}.   The Kondo effect under the coexistence of spin and isospin exchange for heavy  hadrons ($\Sigma_c, \Sigma_c^*, \bar{D}, D^*$) has also been explored \cite{Yasui:2019ogk}.
The spectroscopic properties of singly  and doubly charmed baryons have been studied more recently in nuclear medium using the QCD sum rules
\cite{Turkan:2021dvu,Azizi:2019yoq}.
The medium modification of heavy baryon  masses is studied using the SU(3) chiral soliton model \cite{Won:2021pwb}
and pion mean field approach \cite{Ghim:2022zob}.

Apart from medium modification of masses and decay widths of hadrons, another set of properties such as  magnetic moments and electromagnetic form factors which  play a crucial role in probing the dynamics of internal structure of hadrons should also be investigated at finite density and temperature. 
 The European Muon Collaboration (EMC) effect suggested that the internal structure of nucleons is altered by the nuclear medium \cite{emc1}. 
 The scattering of polarized ($\vec{e},e^{'}p$) off $^{16}$O and $^{4}$He at Mainz Microtron (MAMI) and experiments at Jafferson lab \cite{ Strauch2003,Paolone2010,Malace2011} also indicated the modifications of the 
 electromagnetic properties of hadrons   in the nuclear medium.
The magnetic moments and electromagnetic form factors of light 
 \cite{Contreras2004,Araujo2004,Jones2000,Jun2007,lksharam2007,Sahu2002,Slaughter2011,Dahiya2003,JDey2000,Aliev2000}
and heavy baryons have been studied extensively in the free space.
The magnetic moments of heavy baryons have been calculated, for example, using quark model \cite{Jaffe_Patten_1973, Zenczykowski_1994}, MIT bag model \cite{Johnson_1979}, heavy quark effective theory \cite{Hosaka_1992}, chiral perturbation theory \cite{Bos_1999}, QCD sum rules \cite{Aliey_2010}, non-relativistic hyper central model \cite{Aiello:1996plb}, relativistic quark model \cite{Ebert_2008} and lattice QCD \cite{Alexandrou_2007} etc.

The literature on the in-medium modifications of these electromagnetic properties is limited to few models.
The magnetic moments of  octet
baryons in the dense matter have been investigated  using quark meson model (QMC) 
\cite{ryu3_plb2009,ryu_G2010}.
 The electromagnetic form factors of
baryons in the nuclear medium have been studied using
covariant spectator quark model \cite{Ramalho2013} and the combined approach of QMC and light front approach \cite{Araujo2018}.
The QMC model has also been used to calculate the medium dependent magnetic moments of the low-lying
charm and bottom baryons along-with octet and decuplet baryons in the symmetric nuclear medium considering contribution of valence quarks only (without configuration mixing)  \cite{Tsushimaptep2022} while to the best our knowledge no studies have been performed to calculate impact of medium comprising of nucleons and hyperons with finite isospin asymmetry on the masses and magnetic moments of heavy charmed baryons.

In the present work, we shall investigate the modifications in the masses and magnetic moments of  
spin$-{\frac{1}{2}}^+$
and ${\frac{3}{2}}^+$ singly and doubly charmed baryons 
in  strange hadronic medium (composed of nucleons and hyperons) at finite density and temperature using the combined approach of chiral SU(3) quark mean field model (CQMF) and chiral constituent quark model.
The chiral constituent quark model was initially used to study the magnetic moments of octet and decuplet baryons in the free space in SU(3) sector and was generalized to SU(4) case to calculate the magnetic moments of
spin$-{\frac{1}{2}}^+$
and spin$-{\frac{3}{2}}^+$ charmed baryons \cite{dahiya2010}.
The chiral constituent quark model takes into account the contributions from valence quarks, sea quarks spin polarization and orbital angular moments contribution of sea quarks for calculating the magnetic moments of baryons, considering configuration mixing \cite{dahiya2010,aarti}. 
We shall simulate the effects of finite density and temperature on the masses and magnetic moments of charmed baryons  using the CQMF model through the medium modified constituent quark masses \cite{wang_nuc2001}.
This model  has been used to study the modification in magnetic moments of octet and decuplet baryons in nuclear and strange hadronic matter \cite{harpreet_cpc2017,harpreet_epjp2019,harpreet_epja2018,harpreet_epjp2020,kumar2024}.  

The present paper has been organized as follows: In Sec. \ref{sec:chiral_mean_field} the details of the chiral SU(3) quark mean field model are given while in Sec. \ref{sec:magnetic} the detailed framework used to calculate the magnetic moments of spin$-\frac{1}{2}^+$ and spin$-\frac{3}{2}^+$ charmed baryons in the  chiral constituent quark model is explained. In Sec. \ref{sec:results}, the results and discussions of the current work are presented and finally the work is summarized  in Sec. \ref{sec:summary}.

	\section{Chiral SU(3) quark mean field model for asymmetric strange hadronic matter} \label{sec:chiral_mean_field}
	In this section we  describe the CQMF model which is used to calculate the medium modifications of 
	masses and magnetic moments of spin$-\frac{1}{2}^+$
and $\frac{3}{2}^+$ charmed baryons having at least one light $u, d$ or strange $s$ quark. CQMF model respects the low energy properties of QCD, i.e., model incorporates the 
spontaneous and explicit breaking of the chiral symmetry through the appropriate Lagrangian densities written in terms of non-strange scalar isoscalar  field $\sigma$,  strange scalar isoscalar field $\zeta$ and scalar isovector field $\delta$. 
The model also takes into account the broken scale invariance property of QCD
through the dilaton field $\chi$ incorporated using a logarithmic term in the Lagrangian density.
In the CQMF model, constituent quarks act as the fundamental degrees of freedom and these are confined inside the baryons through a confining potential.
The properties of constituent quarks and hence baryons
 are modified through the exchange of scalar fields $\sigma,\zeta$ and $\delta$
and vector field $\omega, \rho $ and $\phi$. Generally, as we shall see below, the scalar fields $\sigma,\zeta$ and $\delta$ appear in the definition of effective mass $m_i^*$  and vector fields 
$\omega, \rho $ and $\phi$ in the effective energies $e_i^*$ of quarks confined inside the baryons.

As mentioned before, in the present work we shall calculate the masses and magnetic moments of charmed baryons in the strange hadronic matter with finite temperature and isospin asymmetry. The strange hadronic matter considered in this work comprised of hyperons belonging to the octet of baryon in addition to nucleons, i.e., we consider $p,n,\Lambda, \Sigma^{\pm,0}, \Xi^{-,0}$.
The grand canonical potential
of CQMF describing the thermodynamics of strange hadronic medium  is   written as
\cite{wang_nuc2001}
\begin{equation}
		\Omega = -\frac{T}
		{(2\pi)^3} \sum_{i} \gamma_i
		\int_0^\infty d^3k\biggl\{{\rm ln}
		\left( 1+e^{- [ E^{\ast}_i(k) - \nu_i^* ]/T}\right) \\
		+ {\rm ln}\left( 1+e^{- [ E^{\ast}_i(k)+\nu_i^* ]/T}
		\right) \biggr\} -{\cal L}_{M}-{\cal V}_{\text{vac}}, 
		\label{Eq_therm_pot1}  
\end{equation}
where $T$ is the temperature of the medium, $\gamma_i = 2$ is the degeneracy factor and summation $i$ is over all the members of octet baryons. 
Further,	$E^{\ast }(k)=\sqrt{M_i^{\ast 2}+k^{2}}$ is the effective energy of the baryon defined in terms of effective mass $M_i^{\ast}$.
In the CQMF model, 
	the effective mass, $M_i^{\ast}$ of baryon is written in terms of effective energy ${\cal E}_i^*$ and spurious center of mass momentum $p_{i \, \text{cm}}^*$ \cite{barik1,barik2} through the relation
	\begin{align}
		M_i^*=\sqrt{{\cal E}_i^{*2}- \langle p_{i \, \text{cm}}^{*2}\rangle} = \sqrt{\left(\sum_j n_{ji}e_j^*+E_{i \, \text{spin}}\right)^2- \langle p_{i \, \text{cm}}^{*2}\rangle}. \label{baryonmass}
	\end{align} 
In the above equation $n_{ji}$ represents the number of quarks of type $j$ in the $i^{th}$ baryon
	and $e_j^*$ is the energy of that quark.
Also, the term $E_{i \, \text{spin}}$ 
	contributes as a correction to baryon energy due to spin-spin interaction and is fitted to obtain correct vacuum masses of octet baryons.
	The effective chemical potential $\nu_i^*$ of baryons is related to the chemical potential $\nu_i$ through  \cite{wang_nuc2001}
	\begin{align}
		\nu_i^* = \nu_i - g_{\omega}^i\omega -g_{\rho}^i I^{3i} \rho-g_{\phi}^i\phi.
	\end{align}
In Eq. (\ref{Eq_therm_pot1}), the term $
	{\cal L}_{M} \, = 
	{\cal L}_{X} \,+\, {\cal L}_{V} \,+\, {\cal L}_{\chi SB}\,
	$ is comprised of contributions
due to self-interactions of scalar mesons ${\cal L}_{X}$, the vector meson self-interactions, $ {\cal L}_{V}$, and the explicit symmetry breaking term, ${\cal L}_{\chi SB}$ of the model. Individually, the expressions of these terms are written as
\begin{align}
		{\cal L}_{X} =& -\frac{1}{2} \, k_0\chi^2
		\left(\sigma^2+\zeta^2+\delta^2\right)+k_1 \left(\sigma^2+\zeta^2+\delta^2\right)^2
		+k_2\left(\frac{\sigma^4}{2} +\frac{\delta^4}{2}+3\sigma^2\delta^2+\zeta^4\right)\nonumber \\ 
		&+k_3\chi\left(\sigma^2-\delta^2\right)\zeta 
		-k_4\chi^4-\frac14\chi^4 {\rm ln}\frac{\chi^4}{\chi_0^4} +
		\frac{\xi}
		3\chi^4 {\rm ln}\left(\left(\frac{\left(\sigma^2-\delta^2\right)\zeta}{\sigma_0^2\zeta_0}\right)\left(\frac{\chi^3}{\chi_0^3}\right)\right), \label{scalar0}
	\end{align}    
	\begin{equation}
		{\cal L}_{V}=\frac{1}{2} \,
		\left( \frac{\chi}{\chi_0}\right)^2 \left(
		m_\omega^2\omega^2+m_\rho^2\rho^2+m_\phi^2\phi^2\right)+g_4\left(\omega^4+6\omega^2\rho^2+\rho^4+2\phi^4\right), 
		\label{vector}
	\end{equation}
and 
\begin{equation}\label{L_SB}
	{\cal L}_{\chi SB}=	\left( \frac{\chi}{\chi_0}\right)^2\left[m_\pi^2f_\pi\sigma +
		\left(
		\sqrt{2} \, m_K^2f_K-\frac{m_\pi^2}{\sqrt{2}} f_\pi\right)\zeta\right],
	\end{equation}
	respectively.
	Through the logarithmic terms   of 
	Eq. (\ref{scalar0}), the trace anomaly property of QCD leading  to the trace of energy momentum tensor proportional to fourth power of dilaton field, $\chi$,  is introduced in the CQMF model \cite{papag_1999}.
	The order of magnitude about which the value of parameter  $\xi$ is taken into the calculations is generally determined from 
	QCD $\beta$-function at one loop level for three colors and three flavors \cite{papag_1999}. 
	The parameters $k_0, k_1, k_2, k_3$ and $k_4$   are calculated using $\pi$ meson mass, $m_{\pi}$, $K$ meson mass, $m_K$, and the average mass of $\eta$ and $\eta^{'}$ mesons \cite{wang_nuc2001}. 
Also, $\sigma_0$ and $\zeta_0$ are the vacuum expectation values of scalar fields $\sigma$ and $\zeta$, respectively and are related to   pion and kaon decay constants through relations, $
\sigma_0= -f_{\pi}$ and $\zeta_0= \frac{1}{\sqrt{2}}\left( f_{\pi}-2f_{K}\right)$.

As discussed earlier, in the CQMF model, quarks confined inside the baryons interact through the exchange of scalar and vector mean fields.  For the quark field $\Psi_{ji}$ the Dirac equation is written as  \cite{wang_nuc2001}
	\begin{equation}
		\left[-i\vec{\alpha}\cdot\vec{\nabla}+\chi_c(r)+\beta m_j^*\right]
		\Psi_{ji}=e_j^*\Psi_{ji}. \label{Dirac}
	\end{equation} 
In the above equation, $\chi_c =  \frac14 k_{c} \, r^2(1+\gamma^0)$ is the confining potential through which the quarks are confined inside the baryons in the CQMF model. The effective quark mass $m_{j}^*$ and effective energy $e_j^*$ are  defined in terms of scalar and vector  fields through relations
	\begin{equation}
		m_j^*=-g_\sigma^j\sigma - g_\zeta^j\zeta - g_\delta^j I^{3j} \delta + m_{j}^0, \label{qmass}
	\end{equation}
and
\begin{equation}
e_j^*=e_j-g_\omega^j\omega-g_\rho^j I^{3j}\rho -g_\phi^j\phi\,.
\label{eq_eff_Equark1}
\end{equation}
The values of confining parameter $k_c$ and couplings are fitted to the  binding
energy $-16.0$ MeV at nuclear saturation density $\rho_0 = 0.16$ fm$^{-3}$.
This leads to the vacuum mass for the light quarks $u$ and $d$ equal to $256$ MeV.
The value of $m_{j}^0$ is zero for non-strange `$u$' and `$d$' quarks whereas for the strange `$s$'  quark this is  equal to $77$ MeV so as
to  obtain the vacuum constituent strange quark mass at around $477$ MeV.
To obtain the density and temperature dependent values of 
scalar and vector fields, 
the grand canonical potential defined by Eq. (\ref{Eq_therm_pot1}) is minimized with respect these fields, i.e., 
	\begin{align}
		\frac{\partial \Omega}{\partial \sigma} = 
		\frac{\partial \Omega}{\partial \zeta} =
		\frac{\partial \Omega}{\partial \delta} =
		\frac{\partial \Omega}{\partial \chi} =
		\frac{\partial \Omega}{\partial \omega} =
		\frac{\partial \Omega}{\partial \rho} =
		\frac{\partial \Omega}{\partial \phi} =
		0.
		\label{eq:therm_min1}
	\end{align}
The coupled system of equations obtained for the scalar and vector fields are solved 
for given value of isospin asymmetry parameter, $I_a = -\frac{\Sigma_i I_{3i} \rho_{i}}{\rho_{B}}$,  and the strangeness fraction, $f_s = \frac{\Sigma_i \vert s_{i} \vert \rho_{i}}{\rho_{B}}$.
Here,  $I_{3i}$ and $\vert s_{i} \vert$ refer to the third component of isospin quantum number and number of strange quarks in $i^{th}$ baryon, respectively.  
 
	\section{Magnetic Moments  of  spin$-\frac{1}{2}^+$
	and $\frac{3}{2}^+$ charmed baryons}
	\label{sec:magnetic}
We now discuss the chiral constituent quark model extended to SU(4) sector to investigate the magnetic moments of low lying 
spin$-\frac{1}{2}^+$
	and $\frac{3}{2}^+$ charmed baryons 
 \cite{dahiya2010,aarti}.
 In the SU(4) case, the baryons whose quark contents are out of four flavors,
 $u, d, s$ or $c$ quarks, are 
 classified into symmetric, mixed symmetric/antisymmtic and singlets  as: $\mathbf{4} \times \mathbf{4} \times \mathbf{4} = \mathbf{20}_{S} + \mathbf{20}_{MS} + \mathbf{20}_{MA} + \mathbf{\bar{4}}$. Further, mixed symmetric (MS) and mixed antisymmetric (MA) $\mathbf{20}-$plet give us 
 spin$-\frac{1}{2}^+$ $\mathbf{8} + \mathbf{6} + \mathbf{\bar{3}} + \mathbf{3}$, baryon states.  Here, $\mathbf{8}$ belongs to the octet of baryons, $p,n,\Sigma^{\pm,0}, \Xi^{-,0}$,
 composed of $u,d$ and $s$ quarks, $\mathbf{6}$ and $\mathbf{\bar{3}}$ have the baryons, $\Sigma_c^{++},
 \Sigma_c^{+},\Sigma_c^{0},\Xi_c^{\prime +}, \Xi_c^{\prime 0}, \Omega_c^0$,
  and $\Lambda_c^+, \Xi_c^+, \Xi_c^0$, respectively, 
 with single charm quark. Also, $\mathbf{3}$ have baryons $\Xi_{cc}^{++},\Xi_{cc}^{+},\Omega_{cc}^{+}$  with double charm quark. 
 From the $\mathbf{20}_S-$plet we shall obtain spin$-\frac{3}{2}^+$ $\mathbf{10}+\mathbf{6}+\mathbf{3}+\mathbf{1}$, baryon states, with $\mathbf{10}$ being the  decuplet  baryons having $u,d$ or $s$ quark, i.e., $\Delta^{++,+,0,-}, \Sigma^{*\pm,0},
 \Xi^{*-,0}, \Omega^-$.
Also, $\mathbf{6}$
belongs to the singly charmed,  $\Sigma_c^{*++},
 \Sigma_c^{*+},\Sigma_c^{*0},\Xi_c^{* +}, \Xi_c^{* 0}, \Omega_c^{*0}$,
$\mathbf{3}$ has
doubly charmed,
$\Xi_{cc}^{*++},\Xi_{cc}^{*+},\Omega_{cc}^{*+}$, and  $\mathbf{1}$ has  triply charmed
$\Omega_{ccc}^{*++} $ baryon resonances, respectively.
 In the present investigation, we shall calculate the masses and  magnetic moments of 
 spin$-\frac{1}{2}^+$
 	and $\frac{3}{2}^+$ singly and doubly charmed baryons.
 
In the chiral constituent quark model,   model, the emission of Goldstone bosons  takes place from the internal constituent quarks. The emitted Goldstone bosons
splits into the quark-antiquark ($q\bar{q}$) pairs, i.e., the process, $q_{\pm} \rightarrow GB^0 + q^{\prime}_{\mp} \rightarrow (q\bar{q}^{\prime}) + q^{\prime}_{\mp}$
takes place 
	\cite{{chengsu3},{cheng1},{song}}.
In the SU(4) case, the interactions between the quarks and 16 Goldstone bosons (15-plet and one singlet) is described through the  Lagrangian density  ${\cal L} = g_{15} \bar{q} \Phi q$  .  Here, the field $\Phi$
belongs to the GBs and can be expressed as
	\begin{equation}
	\small
		\Phi=\left(\begin{array}{cccc}\frac{\pi^o}{\sqrt{2}}+\beta \frac{\eta}{\sqrt{6}}+\zeta^{'} \frac{\eta^{\prime}}{4\sqrt{3}} - \gamma \frac{\eta_c}{4} & \pi^{+} & \alpha K^{+}  & \gamma\bar{D}^0\\
			 \pi^{-} & -\frac{\pi^o}{\sqrt{2}}+\beta \frac{\eta}{\sqrt{6}}+\zeta^{'} \frac{\eta^{\prime}}{4\sqrt{3}} - \gamma \frac{\eta_c}{4} & \alpha K^0 & \gamma D^- \\ 
			 \alpha K^{-} & \alpha \bar{K}^0 & -\beta \frac{2 \eta}{\sqrt{6}}+\zeta^{'} \frac{\eta^{\prime}}{4\sqrt{3}} - \gamma \frac{\eta_c}{4}& \gamma D_{s}^{-} \\
		\gamma D^0 & \gamma D^{+} & \gamma D_{s}^{+} & - \zeta^{'} \frac{3 \eta^{\prime}}{4\sqrt{3}} + \gamma \frac{3\eta_c}{4}\\

\end{array}\right).
	\end{equation}
The breaking  of SU(4) symmetry is introduced by considering the hierarchy $m_c> m_s> m_{u, d}$.
	The parameter $a=\left|g_{15}\right|^2$ represents the transition probability of chiral fluctuation of the splitting $u(d) \rightarrow$ $d(u)+\pi^{+(-)}$.  Also, $a \alpha^2, a \beta^2, a \zeta^{'2}$, and $a \gamma^2$ represent the probabilities of transitions of $u(d) \rightarrow s+K^{-(0)}, u(d, s) \rightarrow$ $u(d, s)+\eta$ and $ u(d, s) \rightarrow u(d, s)+\eta^{\prime}$ and $u(d) \rightarrow c + \bar{D}^0(D^-)$ respectively
	\cite{dahiya2010,aarti}.
	
	The magnetic moment of baryons have been calculated using the chiral constituent quark model, considering the contributions of valence  and sea quark spin polarizations as well as the orbital angular momentum of sea quark 	\cite{{chengsu3},{cheng1},{song}}.
	To obtain the net magnetic moment of charmed baryons in the strange medium, the modifications in all these three contributions will be considered in the present calculations.
Thus, we write
	\begin{align}
		\mu_{B}^*= \mu_{\text{B,val}}^*+\mu_{\text{B,sea}}^*+\mu_{\text{B,orbital}}^*, \label{magtotal}
	\end{align}
	where $\mu_{\text{B, val}}^*$, $\mu_{\text{B, sea}}^*$ and $\mu_{\text{B, orbital}}^*$ represent the effective  contribution from valence quarks, sea quarks and orbital angular momentum of sea quarks, respectively. 
The values  $\mu_{\text{B, val}}^*$ and $\mu_{\text{B, sea}}^*$ are 
	calculated using  the  spin polarizations of valence quarks, $\Delta q_{\mathrm{val}}$ and sea quarks,  $\Delta q_{\mathrm{sea}}$, and the effective  quark magnetic moment $\mu_q^*$, i.e.,
	\begin{align}
		\mu_{\mathrm{B,val}}^* & =\sum_{q=u, d, s} \Delta q_{\mathrm{val}} \mu_q^*, 
		\label{mu_val1}
		\\
		\mu_{\mathrm{B,sea}}^* & =\sum_{q=u, d, s} \Delta q_{\mathrm{sea}} \mu_q^*.
		\label{mu_sea1} 
\end{align}	
The expressions of spin polarizations, $\Delta q_{\mathrm{val}}$ and  $\Delta q_{\mathrm{sea}}$ considering the configuration mixing can be found in 
Ref.\cite{dahiya2010}.  To calculate the medium modified magnetic moments of quarks, we shall use the following relations
	\begin{equation}
	\mu_d^* = -\frac{1}{2} \mu_u^* =-\left(1-\frac{\Delta M}{M_B^*}\right),~~  \mu_s^*=-\frac{m_u^*}{m_s^*}\left(1-\frac{\Delta M}{M_B^*}\right),~~
	\mu_c^{*}=\frac{2m_u^*}{m_c}\left(1-\frac{\Delta M}{M_B^*}\right).  \label{magandmas}  
	\end{equation} 
Above expressions for the magnetic moments of quarks are motivated to incorporate the impact of quark confinement and relativistic corrections more consistently into the calculations of magnetic moments of baryons    
	and are known as mass adjusted magnetic moments of constituent quarks \cite{aarti}.
	Here, $M_B^*$ is the effective mass of baryon calculated using Eq. (\ref{baryonmass}). Also, 
	$\Delta M= M_{\rm vac}-M_{B}^{*}$, where $M_{\rm vac}$ is the vacuum baryon mass.
	The magnetic moment due to orbital angular momentum of sea quarks is calculated using
	\begin{align}
			\mu_{\text {B,orbit }}^*	  =\sum_{q=u, d, s} \Delta q_{\mathrm{val}} \mu^*\left(q_{+} \rightarrow q_{-}^{\prime}\right).
			\label{mu_seaorbit1}
		\end{align}
	In the above equation, $\mu^*\left(q_{+} \rightarrow q_{-}^{\prime}\right)$ represents the orbital moment for
	each chiral fluctuation. These are   further multiplied by the probability for all such chiral fluctuations to take place, originated from given valence quark
	\cite{Dahiya2003}.
For the orbital moments of four flavors $u,d,s$ and $c$, following expressions in terms of effective quark masses can be
obtained
 \cite{dahiya2010}
	\begin{align}
		{\left[\mu^*\left(u_{+} \rightarrow\right)\right]=} & a\left[\frac{3 m_u^{*2}}{2 M_\pi\left(m_u^{*}+M_\pi\right)}-\frac{\alpha^2\left(M_K^2-3 m_u^{*2}\right)}{2 M_K\left(m_u^{*}+M_K\right)}\right.  +\frac{\beta^2 M_\eta}{6\left(m_u^{*}+M_\eta\right)} \nonumber\\ &
		\left.+\frac{\zeta^{'2} M_{\eta^{\prime}}}{48\left(m_u^{*}+M_{\eta^{\prime}}\right)} + \frac{\gamma^{2} M_{\eta_{c}}}{16\left(m_u^{*}+M_{\eta_{c}}\right)} + \frac{\gamma^2 M_D}{m^*_u + M_D}  \right] \mu_N, 
	\end{align} 
	
	\begin{align}
		{\left[\mu^*\left(d_{+} \rightarrow\right)\right]=} & a \frac{m_u^*}{m_d^*}\left[\frac{3\left(M_\pi^2-2 m_d^{*2}\right)}{4 M_\pi\left(m_d^{*2}+M_\pi\right)}
		-\frac{\alpha^2 M_K}{2\left(m_d^*+M_K\right)} 
		+ \frac{\gamma^2 \left(2M_D^2-3m^*_d)\right)}{2M_D\left(m^*_d + M_D\right)}\right.\nonumber \\ 
		& -\frac{\beta^2 M_\eta}{12\left(m_d^*+M_\eta\right)}  \left.-\frac{\zeta^{'2} M_{\eta^{\prime}}}{96\left(m_d^*+M_{\eta^{\prime}}\right)} + \frac{\gamma^2M_{\eta_c}}{32\left(m^*_d + M_D\right)}\right] \mu_N,
	\end{align}

	\begin{align}
		{\left[\mu^*\left(s_{+} \rightarrow\right)\right]=} & a \frac{m_u^*}{m_s^*}\left[\frac{\alpha^2\left(M_K^2-3 m_s^{*2}\right)}{2 M_K\left(m_s^*+M_K\right)}  -\frac{\beta^2 M_\eta}{3\left(m_s^*+M_\eta\right)} \right. 
		+\frac{\gamma^2 \left(2M_{D_s}^2-3m^{*2}_s)\right)}{2M_D\left(m^*_s + M_{D_s}^2\right)} \nonumber\\& \left.-\frac{\zeta^{'2} M_{\eta^{\prime}}}{96\left(m_s^*+M_{\eta^{\prime}}\right)}  - \frac{\gamma^2M_{\eta_c}}{32\left(m^*_s + M_D\right)} \right] \mu_N,
	\end{align}
	
	\begin{align}
		{\left[\mu^*\left(c_{+} \rightarrow\right)\right]=} & a \frac{m_u^*}{m_c}\left[
		\frac{\gamma^2\left(M^2_D + 3m_c^2\right)}{2M_D\left(m_c + M^2_D\right)}
	- \frac{\gamma^2\left(M^2_{D_s} + 3m_c^2\right)}{2M_d\left(m_c + M^2_{D_s}\right)} 
	+ \frac{3\zeta^{\prime 2}M_{\eta^{\prime}}}{16\left( m_c + M_{\eta^{\prime}}\right)}
	 +\frac{9\gamma^2M_{\eta_c}}{16(m_c+M_{\eta_c})}\right] \mu_N,
	\end{align}
	where $\mu_N$ is nuclear magneton.
	%
	
	
\section{Numerical results} \label{sec:results}
In this section, we shall discuss the numerical results on the medium modifications of masses and magnetic moments of
spin$-\frac{1}{2}^{+}$ and 
$-\frac{3}{2}^{+}$ charmed baryons in the strange hadronic medium.
The magnetic moments of singly and doubly charmed baryons
are calculated using the chiral constituent quark model and medium modifications are induced through the effective  constituent quark masses which are calculated using the chiral SU(3) quark mean field model at finite density and temperature.
In Table \ref{table1_para}, we have given values of different parameters used in the present calculations. 
%
The relations to obtain the coupling of hyperons with scalar and vector fields are given in Table \ref{table_coup_vec}.  The parameters 	$ x_\Lambda,  x_\Sigma$ and  $x_\Xi$ appearing in  expressions of coupling of hyperons with vector mesons are fitted to obtain the  hyperons potentials  $U_\Lambda = -30$ MeV,
	$U_\Sigma = 30$ MeV and $U_\Xi = -18$ MeV   \cite{millener2001,yamamoto1988,
		mares1995,bart1999,fukuda1998}. The value of  $g_\phi^N$ which gives the coupling of $\phi$ meson with nucleons is set to zero.
	\begin{table}
		\begin{tabular}{|c|c|c|c|c|} 
			\hline 
			$k_0$ & $k_1$ & $k_2$ & $k_3$ & $k_4$  \\ 
			\hline 
			$4.94$ & $2.12$ & $-10.16$ & $-5.38$ & $-0.06$  \\ 
			\hline
			$\sigma_0$ (MeV) & $\zeta_0$ (MeV) & $\chi_0$ (MeV) & $\xi$ & $\rho_0$ ($\text{fm}^{-3}$)  \\ 
			\hline 
			$-92.8$ & $-96.5$ & $254.6$ & 6/33 & $0.16$  \\ 
			\hline
			$g_{\sigma}^u=g_{\sigma}^d$ & $g_{\sigma}^s = g_{\zeta}^u=g_{\zeta}^d$ & $g_{\delta}^u$ & $g_{\zeta}^s = g_s$ & $g_4$\\ 
			\hline 
			2.72 & 0 & 2.72 & 3.847
			& 37.4 \\ 
			\hline
			$g_{\delta}^p = g_{\delta}^u$  & $g_{\omega}^N = 3g_{\omega }^u$  & $g_{\rho}^p$  & $m_{\pi}$ (MeV) & $m_K$ (MeV) \\
			\hline
			2.72 & 9.69 & 8.886 & 139 & 494  \\ 
			\hline 
		\end{tabular}
		\caption{Values of various parameters used in the present work \cite{wang_nuc2001}.} \label{table1_para}
	\end{table}
\begin{table}
\begin{tabular}{|c|c|c|}
\hline
$g_\sigma^\Lambda = 2g_\sigma^u + g_\sigma^s$ & 
$g_\zeta^\Lambda = 2g_\zeta^u+g_\zeta^s$ &  
	$g_\delta^\Lambda = 0$\\
\hline
$g_\sigma^\Sigma = 2g_\sigma^u + g_\sigma^s$ & 
	$g_\zeta^\Sigma = 2g_\zeta^u+g_\zeta^s$ & 
		$g_\delta^\Sigma = g_\delta^p$ \\
\hline
$g_\sigma^\Xi = g_\sigma^u + 2g_\sigma^s$ & 
$g_\zeta^\Xi = g_\zeta^u+2g_\zeta^s$ &  
		$g_\delta^\Xi = g_\delta^p$\\
\hline
$g_\omega^\Lambda = x_\Lambda \frac{2}{3}g_\omega^N$ &
$g_\rho^\Lambda = 0$
		&
	$g_\phi^\Lambda=-\frac{\sqrt{2}}{3}g_\omega^N$ \\
	\hline
$g_\omega^\Sigma = x_\Sigma \frac{2}{3}g_\omega^N$ &
$g_\rho^\Sigma = \frac{2}{3}g_\omega^N$ &
$g_\phi^\Sigma=-\frac{\sqrt{2}}{3}g_\omega^N$ \\
\hline
$g_\omega^\Xi = x_\Xi \frac{1}{3}g_\omega^N$ &		$g_\rho^\Xi = \frac{2}{3}g_\omega^N$ &
		$g_\phi^\Xi=-\frac{2\sqrt{2}}{3}g_\omega^N$ \\
\hline
\end{tabular}
\caption{Coupling constants of hyperons with scalar and vector fields.}
\label{table_coup_vec}
\end{table}
In the  following  Sec. \ref{sec_quark_de_mass}, we shall discuss the medium modifications in the masses of constituent quarks and shall discuss the impact of finite density and temperature on the in-medium masses of spin$-\frac{1}{2}^+$ and spin$-\frac{3}{2}^{+}$ charmed baryons. Numerical results on the effective  magnetic moments of charmed baryons are discussed in Sec. \ref{sec_deculet_moments}.
	\subsection{Effective  quark and charmed baryon masses}
	\label{sec_quark_de_mass}
The impact of finite baryon density $\rho_B$, temperature, $T$, isospin asymmetry, $I_a$ and the strangeness fraction, $f_s$,  of the hadronic medium, on the effective masses of charmed baryons is calculated in the present work using Eq. (\ref{baryonmass}).
The procedure is as follows:
we first solve the coupled system of scalar fields $\sigma, \zeta$ and $\delta$, dilaton field $\chi$ and the vector fields 
$\omega, \rho$ and $\phi$, in the isospin asymmetric strange matter comprising of nucleons ($p,n$) and hyperons ($\Lambda, \Sigma^{\pm,0}, \Xi^{-,0}$) and the effective  values of  masses, $m_j^*$ and  energies,
$e_j^*$   are obtained 
using Eqs. (\ref{qmass}) and (\ref{eq_eff_Equark1}), respectively for $u,d$ and $s$ quarks. Now, using these effective
quark masses and energies, effective masses of the charmed baryons are calculated using Eq. (\ref{baryonmass}), fitting $E_{\text{spin}}$ to obtain their corresponding vacuum masses. Note that for a given charmed baryons, the medium modifications are simulated through their constituents  
$u, d$ or $s$ quarks only, whereas the  mass, $m_c$ of  heavy charm quark is not modified within the medium.

\begin{figure}[ht!] 
\includegraphics[width= 12 cm, height=14 cm]{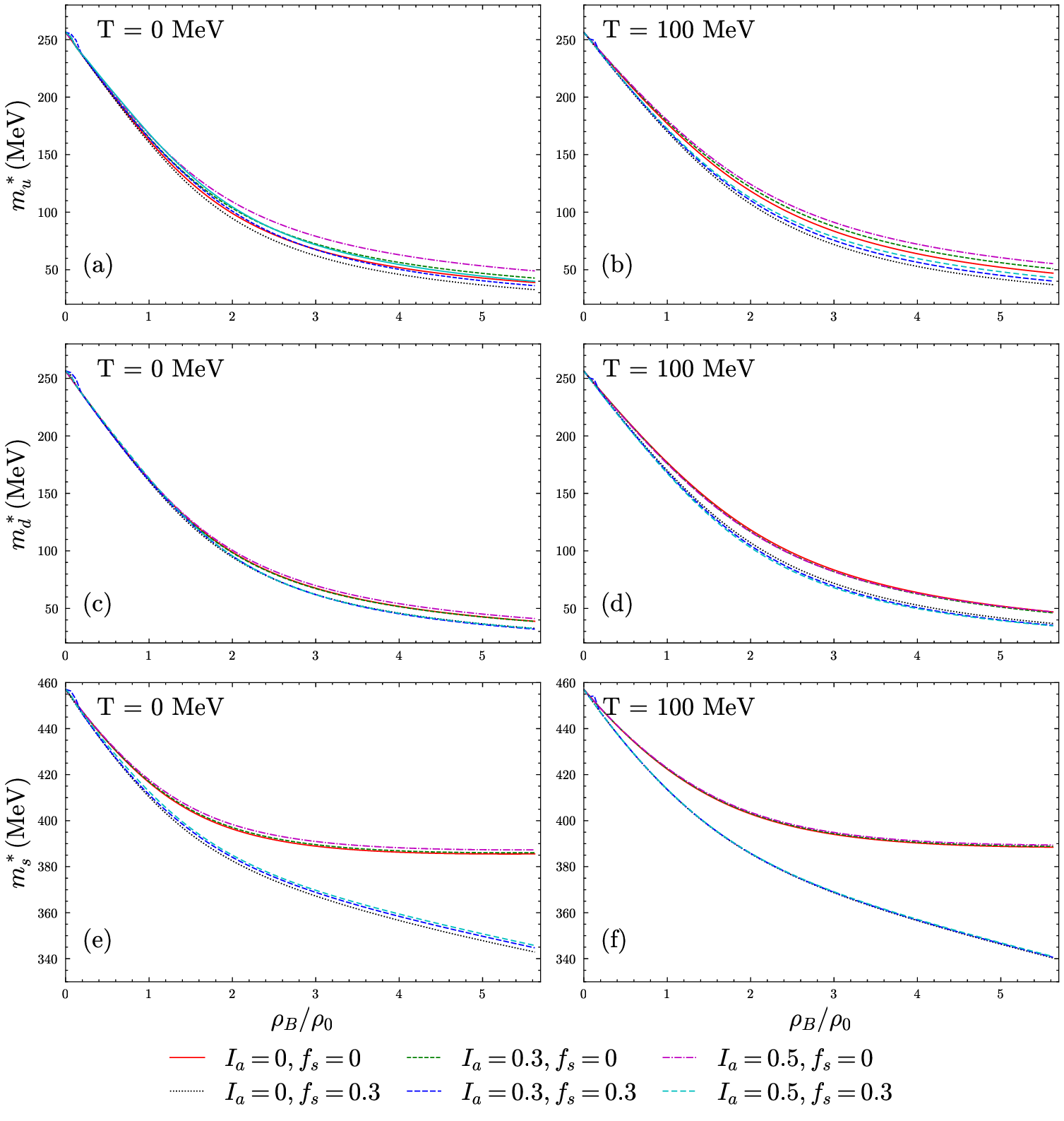}\hfill
\caption{Effective masses of constituent quarks $u, d$ and $s$ are shown as a function of baryon density $\rho_B$ (in units of  $\rho_0$) at temperatures $T = 0$ [subplots (a), (c), (e)] and $100$ [subplots (b), (d), (f)] MeV. Results are shown for isospin asymmetry $I_a = 0,\ 0.3,\ 0.5$ and strangeness fractions $f_s = 0,\ 0.3$.}
		\label{fig_massquarks}
	\end{figure}

For a thorough understanding of the  effective  masses of different charmed baryons, 
we first discuss the behavior of constituent quark masses for $u$, $d$ and $s$ quarks in the nuclear $(f_s = 0)$ and strange medium $(f_s \neq 0)$. 
The effective quark masses $m_u^*, m_d^*$ and $m_s^*$  with respect to baryon density ratio $\rho_B/\rho_0$ are shown in Fig. \ref{fig_massquarks}, at temperatures $T = 0$ (in subplots (a), (c) and (e)) and  $100$ MeV (in subplots (b), (d) and (f)), respectively.
In each subplot, the results of
nuclear medium, $f_s = 0$
are compared with the
strange matter, $f_s = 0.3$, for isospin asymmetry $I_a = 0,\ 0.3$ and $0.5$.
The effective  masses of quarks are observed to decrease with an increase in the baryon density $\rho_B$ of nuclear and strange medium. As expected, in comparison to the strange quark $s$, the light quarks $u$ and $d$  undergo comparatively higher mass modifications as a function of density. 
As can be seen from 
Eq. (\ref{qmass}), the effective quark masses are linearly proportional to the scalar fields, $\sigma, \zeta$ and $\delta$. As an example,  for the light $u$ or $d$ quark, $g_\zeta^{u} = g_\zeta^{d} = 0$,  and $m_u^{*} \propto \sigma$, whereas, for the strange $s$ quark, $g_\sigma^s =0$ and therefore, 
$m_s^{*} \propto \zeta$.  This implies that the light quark condensates are modified more significantly in comparison to the strange quark condensates as a function of density of nuclear and strange matter which result in the relatively higher change in the masses $m_u^*$ and $m_d^{*}$ as compared to $m_s^{*}$.
In the isospin symmetric medium ($I_a = 0$), the $\delta$ field is zero and $m_u^{*} = m_d^{*}$, whereas for a finite value of $I_a$, because of non-zero value of $\delta$, the mass splitting takes place among $u$ and $d$ quarks (see the third term of Eq. (\ref{qmass})). As a function of isopsin asymmetry $I_a$ (keeping $\rho_B, T, f_s$ fixed), the effective  mass $m_d^{*}$ is observed to decrease whereas the  mass $m_u^{*}$ increases. An increase of strangeness fraction in the medium causes further decrease in the effective masses of quarks. The mass of strange $s$ quark undergoes higher change as a function of $f_s$ in comparison to $u$ and $d$ quarks. As the temperature of nuclear and strange matter is increased from zero to finite value, the effective quark masses are observed to increase, however, at sufficiently high temperature they will decrease as expected for the restoration of chiral symmetry at high temperatures.

\begin{figure}[ht!] 
		\includegraphics[width= 16 cm, height=18 cm]{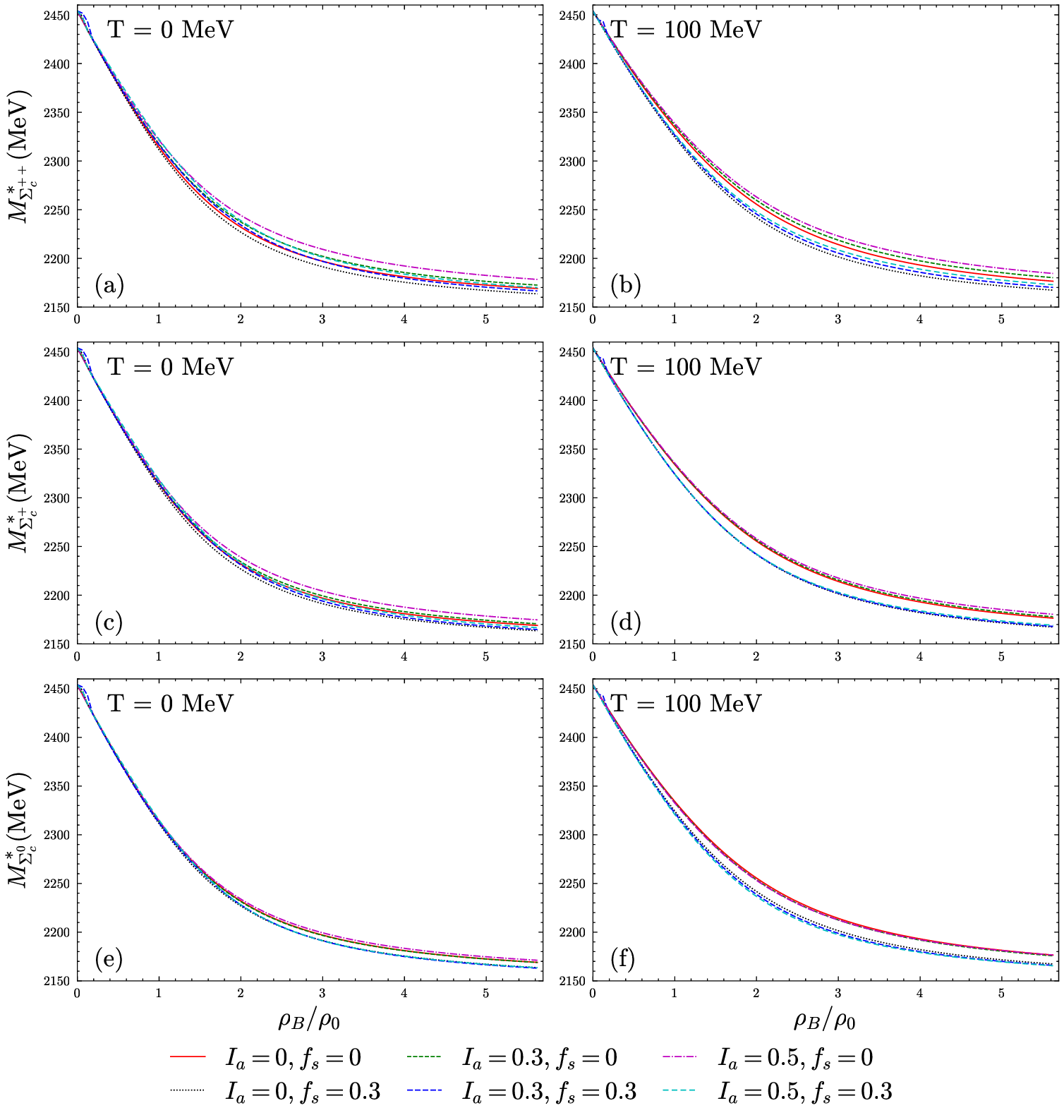}\hfill
		\caption{Effective masses of 
		 spin$-\frac{1}{2}^+$
	singly charmed,	  $\Sigma^{++}_{c},\ \Sigma^+_c $ and $\Sigma^0_c$ baryons are shown as a function of baryon density $\rho_B$ (in units of $\rho_0$) at temperatures $T = 0$ [subplots (a), (c), (e)] and $100$ [ subplots (b), (d), (f)] MeV. Results are shown for isospin asymmetry $I_a = 0,\ 0.3,\ 0.5$ and strangeness fractions $f_s = 0,\ 0.3$.}
		\label{fig_massSigmaC}
	\end{figure}

\begin{figure}[ht!] 
		\includegraphics[width= 16 cm, height=18 cm]{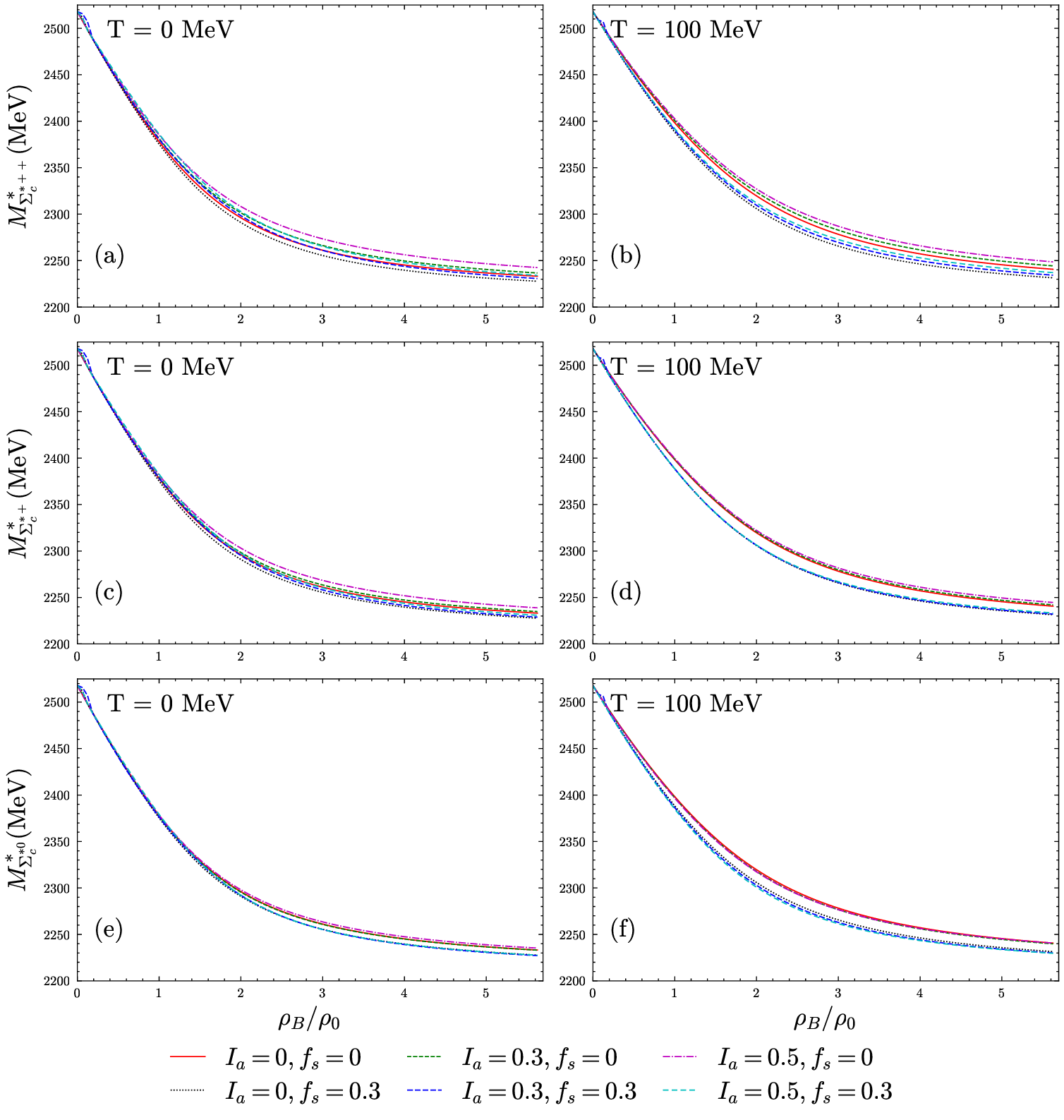}\hfill
		\caption{Effective masses of
		 spin$-\frac{3}{2}^+$
	singly charmed, $\Sigma^{*++}_c,\ \Sigma^{*+}_c$ and $\Sigma^{*0}_c$ resonance baryons are shown as a function of baryon density $\rho_B$ (in units of    $\rho_0$) at temperatures $T = 0$ [subplots (a), (c), (e)] and $100$ [subplots (b), (d), (f)] MeV. Results are shown for isospin asymmetry $I_a = 0,\ 0.3,\ 0.5$ and strangeness fractions $f_s = 0,\ 0.3$.}
		
		\label{figmassSigmaStar}
	\end{figure}

In Figs. \ref{fig_massSigmaC} and \ref{figmassSigmaStar},  the variations of effective  masses of singly charmed spin$-\frac{1}{2}^+$ $\left(\Sigma_c^{++},\Sigma_c^{+}, \Sigma_c^{0}\right)$  and $\frac{3}{2}^+$ $\left(\Sigma_c^{*++},\Sigma_c^{*+}, \Sigma_c^{*0}\right)$  baryons as a function of density $\rho_B$ of the
 medium are shown.
The behavior of effective  constituent quarks masses is reflected in the trends of masses of the baryons as well. The masses of all the singly charmed $\Sigma_c$ baryons decrease with the increase in the density with the rate of decrease being higher at the lower density values.
The values of mass shift of spin$-\frac{1}{2}^+$ charmed baryons are given in Tables \ref{tab:masses_1half2T0}
  and \ref{tab:masses_1half2T100}, at temperatures $T = 0$ and 100 MeV, respectively.
Tables \ref{tab:masses_3half2T0}
  and \ref{tab:masses_3half2T100}  
  have the values of mass shift for  spin$-\frac{3}{2}^+$ charmed baryons.
 It is also observed that with the increase in isospin asymmetry the mass of $\Sigma_c^{++}$ and $\Sigma_c^{*++}$ baryons increases, at a given value of $\rho_B$, $T$ and $f_s$. This is because  both $\Sigma_c^{++}$ and $\Sigma_c^{*++}$ are composed of two $u-$quarks in addition to a $c-$quark and the mass of $u-$quarks is observed to be increasing with isospin asymmetry, $I_a$, as seen in Fig. \ref{fig_massquarks}. As the number of $d-$quarks are increased by replacing the $u-$quarks the increase in masses becomes slower with the increase in $I_a$ values which explains the trend observed in $\Sigma_c^{+}, \Sigma_c^{*+}, \Sigma_c^{0}, $ and $\Sigma_c^{*0}$. 
	In the symmetric matter, ($I_a =0$), at baryonic density $\rho_B = 3\rho_0$, $T=0$  and $f_s$ values $0(0.3)$, the effective masses of  singly charmed $\frac{1}{2}^+$ baryons $(\Sigma_c^{++}, \Sigma_c^{+}, \Sigma_c^{0})$ are observed to be 2197.00 (2191.33) MeV while those of $\frac{3}{2}^+$ baryons $(\Sigma_c^{*++}, \Sigma_c^{*+}, \Sigma_c^{*0})$ are found to be 2261.14(2255.47) MeV, respectively. With an increase in the value of $I_a$ splitting in the masses of singly charmed $\frac{1}{2}^+$ and $\frac{3}{2}^+$ baryon multiplet takes place. In asymmetric matter with $I_a = 0.3$, at $\rho_B = 3\rho_0$, $T=0$  and $f_s$ values $0(0.3)$, the effective masses
	(in MeV) of  $\Sigma_c^{++}, \Sigma_c^{+}, \Sigma_c^{0}$ are observed to be 2202.07(2196.83), 2199.35(2193.99) and 2196.62(2191.16), while for $\Sigma_c^{*++}, \Sigma_c^{*+}, \Sigma_c^{*0}$ the mass values are found to be 2266.21(2260.97), 2263.48(2258.13) and 2260.76 (2255.30), respectively.
As one can observe, in
	case of $\Sigma_c^{0}$
	and $\Sigma_c^{*0}$ baryons,
	which are composed of two light $d$ quarks,
	the effective mass decreases slightly as a 
	function of isospin asymmetry $I_a$ of the medium, i.e., more negative mass shift is observed for these baryons as compared to partners of this $\Sigma_c$ isospin multiplet (see Tables also). 
	From Figs. \ref{fig_massSigmaC} and \ref{figmassSigmaStar}, it is also observed that the increase of temperature from $T = 0$ to $T=100$ MeV causes suppressed decrease in the mass of singly charmed $\frac{1}{2}^+$ and $\frac{3}{2}^+$ baryons, i.e., lower mass drop is observed from the vacuum value, at finite temperature.
For given density $\rho_B$, temperature $T$ and isospin asymmetry $I_a$ of the medium, the increase in strangeness fraction causes further drop in the masses of both spin$-\frac{1}{2}^+$ and
$\frac{3}{2}^+$  charmed baryons.

\begin{figure}[ht!] 
 	\includegraphics[width= 16 cm, height=20 cm]{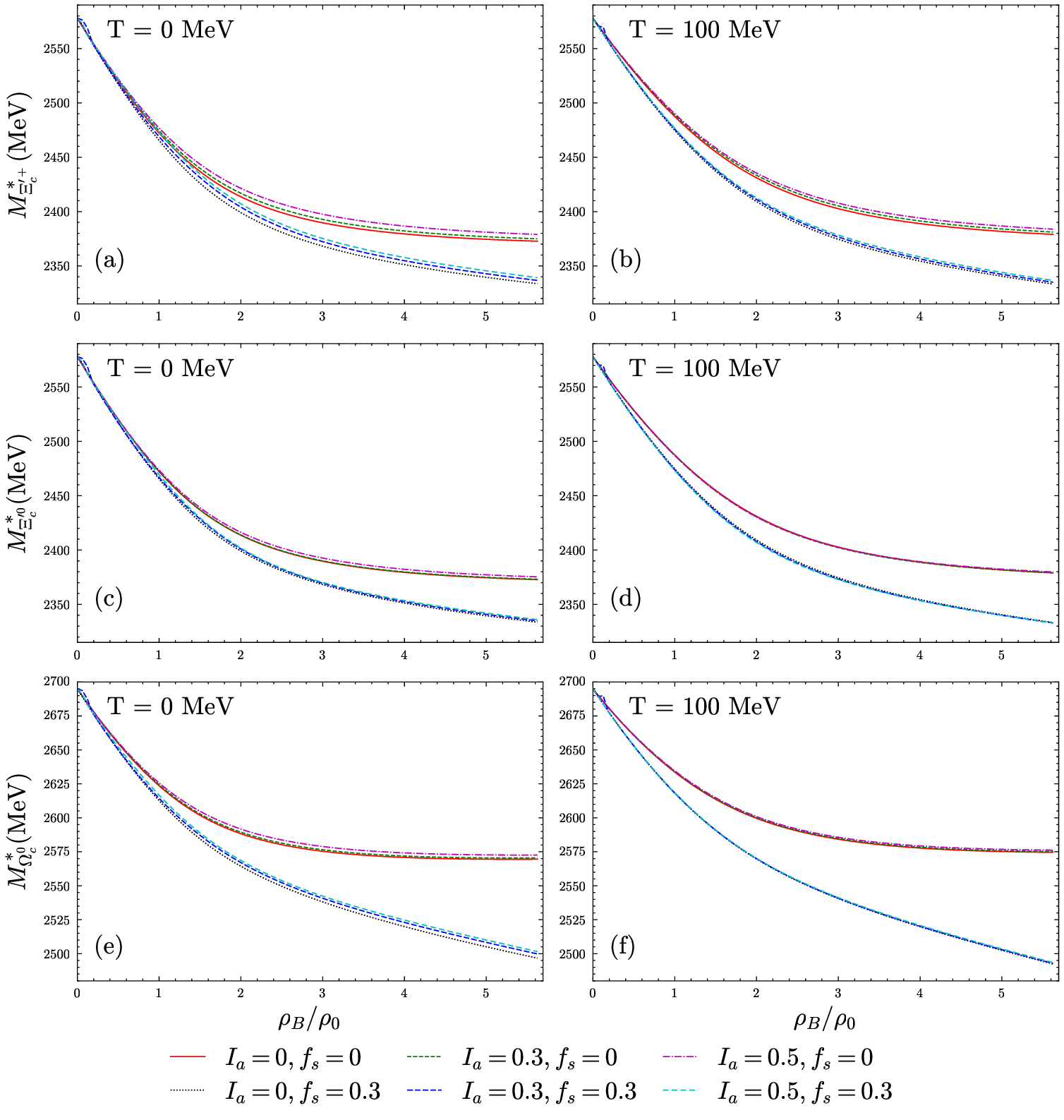}\hfill
 	\caption{Effective masses of  spin$-\frac{1}{2}^+$
 		singly charmed,  $\Xi^{\prime +}_c,\ \Xi^{\prime0}_c$ and $\Omega^0_c$ baryons are shown as a function of baryon density $\rho_B$ (in units   density $\rho_0$) at temperatures $T = 0$ [subplots (a), (c), (e)] and $100$ [subplots (b), (d), (f)] MeV. Results are shown for isospin asymmetry $I_a = 0,\ 0.3,\ 0.5$ and strangeness fractions $f_s = 0,\ 0.3$.}
 	\label{fig_charmed_half_primeXipXi0_Omega0}
 \end{figure}

\begin{figure}[ht!] 
 	\includegraphics[width= 16 cm, height=20 cm]{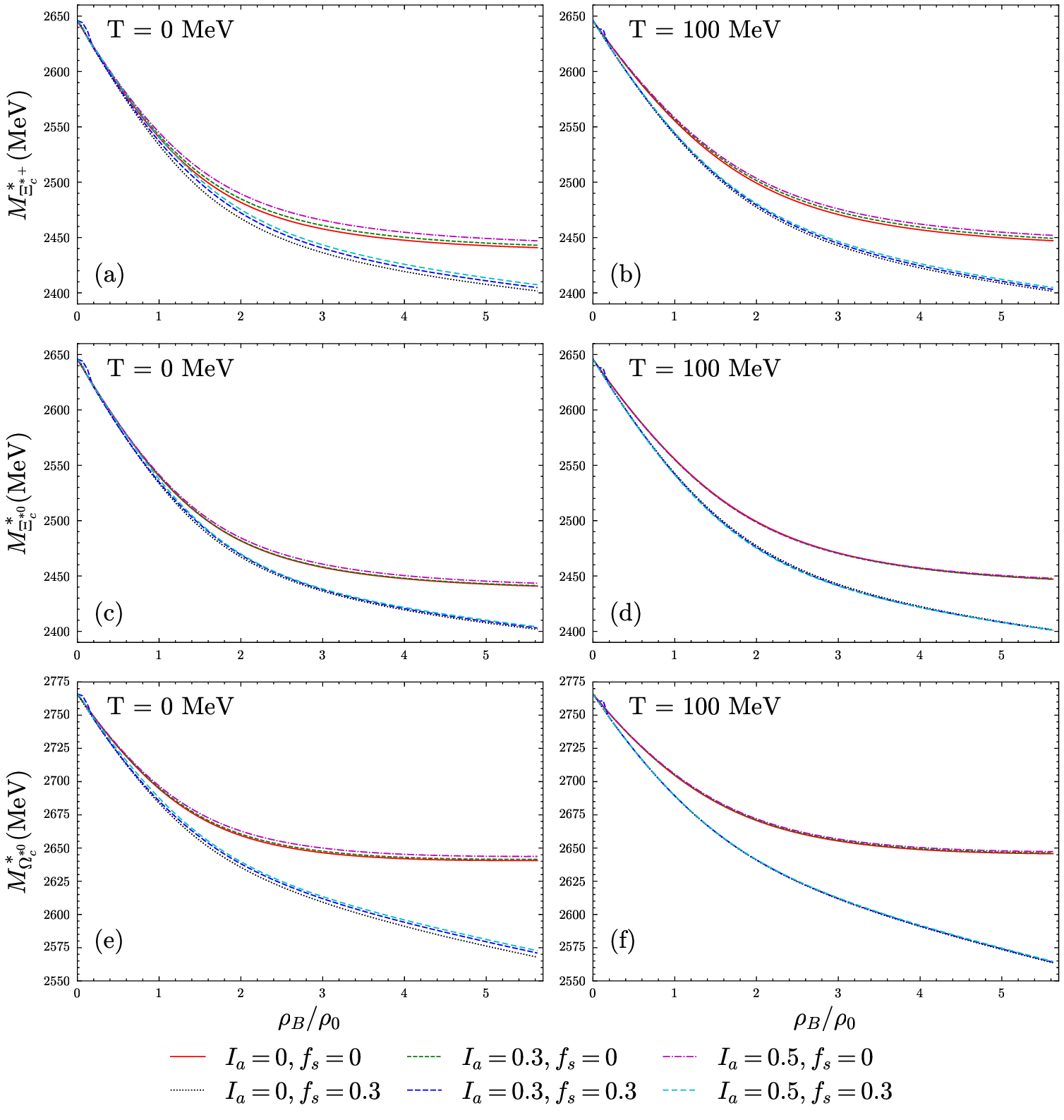}\hfill
 	\caption{Effective masses  of 
 	spin$-\frac{3}{2}^+$
 		singly charmed,  $\Xi^{*+}_{c},\ \Xi^{*0}_{c}$ and $\Omega^{*0}_{c}$
 		resonance baryons are shown as a function of baryon density $\rho_B$ (in units of   $\rho_0$) at temperatures $T = 0$ [subplots (a), (c), (e)] and $100$ [subplots (b), (d), (f)] MeV. Results are shown for isospin asymmetry $I_a = 0,\ 0.3,\ 0.5$ and strangeness fractions $f_s = 0,\ 0.3$.}
 	\label{fig_charmed_3by2_XiOmega}
 \end{figure}

\begin{figure}[ht!] 
	\includegraphics[width= 16 cm, height=20 cm]{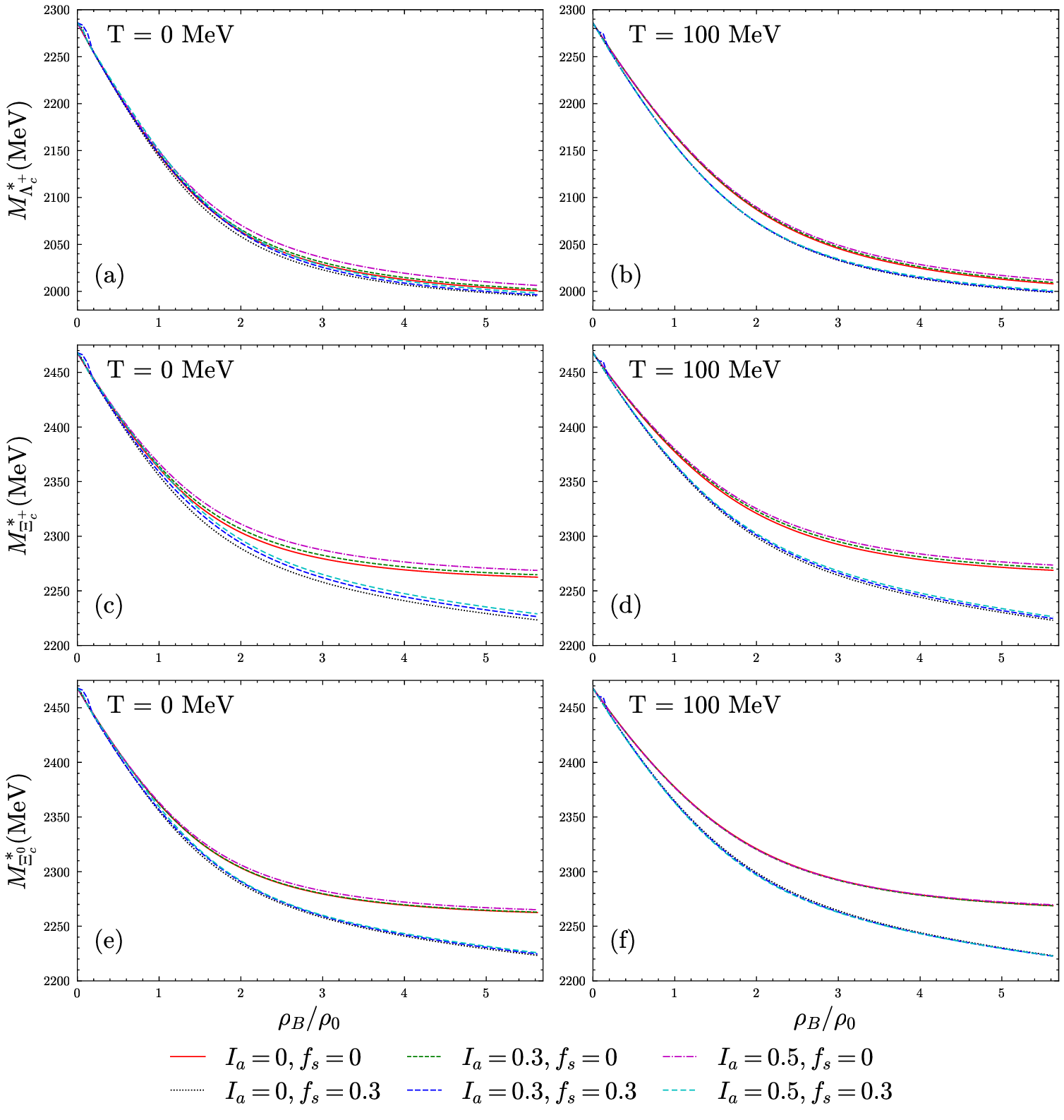}\hfill
	\caption{Effective masses 
			of spin$-\frac{1}{2}^+$
		singly charmed,  $\Lambda^+_c,\ \Xi^{+}_c$ and $\Xi^0_c$ baryons are shown as a function of baryon density $\rho_B$ (in units of   $\rho_0$) at temperatures $T = 0$  
	[subplots (a), (c), (e)] and $100$ [subplots (b), (d), (f)] MeV. Results are shown for isospin asymmetry $I_a = 0,\ 0.3,\ 0.5$ and strangeness fractions $f_s = 0,\ 0.3$.}
	\label{fig_charmed_half_XipXi0_Lambdap}
\end{figure}

\begin{figure}[ht!] 
 	\includegraphics[width= 16 cm, height=20 cm]{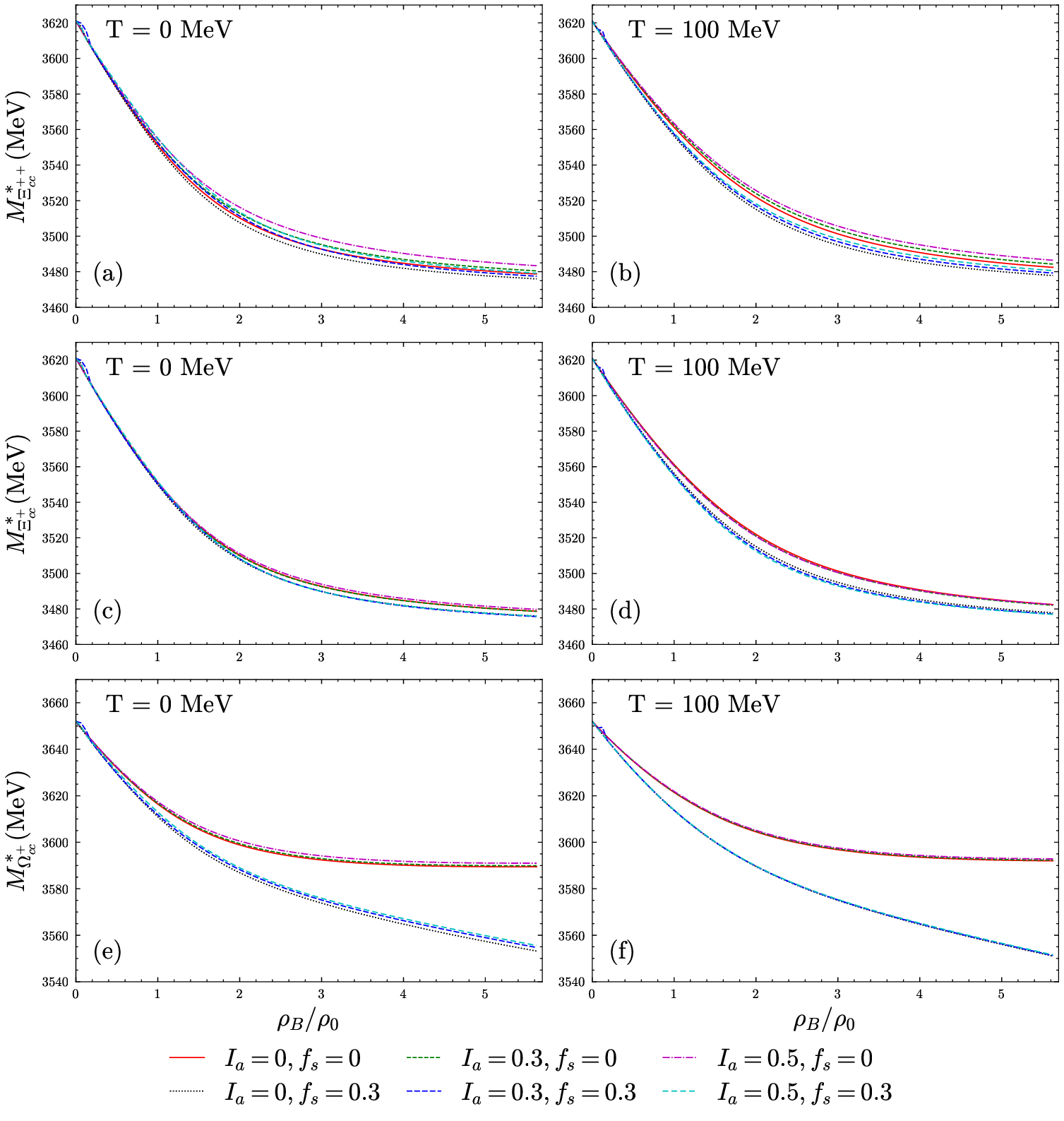}\hfill
 	\caption{Effective masses of spin$-\frac{1}{2}^+$
 	doubly charmed, $\Xi^{++}_{cc},\ \Xi^{+}_{cc}$ and $\Omega^{+}_{cc}$ baryons are shown as a function of baryon density $\rho_B$ (in units of $\rho_0$) at temperatures $T = 0$ 
 	[subplots (a), (c), (e)] and $100$ [subplots (b), (d), (f)] MeV. Results are shown for isospin asymmetry $I_a = 0,\ 0.3,\ 0.5$ and strangeness fractions $f_s = 0,\ 0.3$.}
 	\label{fig_charmed_1by2_Sigmastar}
 \end{figure}

\begin{figure}[ht!] 
 	\includegraphics[width= 16 cm, height=20 cm]{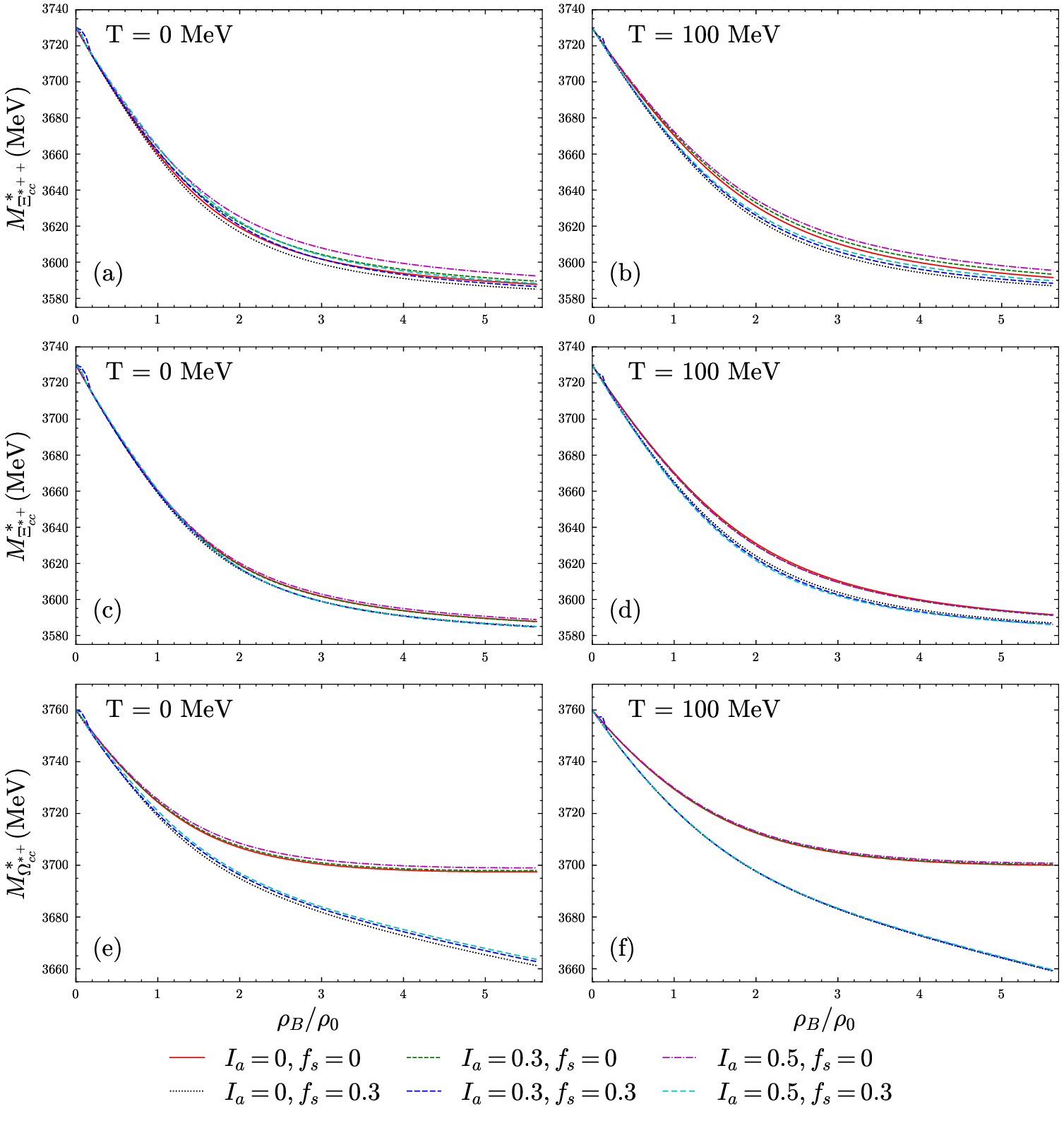}\hfill
 	\caption{Effective masses of spin$-\frac{3}{2}^+$
 	 	doubly charmed,  $\Xi^{*++}_{cc}, \Xi^{*+}_{cc}$ and $\Omega^{*+}_{cc}$ baryon resonances are shown as a function of baryon density $\rho_B$ (in units $\rho_0$) at temperatures $T = 0$ [subplots (a), (c), (e)] and $100$ [subplots (b), (d), (f)] MeV. Results are shown for isospin asymmetry $I_a = 0,\ 0.3,\ 0.5$ and strangeness fractions $f_s = 0,\ 0.3$.}
 	\label{fig_charmed_3by2_Sigmastar}
 \end{figure}

In Figs. \ref{fig_charmed_half_primeXipXi0_Omega0} and 	\ref{fig_charmed_3by2_XiOmega},
we have plotted the effective masses of singly charmed
spin$-\frac{1}{2}^+$
 		  $\Xi^{\prime +}_c, \Xi^{\prime0}_c$ and $\Omega^0_c$ 
 		and spin$-\frac{3}{2}^+$
 	 $\Xi^{* +}_c, \Xi^{*0}_c$ and $\Omega^{*0}_c$ baryons, respectively.
 The trend in the variation
 of the effective masses of these baryons is same as was observed in the case of
 $\Sigma_c$  and  $\Sigma_c^*$.
 For a given value of $\rho_B$, $T$, $I_a$ and $f_s$ of the medium, 
 the magnitude of mass shift decreases as
 we move towards the baryons with more strange quarks. For example, in symmetric cold nuclear matter, at $\rho_B = \rho_0$, mass shift for $\Sigma_c$,  $\Xi^{\prime}_c$ 
 and  $\Omega^{0}_c$
 are observed to be $-140.127,-105.757,-71.383$ MeV, respectively and thus follow the relation,  $|\Delta M_{\Sigma_c}|> |\Delta M_{\Xi^{\prime}_c}|> |\Delta M_{\Omega^{0}_c}|$.
Fig. \ref{fig_charmed_half_XipXi0_Lambdap} shows the variation of effective masses of singly charmed spin$-\frac{1}{2}^+$ $\Lambda^+_c$, $\Xi^{+}_c, \Xi^0_c$ baryons (which belongs to anti-triplet $\mathbf{\bar{3}}$)
as a function of baryon density $\rho_B$. The effective masses of doubly charmed spin$-\frac{1}{2}^+$
$\Xi^{++}_{cc}, \Xi^{+}_{cc}$ and $\Omega^{+}_{cc}$ and spin$-\frac{3}{2}^+$
$\Xi^{*++}_{cc}, \Xi^{*+}_{cc}$ and $\Omega^{*+}_{cc}$ baryons are plotted in Figs. \ref{fig_charmed_1by2_Sigmastar} and \ref{fig_charmed_3by2_Sigmastar}, respectively.
Since the charm quarks are not modified in the medium as expected, compared to 
singly charmed,
the magnitude of mass-shift are observed to be smaller for the doubly charmed  baryons.
Among the given multiplet of doubly charmed baryons, the one having strange $s$ quark has smaller mass shift at given baryon density. Thus, we observe 
the relation
$|\Delta  M_{\Omega^{*+}_{cc}}|< 
 |\Delta M_{\Xi^{*++}_{cc}}| $.
For given baryon density, temperature and isospin asymmetry of the medium,	an increase in the strangeness fractions causes a  decrease in the masses of baryons composed of one or more strange quarks. For example, 
in symmetric matter, at $\rho_B = 3\rho_0$,
for
 spin$-\frac{1}{2}^+$
 $\Sigma_c$, $\Xi_c^{\prime}$ and $\Omega_c^0$ baryons, increase of $f_s$ from zero to $0.3$ causes a decrease of $5.67$,
  $21.42$  and $37.30$ MeV in the effective masses, respectively.
  
We are now briefly going to discuss the results obtained on the in-medium masses of charmed baryons studied using different approaches in the literature. In the QMC model calculations of Ref. \cite{Tsushima:2002cc} in the nuclear matter at zero temperature, the masses of $\Lambda_c$, $\Sigma_c$ and $\Xi_c$ baryons, expressed in terms of scalar mean field $\sigma$, were observed to decrease with the baryon density. Within the QCD sum rule calculations of Ref. \cite{Wang:2011hta} a positive mass shift of 51 MeV for $\Lambda_c$ baryon, whereas in Ref. \cite{Wang:2011yj}, for the $\Sigma_c$ baryon a negative mass shift of $-123$ MeV is observed  at nuclear saturation density of the cold symmetric nuclear matter.
For the doubly charmed $\Xi_{cc}$ and $\Omega_{cc}$ baryons within QCD sum rules, large values of mass shift, $-1.11$ and $-0.33$ GeV respectively, are obtained in Ref. \cite{Wang:2012xk}.
Parity projected QCD sum rules were applied to investigate the in-medium properties of 
$\Lambda_c$ charmed baryon in Ref \cite{Ohtani:2017wdc}.
The four-quark condensates were observed to give the dominant contributions to the in-medium properties and their density dependence was calculated using the factorization hypothesis (which gives repulsive mass shift to $\Lambda_c$ baryon) and perturbative chiral quark model (which gives repulsive mass shift $\sim 20$ MeV to $\Lambda_c$ baryon). 
In Ref. \cite{Azizi:2018dtb},  for spin$-\frac{3}{2}$
$\Sigma_c^*$ attractive mass shift of approximately 38 MeV is observed, whereas $\Xi_c^*$ and $\Omega_c^*$ remain  almost unaffected.
Using the heavy quark effective field theory, the negative mass shift having values $-15$ and $-14$ MeV were observed
for $\Sigma_c$ and $\Sigma_c^*$, respectively, in Ref. \cite{Yasui:2018sxz}.
In Ref. \cite{Won:2021pwb},
using the chiral soliton model the density dependence of singly charmed heavy spin$-\frac{1}{2}$ baryons was studied and the density dependence was parameterized in terms of some parameter $C_4$. As one can see from Table I of Ref. \cite{Won:2021pwb}, for the values of parameter $C_4 = -0.1 (0.1)$, the mass shift values $-166.91  (209.34)$, $-132.30 (241.25)$, $-86.50 (289.57)$, $-69.89 (304.43)$ and $-54.23 (318.46)$ MeV are quoted for $\Lambda_c, \Xi_c, \Sigma_c, \Xi_c^{\prime}$ and $\Omega_c$, respectively, at nuclear saturation density $\rho_0$ of cold symmetric nuclear matter. 
Only in one recent Ref. \cite{Ghim:2022zob}, the mass modifications of singly charmed and singly bottom spin$-\frac{1}{2}^+$ and spin$-\frac{3}{2}^+$ baryons were studied in asymmetric nuclear matter at zero temperature (see Table I of Ref. \cite{Ghim:2022zob} for all values).

	

		\begin{sidewaystable}[h!]
	\centering

\begin{tabular}{|c|c|c|c|c|c|c|c|c|c|}
	\hline
	\multirow{3}{*}{Baryon} & \multirow{3}{*}{Vacuum Mass   
		} & \multicolumn{4}{c|}{$\rho_{B}=\rho_{0}$} & \multicolumn{4}{c|}{$\rho_{B}=3\rho_{0}$} \\ \cline{3-10}
	&  & \multicolumn{2}{c|}{$f_s=0$} & \multicolumn{2}{c|}{$f_s=0.3$} & \multicolumn{2}{c|}{$f_s=0$} & \multicolumn{2}{c|}{$f_s=0.3$} \\ \cline{3-10}
	& (MeV)   & $I_a=0$ & $I_a=0.3$ &  $I_a=0$ &  $I_a=0.3$ &  $I_a=0$ &  $I_a=0.3$ &  $I_a=0$ &  $I_a=0.3$ \\ 
\hline $M_{ \Sigma_c^{++} }^*$ &  2454.0 &  $-140.127$ &  $-136.767$ &  $-142.737$ &  $-137.562$ &  $-256.998$ &  $-251.929$ &  $-262.67$ &  $-257.17$ \\
\hline $M_{ \Sigma_c^{+} }^*$ &  2454.0 &  $-140.127$ &  $-138.808$ &  $-142.737 $&  $-139.648 $&  $-256.998 $&  $-254.655 $&  $-262.67 $&  $-260.006$ \\
\hline $M_{ \Sigma_c^{0} }^*$ &  2454.0 &  $-140.127 $&  $-140.85 $&  $-142.737 $&  $-141.734 $&  $-256.998 $&  $-257.381 $&  $-262.67 $&  $-262.843$ \\
\hline $M_{ \Xi_c^{\prime+} }^*$ &  2578.0 &  $-105.757 $&  $-103.72 $&  $-112.523 $&  $-109.105 $&  $-188.323 $&  $-185.227 $&  $-209.811 $&  $-205.708$ \\
\hline $M_{ \Xi_c^{\prime0} }^*$ &  2578.0 &  $-105.757 $&  $-105.762 $&  $-112.523 $&  $-111.191 $&  $-188.323 $&  $-187.954 $&  $-209.811 $&  $-208.547$ \\
\hline $M_{ \Omega_c^{0} }^*$ &  2695.0 &  $-71.383 $&  $-70.67 $&  $-82.302 $&  $-80.641 $&  $-119.6 $&  $-118.479 $&  $-156.891 $&  $-154.19$ \\
\hline $M_{ \Xi_c^{+} }^*$ &  2468.0 &  $-105.869 $&  $-103.83 $&  $-112.642 $&  $-109.221 $&  $-188.516 $&  $-185.417 $&  $-210.03 $&  $-205.924$ \\
\hline $M_{ \Xi_c^{0} }^*$ &  2468.0 &  $-105.869 $&  $-105.873 $&  $-112.642 $&  $-111.309 $&  $-188.516 $&  $-188.147 $&  $-210.03 $&  $-208.765$ \\
\hline $M_{ \Lambda_c^{+} }^*$ &  2286.0 &  $-140.378 $&  $-139.056 $&  $-142.992 $&  $-139.898 $&  $-257.441 $&  $-255.094 $&  $-263.12 $&  $-260.453$ \\
\hline $M_{ \Xi_{cc}^{++} }^*$ &  3621.0 &  $-69.885 $&  $-68.208 $&  $-71.187 $&  $-68.605 $&  $-128.259 $&  $-125.723 $&  $-131.097 $&  $-128.345$ \\
\hline $M_{ \Xi_{cc}^{+} }^*$ &  3621.0 &  $-69.885 $&  $-70.245 $&  $-71.187 $&  $-70.686 $&  $-128.259 $&  $-128.45 $&  $-131.097 $&  $-131.184$ \\
\hline $M_{ \Omega_{cc}^{+} }^*$ &  3652.0 &  $-35.58 $&  $-35.225 $&  $-41.022 $&  $-40.195 $&  $-59.611 $&  $-59.053 $&  $-78.195 $&  $-76.849$ \\
\hline
\end{tabular}
\caption{The change in effective masses of spin$-\frac{1}{2}^+$ charmed baryons with respect to vacuum values are given at $\rho_B = \rho_0$ and $\rho_{B}=3\rho_0$ for isospin asymmetry $I_a = 0,\ 0.3$ and strangeness fractions $f_s = 0,\ 0.3$ at temperatures $T = 0$ MeV.}
\label{tab:masses_1half2T0}
\end{sidewaystable}


		\begin{sidewaystable}[h!]
	\centering

\begin{tabular}{|c|c|c|c|c|c|c|c|c|c|}
	\hline
	\multirow{3}{*}{Baryon} & \multirow{3}{*}{Vacuum Mass   
		} & \multicolumn{4}{c|}{$\rho_{B}=\rho_{0}$} & \multicolumn{4}{c|}{$\rho_{B}=3\rho_{0}$} \\ \cline{3-10}
	&  & \multicolumn{2}{c|}{$f_s=0$} & \multicolumn{2}{c|}{$f_s=0.3$} & \multicolumn{2}{c|}{$f_s=0$} & \multicolumn{2}{c|}{$f_s=0.3$} \\ \cline{3-10}
	&  (MeV)  & $I_a=0$ & $I_a=0.3$ &  $I_a=0$ &  $I_a=0.3$ &  $I_a=0$ &  $I_a=0.3$ &  $I_a=0$ &  $I_a=0.3$ \\  
\hline $M_{ \Sigma_c^{++} }^*$ &  2454.0 &  $-119.602 $&  $-117.198 $&  $-129.552 $&  $-127.335 $&  $-239.688 $&  $-235.2 $&  $-252.383 $&  $-248.459$\\
\hline $M_{ \Sigma_c^{+} }^*$ &  2454.0 &  $-119.602 $&  $-118.984 $&  $-129.552 $&  $-129.403 $&  $-239.688 $&  $-238.251 $&  $-252.383 $&  $-251.746$\\
\hline $M_{ \Sigma_c^{0} }^*$ &  2454.0 &  $-119.602 $&  $-120.771 $&  $-129.552 $&  $-131.47 $&  $-239.688 $&  $-241.303 $&  $-252.383 $&  $-255.033$\\
\hline $M_{ \Xi_c^{\prime+} }^*$ &  2578.0 &  $-90.402 $&  $-89.012 $&  $-103.145 $&  $-101.957 $&  $-175.191 $&  $-172.644 $&  $-203.394$ &  $-201.261$\\
\hline $M_{ \Xi_c^{\prime0} }^*$ &  2578.0 &  $-90.402 $&  $-90.798 $&  $-103.145 $&  $-104.024 $&  $-175.191 $&  $-175.698 $&  $-203.394 $&  $-204.551$\\
\hline $M_{ \Omega_c^{0} }^*$ &  2694.0 &  $-61.201 $&  $-60.825 $&  $-76.732 $&  $-76.573 $&  $-110.659 $&  $-110.057 $&  $-154.352 $&  $-154.013$\\
\hline $M_{ \Xi_c^{+} }^*$ &  2468.0 &  $-90.497 $&  $-89.106 $&  $-103.255 $&  $-102.065 $&  $-175.372 $&  $-172.823 $&  $-203.608 $&  $-201.474$\\
\hline $M_{ \Xi_c^{0} }^*$ &  2468.0 &  $-90.497 $&  $-90.894 $&  $-103.255 $&  $-104.135 $&  $-175.372 $&  $-175.879 $&  $-203.608 $&  $-204.766$\\
\hline $M_{ \Lambda_c^{+} }^*$ &  2286.0 &  $-119.816 $&  $-119.197 $&  $-129.784 $&  $-129.634 $&  $-240.106 $&  $-238.667 $&  $-252.819 $&  $-252.181$\\
\hline $M_{ \Xi_{cc}^{++} }^*$ &  3621.0 &  $-59.644 $&  $-58.445 $&  $-64.609 $&  $-63.502 $&  $-119.602 $&  $-117.358 $&  $-125.95 $&  $-123.988$\\
\hline $M_{ \Xi_{cc}^{+} }^*$ &  3621.0 &  $-59.644 $&  $-60.227 $&  $-64.609 $&  $-65.565 $&  $-119.602 $&  $-120.409 $&  $-125.95 $&  $-127.276$\\
\hline $M_{ \Omega_{cc}^{+} }^*$ &  3652.0 &  $-30.505 $&  $-30.318 $&  $-38.246 $&  $-38.167 $&  $-55.155 $&  $-54.855 $&  $-76.93 $&  $-76.761$\\
\hline
\end{tabular}
\caption{The change in effective masses of spin$-\frac{1}{2}^+$ charmed baryons with respect to vacuum values are given at $\rho_B = \rho_0$ and $\rho_{B}=3\rho_0$ for isospin asymmetry $I_a = 0,\ 0.3$ and strangeness fractions $f_s = 0,\ 0.3$ at temperatures $T = 100$ MeV.}
\label{tab:masses_1half2T100}
\end{sidewaystable}


		\begin{sidewaystable}[h!]
				\centering
			\begin{tabular}{|c|c|c|c|c|c|c|c|c|c|}
				
				\hline
				\multirow{3}{*}{Baryon} & \multirow{3}{*}{Vacuum Mass   
				} & \multicolumn{4}{c|}{$\rho_{B}=\rho_{0}$} & \multicolumn{4}{c|}{$\rho_{B}=3\rho_{0}$} \\ \cline{3-10}
				&  & \multicolumn{2}{c|}{$f_s=0$} & \multicolumn{2}{c|}{$f_s=0.3$} & \multicolumn{2}{c|}{$f_s=0$} & \multicolumn{2}{c|}{$f_s=0.3$} \\ \cline{3-10}
				&  (MeV)  & $I_a=0$ & $I_a=0.3$ &  $I_a=0$ &  $I_a=0.3$ &  $I_a=0$ &  $I_a=0.3$ &  $I_a=0$ &  $I_a=0.3$ \\

	\hline 
	$M_{ \Sigma_c^{*++} }^*$ &  2518.0 &  $-140.048 $&  $-136.69 $&  $-142.656 $&  $-137.485 $&  $-256.86 $&  $-251.794 $&  $-262.53 $&  $-257.032$\\
	\hline $M_{ \Sigma_c^{*+} }^*$ &  2518.0 &  $-140.048 $&  $-138.73 $&  $-142.656 $&  $-139.569 $&  $-256.86 $&  $-254.518 $&  $-262.53 $&  $-259.867$\\
	\hline $M_{ \Sigma_c^{*0} }^*$ &  2518.0 &  $-140.048 $&  $-140.77 $&  $-142.656 $&  $-141.654 $&  $-256.86 $&  $-257.243 $&  $-262.53 $&  $-262.703$\\
	\hline $M_{ \Xi_c^{*+} }^*$ &  2646.0 &  $-105.697 $&  $-103.661 $&  $-112.458 $&  $-109.042 $&  $-188.219 $&  $-185.124 $&  $-209.692 $&  $-205.592$\\
	\hline $M_{ \Xi_c^{*0} }^*$ &  2646.0 &  $-105.697 $&  $-105.701 $&  $-112.458 $&  $-111.127 $&  $-188.219 $&  $-187.85 $&  $-209.692 $&  $-208.429$\\
	\hline $M_{ \Omega_c^{*0} }^*$ &  2766.0 &  $-71.34 $&  $-70.627 $&  $-82.251 $&  $-80.592 $&  $-119.525 $&  $-118.406 $&  $-156.792 $&  $-154.094$\\
	\hline $M_{ \Xi_{cc}^{*++} }^*$ &  3730.0 &  $-69.862 $&  $-68.186 $&  $-71.164 $&  $-68.583 $&  $-128.223 $&  $-125.687 $&  $-131.06 $&  $-128.309$\\
	\hline $M_{ \Xi_{cc}^{*+} }^*$ &  3730.0 &  $-69.862 $&  $-70.223 $&  $-71.164 $&  $-70.664 $&  $-128.223 $&  $-128.414 $&  $-131.06 $&  $-131.147$\\
	\hline $M_{ \Omega_{cc}^{*+} }^*$ &  3760.0 &  $-35.566 $&  $-35.211 $&  $-41.005 $&  $-40.178 $&  $-59.586 $&  $-59.028 $&  $-78.163 $&  $-76.818$\\
\hline
\end{tabular}
\caption{The change in effective masses of spin$-\frac{3}{2}^+$ charmed baryons with respect to vacuum values are given at $\rho_B = \rho_0$ and $\rho_{B}=3\rho_0$ for isospin asymmetry $I_a = 0,\ 0.3$ and strangeness fractions $f_s = 0,\ 0.3$ at temperatures $T = 0$ MeV.}
\label{tab:masses_3half2T0}
\end{sidewaystable}

		\begin{sidewaystable}[h!]
	\centering
			\begin{tabular}{|c|c|c|c|c|c|c|c|c|c|}
	
	\hline
	\multirow{3}{*}{Baryon} & \multirow{3}{*}{Vacuum Mass   
	} & \multicolumn{4}{c|}{$\rho_{B}=\rho_{0}$} & \multicolumn{4}{c|}{$\rho_{B}=3\rho_{0}$} \\ \cline{3-10}
	&  & \multicolumn{2}{c|}{$f_s=0$} & \multicolumn{2}{c|}{$f_s=0.3$} & \multicolumn{2}{c|}{$f_s=0$} & \multicolumn{2}{c|}{$f_s=0.3$} \\ \cline{3-10}
	&  (MeV)  & $I_a=0$ & $I_a=0.3$ &  $I_a=0$ &  $I_a=0.3$ &  $I_a=0$ &  $I_a=0.3$ &  $I_a=0$ &  $I_a=0.3$ \\
	
	\hline 

\hline $M_{ \Sigma_c^{*++} }^*$ &  2518.0 &  $-140.048 $&  $-117.297 $&  $-129.645 $&  $-127.428 $&  $-256.86 $&  $-235.236 $&  $-252.413 $&  $-248.49$\\
\hline $M_{ \Sigma_c^{*+} }^*$ &  2518.0 &  $-140.048 $&  $-119.083 $&  $-129.645 $&  $-129.495 $&  $-256.86 $&  $-238.287 $&  $-252.413 $&  $-251.776$\\
\hline $M_{ \Sigma_c^{*0} }^*$ &  2518.0 &  $-140.048 $&  $-120.868 $&  $-129.645 $&  $-131.561 $&  $-256.86 $&  $-241.337 $&  $-252.413 $&  $-255.062$\\
\hline $M_{ \Xi_c^{*+} }^*$ &  2646.0 &  $-105.697 $&  $-89.09 $&  $-103.214 $&  $-102.026 $&  $-188.219 $&  $-172.677 $&  $-203.407 $&  $-201.275$\\
\hline $M_{ \Xi_c^{*0} }^*$ &  2646.0 &  $-105.697 $&  $-90.875 $&  $-103.214 $&  $-104.093 $&  $-188.219 $&  $-175.729 $&  $-203.407 $&  $-204.563$\\
\hline $M_{ \Omega_c^{*0} }^*$ &  2766.0 &  $-71.34 $&  $-60.88 $&  $-76.777 $&  $-76.618 $&  $-119.525 $&  $-110.081 $&  $-154.347 $&  $-154.009$\\
\hline $M_{ \Xi_{cc}^{*++} }^*$ &  3730.0 &  $-69.862 $&  $-58.509 $&  $-64.67 $&  $-63.564 $&  $-128.223 $&  $-117.406 $&  $-125.997 $&  $-124.034$\\
\hline $M_{ \Xi_{cc}^{*+} }^*$ &  3730.0 &  $-69.862 $&  $-60.29 $&  $-64.67 $&  $-65.627 $&  $-128.223 $&  $-120.457 $&  $-125.997 $&  $-127.323$\\
\hline $M_{ \Omega_{cc}^{*+} }^*$ &  3760.0 &  $-35.566 $&  $-30.352 $&  $-38.277 $&  $-38.197 $&  $-59.586 $&  $-54.878 $&  $-76.944 $&  $-76.775$\\
\hline
			\end{tabular}
	\caption{The change in effective masses of spin$-\frac{3}{2}^+$ charmed baryons with respect to vacuum values are given at $\rho_B = \rho_0$ and $\rho_{B}=3\rho_0$ for isospin asymmetry $I_a = 0,\ 0.3$ and strangeness fractions $f_s = 0,\ 0.3$ at temperatures $T = 100$ MeV.}
	\label{tab:masses_3half2T100}
\end{sidewaystable}

\subsection{In-medium magnetic moments of singly charmed baryons}
\label{sec_deculet_moments}
In the present section, we have discussed the results on medium modification of  magnetic moments for spin$-\frac{1}{2}^+$ and $\frac{3}{2}^{+}$ charmed baryons. 
The in-medium magnetic moments are
calculated using combined approach of  chiral SU(4) constituent
quark model and chiral SU(3) quark mean field model. As discussed in Sec. \ref{sec:magnetic}, in present calculations the sum of contributions due to valence, sea and orbital angular momentum of sea quarks contribute to the net magnetic moment of a given charmed baryon.
In Tables  \ref{table:comparison_theor_models}
and \ref{table:comparison_theor_models3_2}
the vacuum values of magnetic moments calculated in the present work are compared with literature whereas Tables \ref{table:mub_eta0fs0} and
\ref{table:mub_eta5fs3_3by2}
have values calculated in the medium at finite baryon density.

Looking into literature, in case of charmed baryons, the magnetic moments of only low lying spin$-\frac{1}{2}^+$  $\Sigma_c^{++}, \Sigma_c^{+}, \Sigma_c^{0}$
and $\Lambda_c^{+},\Xi_c^{+}, \Xi_c^{0}$, belonging to the sextet $\mathbf{6}$ and
antisymmetric triplet $\mathbf{\bar{3}}$ have been studied in symmetric nuclear matter at zero temperature using the QMC model  \cite{Tsushimaptep2022},
whereas the magnetic
 moments of doubly charmed spin$-\frac{1}{2}^+$   and singly and doubly charmed spin$-\frac{3}{2}^+$ baryons are studied for the first time at finite density and temperature in the current work. 
As discussed earleir,  in Ref. \cite{Tsushimaptep2022} 
the magnetic moment of charmed baryons are calculated considering the valence quarks only, without configuration mixing \cite{Tsushimaptep2022}, whereas in the present calculations configuration mixing has been taken into account, in addition to 
contribution to magnetic moments from sea quarks.

\begin{figure}[ht!] 
		\includegraphics[width= 16 cm, height=18 cm]{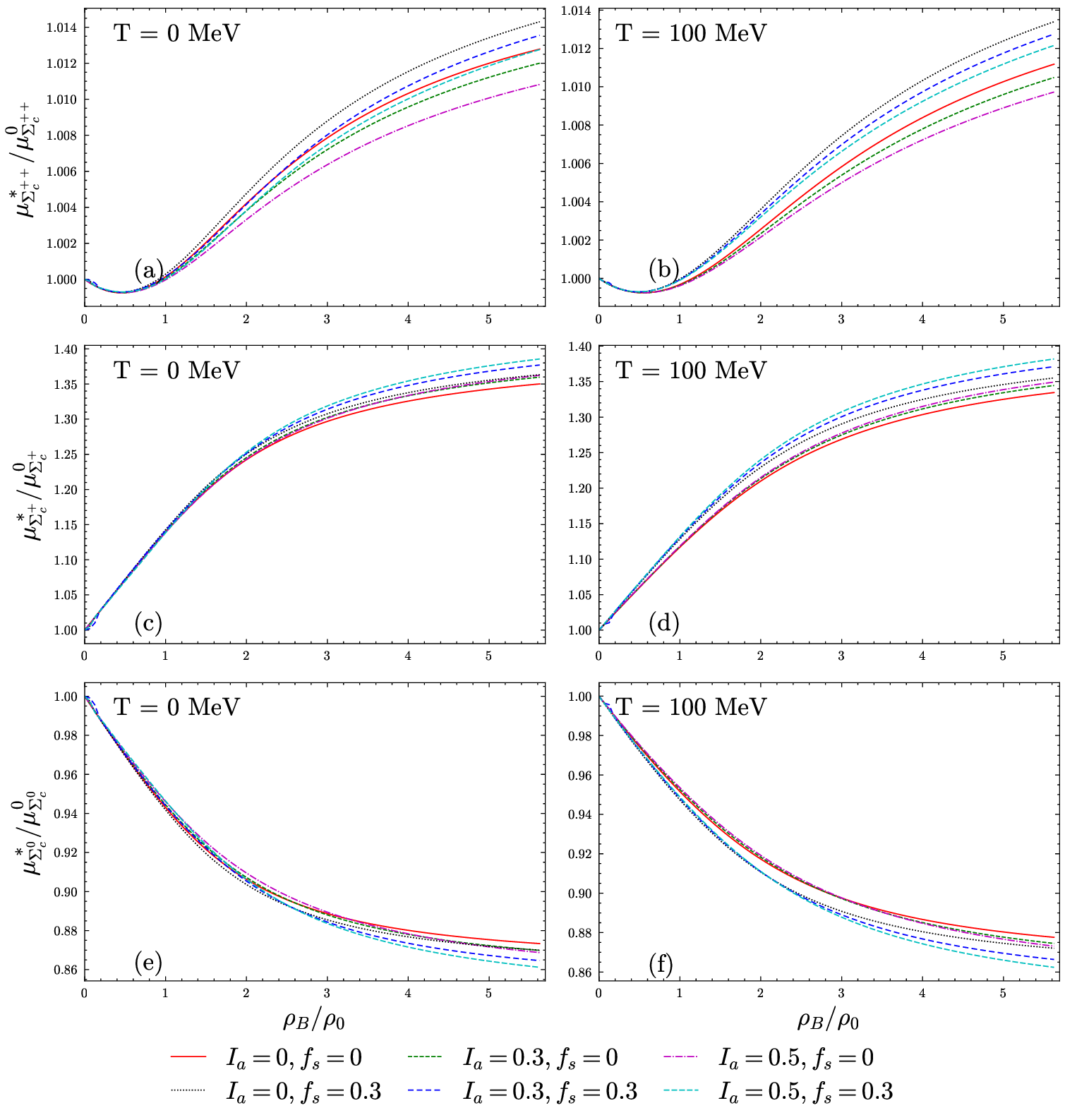}\hfill
		\caption{Effective magnetic moments to vacuum magnetic moments ratios for spin$-\frac{1}{2}^+$ singly charmed $\Sigma_c^{++},\ \Sigma_c^{+}$  and $\Sigma^{0}_c$ baryons are plotted as a function of baryon density $\rho_B$ (in units of $\rho_0$) at temperatures $T = 0$ [in subplots (a), (c) and (e)] and $100$ 
		[in subplots (b), (d) and (f)] MeV. Results are shown for isospin asymmetry $I_a = 0,\ 0.3,\ 0.5$ and strangeness fractions $f_s = 0,\ 0.3$.}
		\label{fig_moment_Sigma_c}
	\end{figure}

\begin{figure}[ht!] 
		
		\includegraphics[width= 18 cm, height=14 cm] {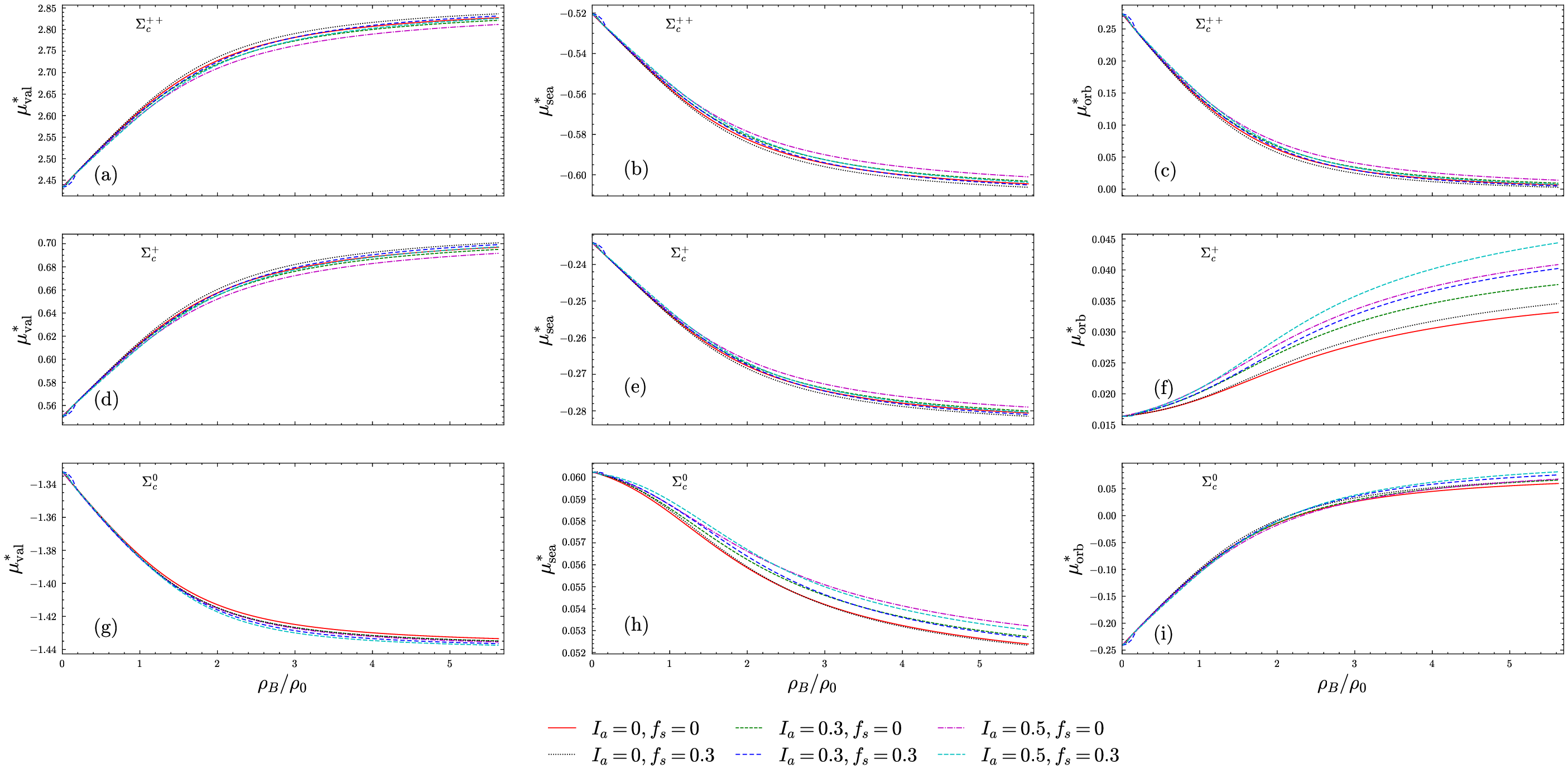}\hfill

		\caption{Contributions of valence, sea and orbital moment of sea quarks ($\mu_{\mathrm{val}}^{*}$,\ $\mu_{\mathrm{sea}}^{*}$,\ $\mu_{orb}^{*}$) towards the total magnetic moment of \spinhalf singly charmed $\Sigma_c^{++}$[in subplots (a), (b) and (c)], $\Sigma_c^{+}$[in subplots (d), (e) and (f)] and $\Sigma_c^{0}$[in subplots (g), (h) and (i)] are shown as a function of baryon density $\rho_B$ (in units of nuclear saturation density $\rho_0$) at temperatures $T = 0$. Results are shown for isospin asymmetry $I_a = 0,\ 0.3,\ 0.5$ and strangeness fractions $f_s = 0,\ 0.3$.}
		\label{fig_moment_Sigma_indi}
		
	\end{figure}

In Fig. \ref{fig_moment_Sigma_c}, we have shown the ratio, in-medium to vacuum value of magnetic moment for spin$-\frac{1}{2}^+$ singly charmed $\Sigma_c$ baryons , $\Sigma_c^{++}, \Sigma_c^{+} $ and $\Sigma_c^{-}$,  as a function of baryon density $\rho_B$ (in units of $\rho_0$).
The results are shown for the isospin symmetric ($I_a = 0$) and  asymmetric ($I_a = 0.3,0.5$)
nuclear ($f_s =0$) and strange ($f_s = 0.3$) hadronic medium.
The magnetic moments of $\Sigma_c^{++}$ baryons first show a very slight decrease and then a small increase with the baryon density of the symmetric nuclear medium. 
%
In the present calculations, as the total magnetic moment of baryons are determined by the contributions coming from valence quarks, sea quarks and orbital moment of sea quarks, hence contributions as a function of density coming from each of the these components have been shown for singly charmed $\Sigma_c$ baryons in 	Fig. \ref{fig_moment_Sigma_indi}. As is observed in subplots $(a), (b), (c)$ of Fig. \ref{fig_moment_Sigma_indi}, the valence contributions dominate over the other two components for   $\Sigma_c^{++}$ baryons. It is also observed that $\mu_{\mathrm val}^*$ shows an increase as a function of baryon density, $\rho_B$ while the value of sea ($\mu_{\mathrm{sea}}^*$) and orbital ($\mu_{\mathrm{\mathrm{orb}}}^*$) contributions decreases. The increase in value of $\mu_{\mathrm{val}}^*$ is almost neutralized by the decrease in other two components, hence a very small increase in total magnetic moment is observed for $\mu_{\Sigma_c^{++}}^*$.
As the baryon density is increased from zero to $\rho_0$ and $3\rho_0$, the in-medium magnetic 
moment of $\Sigma_c^{++}$
increase by $0.32\%$ and $1.37\%$, respectively.
In  the calculations of
 Ref.\cite{Tsushimaptep2022} within the QMC model, the values of magnetic moment of $\Sigma_c^{++}$ are observed
 to increase by $11.4\%$ and $21.8\%$, at $\rho_0$
 and $3\rho_0$, respectively, from the vacuum value. Smaller change observed in our calculations is due to consideration of sea quarks and orbital moment of sea quarks. As discussed above and also can be seen from Table \ref{table:mub_eta0fs0}, sea quarks contribute negatively to the overall magnetic moment and the magnitude of this negative contribution increases with density. If we had considered the valence quarks only, the values of percentage increase will 
 change to $7.44\%$ and $14.48\%$ at $\rho_0$ and $3\rho_0$, respectively.

 For the $\Sigma_c^{+}$ (see Fig. \ref{fig_moment_Sigma_c}), the magnetic moment values increases by $ 13.8\%$
and $29.6\%$ from the vacuum value 
 (highest among the $\Sigma_c$ baryons) at baryon density $\rho_0$ and  $3\rho_0$ of symmetric nuclear matter. For the $\Sigma_c^{0}$, the ratio of in-medium to vacuum magnetic moments  show a decrease of the order of $6\%$ and $11 \%$  at $\rho_0$ and  $3\rho_0$, respectively.
 As observed from the subplots $(d), (e), (f)$ of Fig. \ref{fig_moment_Sigma_indi}, for  $\Sigma_c^{+}$ the valence component, $\mu_{\mathrm{val}}^*$ dominates the other two components. The sea quark contribution $\mu_{\mathrm{sea}}^*$ is significant, though not strong enough to neutralize the effect of $\mu_{\mathrm{val}}^*$, thus showing appreciable amount of impact on value of total magnetic moments. The contribution of the $\mu_{\mathrm{orb}}^*$ is negligibly small in comparison to $\mu_{\mathrm{val}}^*$ and $\mu_{\mathrm{sea}}^*$. 

In QMC model, values of $\mu_{\Sigma_c^{+}}^*$ increases by 16.6 \% and
$31.5\%$ as density is changed from zero to $\rho_0$ and $3\rho_0$, respectively and while in our case increase in the valence magnetic moments, $\mu_{\mathrm{val}}^*$ is $13.8 \%$ and $29.72\%$ at $\rho_0$ and  $3\rho_0$, respectively.
As is observed from subplots $(g), (h), (i)$ of Fig. \ref{fig_moment_Sigma_indi}, for the case $\Sigma_c^{0}$ baryon also, $\mu_{\mathrm{val}}^*$ dominates over orbital and sea components, though a significant contribution also comes from $\mu_{\mathrm{orb}}^*$. With increase in density, the magnitude of $\mu_{\mathrm{val}}^*$ for $\Sigma_c^{0}$ increases while the magnitude of $\mu_{\mathrm{orb}}^*$ decreases, resulting in a slower increase in the magnetic moment ratio with density for these baryons. 
In case of $\Sigma_c^{0}$ baryon, within the QMC model calculations (valence quarks only), the magnitude of in-medium  values of magnetic moment 
increases as a function of $\rho_B$, which is opposite to our case. This is again due to additional contribution of sea quarks in our calculations which change the overall trend of variations. As can be seen from Table \ref{table:mub_eta0fs0}, in our calculations also, the magnitude of magnetic moments predominantly comes from the contribution of valence quarks which increases with baryon density $\rho_B$ though there is clearly a significant contribution also coming from sea quarks and orbital moment of sea quarks in the effective magnetic moments of charmed baryons.

In the current study, for the singly  charmed $\Sigma_c^{++}$ baryons, the increase in isospin asymmetry, $I_a$, shows a marginal increase in the value of magnetic moment with respect to vacuum value, while a substantial increase in values of magnetic moment ratio is observed with the increase in strangeness fraction $f_s$ of the medium. For singly charmed $\Sigma_c^{+}$ and $\Sigma_c^{0}$ baryons, higher increase in the magnitude of magnetic moments is observed both with the increase in $I_a$ and $f_s$.
The impacts of strangeness and isospin asymmetry of the medium on magnetic momentum ratios are more significant at the higher baryonic densities.

\begin{figure}[ht!] 
	\includegraphics[width= 18 cm, height=18 cm]{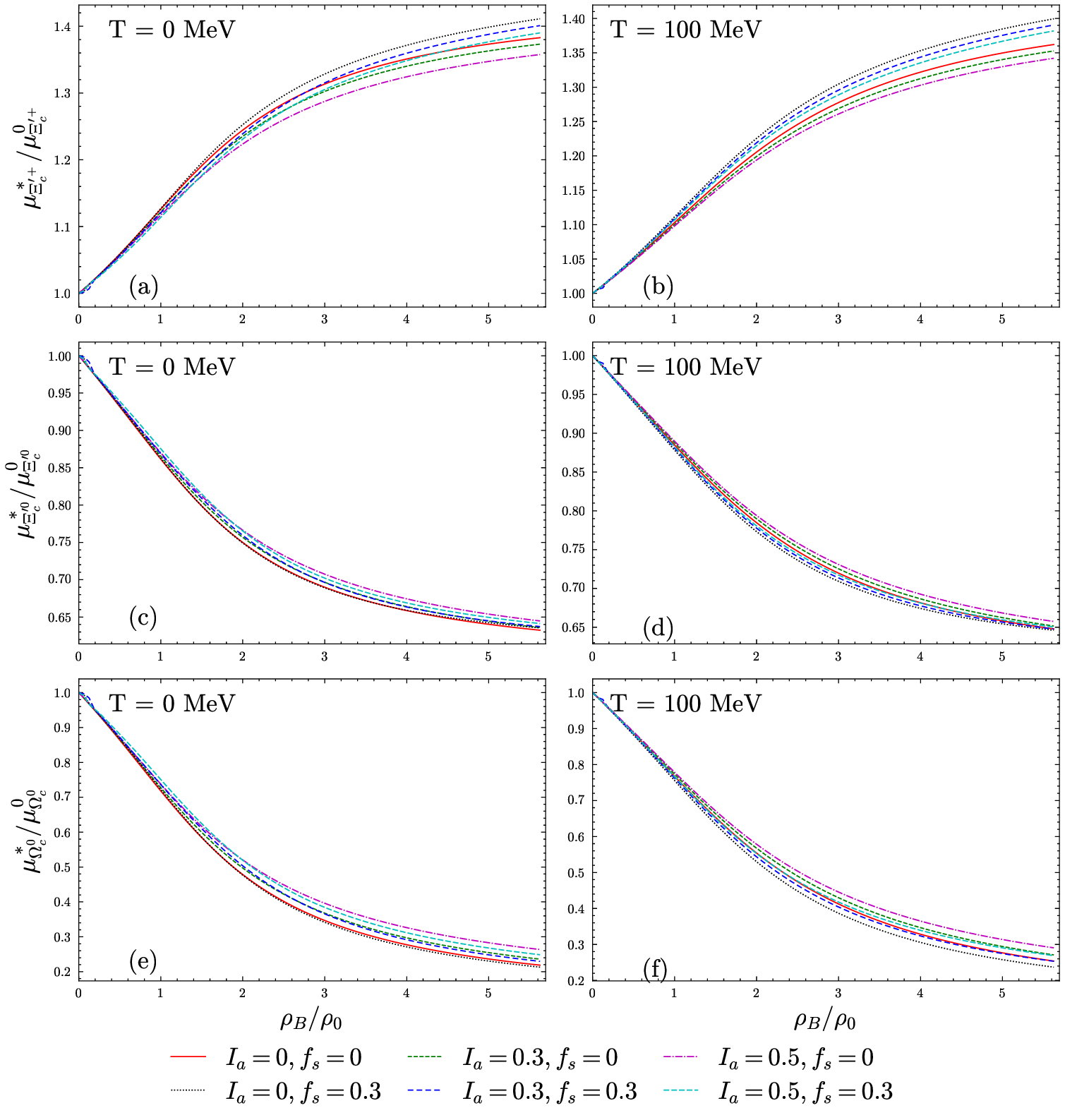}\hfill
	\caption{Effective magnetic moments  for \spinhalf, singly charmed  $\Xi_c^{\prime +},\ \Xi_c^{\prime 0}$ and $\Omega_c^0$ baryons relative to their respective vacuum values are shown as a function of baryon densities $\rho_B$ (in units of nuclear saturation density $\rho_0$) at temperatures $T = 0$ [in subplots (a), (c) and (e)] and $100$ [in subplots (b), (d) and (f)] MeV. Results are shown for isospin asymmetry $I_a = 0,\ 0.3,\ 0.5$ and strangeness fractions $f_s = 0,\ 0.3$.}
	\label{fig_moment_Xiprime_c}
\end{figure}

\begin{figure}[ht!] 
		\includegraphics[width= 18 cm, height=14 cm] {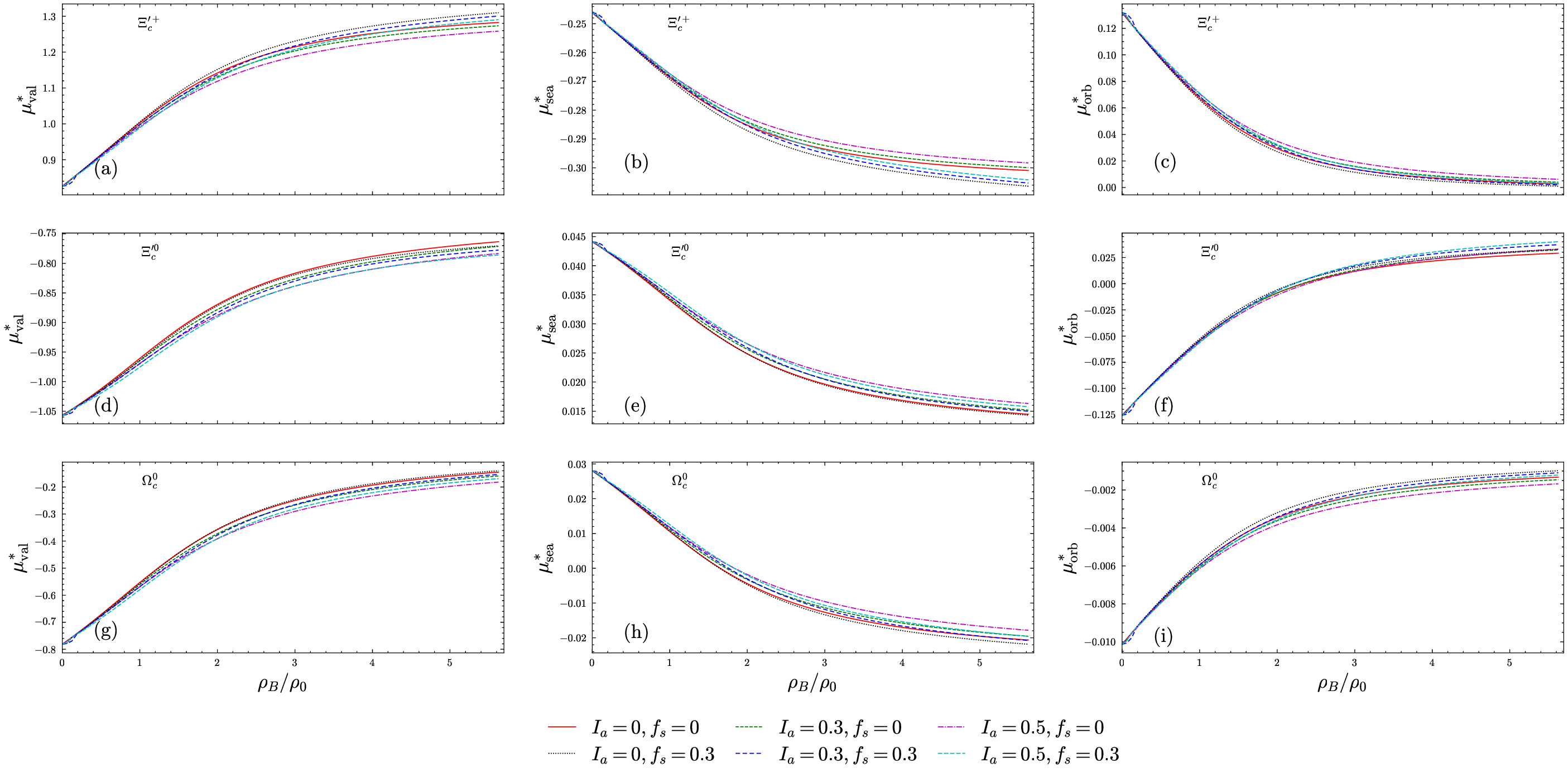}\hfill

		\caption{Contributions of valence, sea and orbital moment of sea quarks ($\mu_{\mathrm{val}}^{*},\ \mu_{\mathrm{sea}}^{*}\ \mu_{orb}^{*}$) towards the total magnetic moment of \spinhalf singly charmed $\Xi_c^{\prime +}$[in subplots (a), (b) and (c)], $\Xi_c^{\prime 0}$[in subplots (d), (e) and (f)] and $\Omega_c^{0}$[in subplots (g), (h) and (i)] are shown as a function of baryon density $\rho_B$ (in units of nuclear saturation density $\rho_0$) at temperatures $T = 0$. Results are shown for isospin asymmetry $I_a = 0,\ 0.3,\ 0.5$ and strangeness fractions $f_s = 0,\ 0.3$.}
		\label{fig_moment_Xiprime_indi}
	\end{figure}

	
The magnetic moments of spin$-\frac{1}{2}^{+}$ baryons, $\Xi^{\prime+}_c$,  $\Xi^{\prime0}_c$  and $\Omega_c^{0}$ as a function of
baryon density $\rho_B$ are shown in Fig. \ref{fig_moment_Xiprime_c}, while the individual contributions due to valence, sea and orbital moment of sea quarks towards their magnetic moments are shown in Fig. \ref{fig_moment_Xiprime_indi}. 
In cold symmetric nuclear medium, at baryon density $3\rho_0$, the ratio in-medium to vacuum value  of magnetic moments shows an overall increase of the order of $31.32\%$, for  $\Xi^{\prime+}_c$, while a decrease of the order of $31.01\%$ and $65.27 \%$  is observed for $\Xi^{\prime0}_c$  and $\Omega_c^{0}$, respectively. It is also observed that with the increase in isospin asymmetry, $I_a$, the impact of medium modifications is reduced, for example, in nuclear medium ($f_s =0$) at temperature $T = 0$ MeV and $\rho_{B} = 3 \rho_0$, value of magnetic moment $\mu_{\Xi^{\prime+}_c}^{*}$  decrease from 0.935 (Table \ref{table:mub_eta0fs0}) to 0.916 (Table \ref{table:mub_eta5fs0}), as $I_a$ changes from zero to $I_a=0.5$.
On other side, for the fixed value of $I_a = 0$,  changing strangeness fraction from $f_s = 0$ to $0.3$ increases value of $\mu_{\Xi^{\prime+}_c}^{*}$ to $0.946$ (Table \ref{table:mub_eta0fs3_1by2}). 
As is the case with magnetic moments for singly charmed 
$\Sigma_c$ baryons, for $\Xi^{\prime+}_c$,  $\Xi^{\prime0}_c$  and $\Omega_c^{*-}$ also, the valence component of the magnetic moment dominates over sea and orbital contributions.

	
\begin{figure}[ht!] 
		\includegraphics[width= 16 cm, height=18 cm]{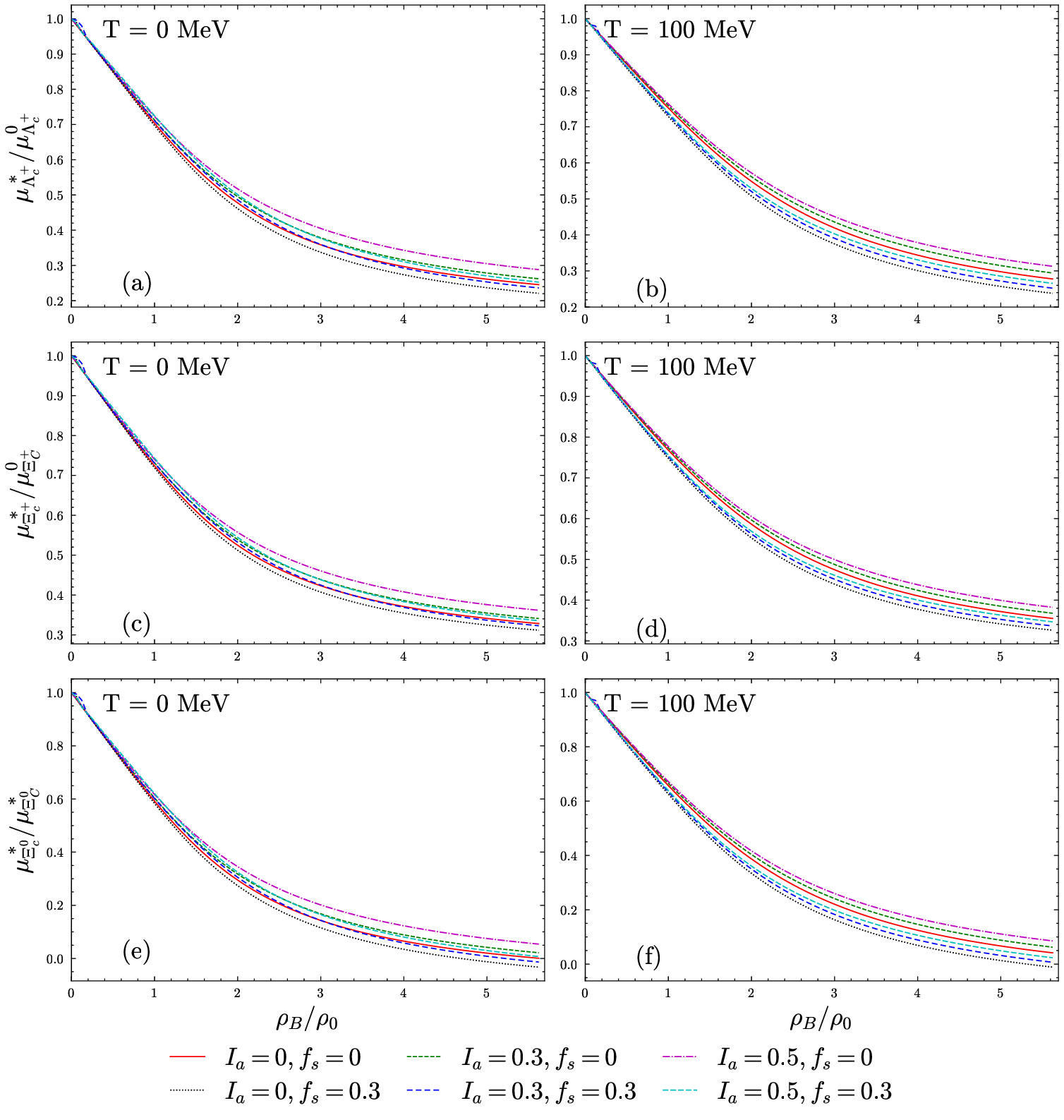}\hfill
		\caption{Effective magnetic moments for \spinhalf, singly charmed  $\Lambda_c^{+},\ \Xi_c^{+}$ and  $\Xi_c^{0}$ baryons relative to their respective vacuum values are shown as a function of baryon densities $\rho_B$ (in units of nuclear saturation density $\rho_0$) at temperatures $T = 0$ [in subplots (a), (c) and (e)] and $100$ [in subplots (b), (d) and (f)] MeV. Results are shown for isospin asymmetry $I_a = 0,\ 0.3,\ 0.5$ and strangeness fractions $f_s = 0,\ 0.3$.}
		\label{fig_moment_Xi_c}

	\end{figure}

\begin{figure}[ht!] 
		\includegraphics[width= 16 cm, height=18 cm]{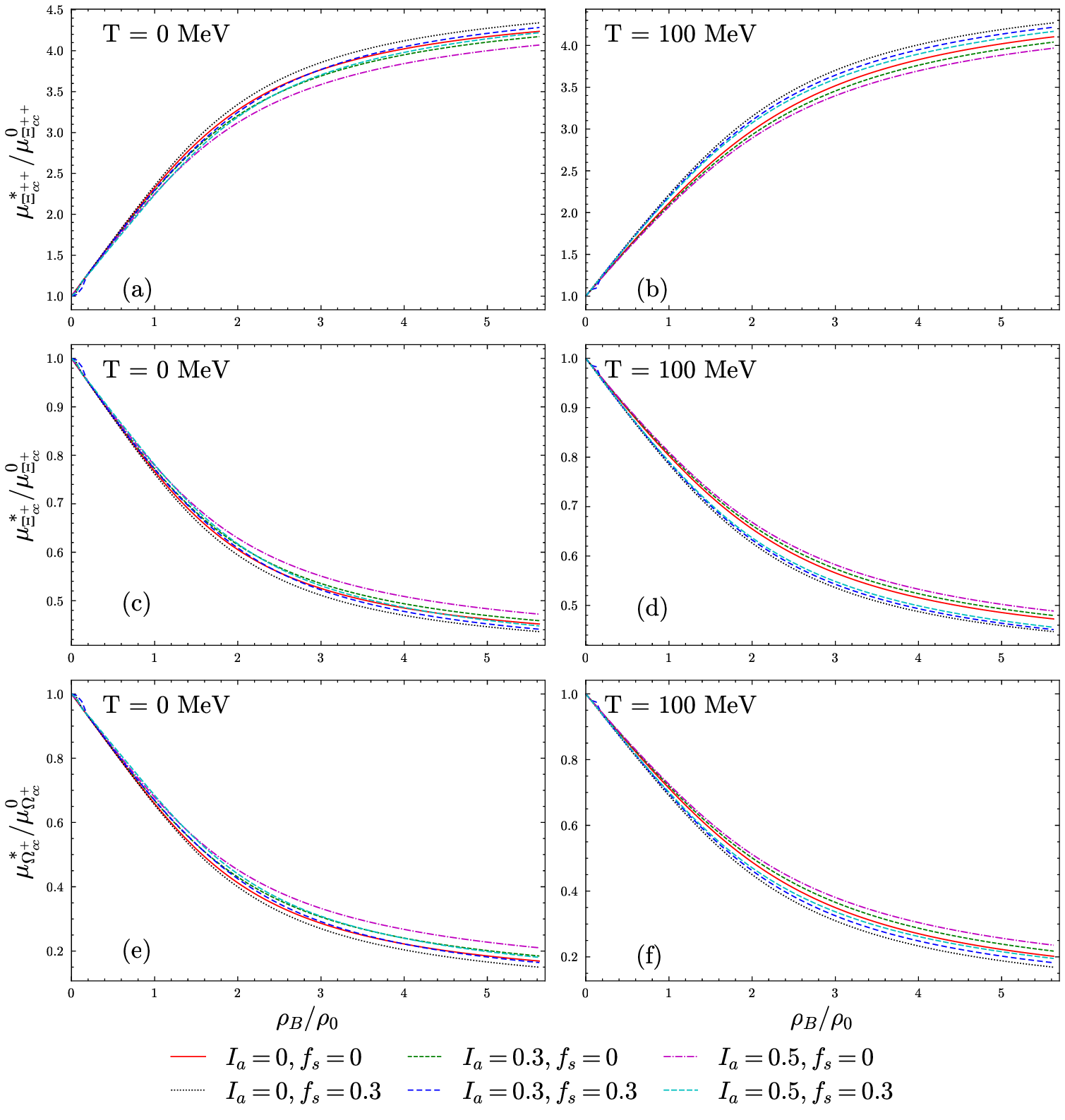}\hfill
		\caption{Effective magnetic moments for \spinhalf, doubly charmed  $\Xi_{cc}^{++},\ \Xi_{cc}^{+}$ and $\Omega_{cc}^{+}$ baryons relative to their respective vacuum values are shown as a function of baryon densities $\rho_B$ (in units of nuclear saturation density $\rho_0$) at temperatures $T = 0$ [in subplots (a), (c) and (e)] and $100$ [in subplots (b), (d) and (f)] MeV. Results are shown for isospin asymmetry $I_a = 0,\ 0.3,\ 0.5$ and strangeness fractions $f_s = 0,\ 0.3$.}
		\label{fig_moment_Xi_cc}
	\end{figure}

In Figs. \ref{fig_moment_Xi_c}
and \ref{fig_moment_Xi_cc}, the magnetic moment ratios are shown for spin$-\frac{1}{2}^{+}$  singly charmed $\Lambda_c^{+}$,
$\Xi_c^{+}$, $\Xi_c^{0}$ and doubly charmed baryons
$\Xi_{cc}^{++}$, 	$\Xi_{cc}^{+}$, $\Omega_{cc}^{+}$, respectively.
As can be seen from
Fig. \ref{fig_moment_Xi_c}, 
the magnitude of magnetic moments of singly charmed baryons $\Lambda_c^{+}$, $\Xi_c^{+}$, $\Xi_c^{0}$
 decreases as a function of baryon density, $\rho_B$ of the medium.
For given density $\rho_B$, the finite isospin asymmetry, $I_a$ of the medium causes an increase
whereas the strangeness fraction $f_s$, 
decreases the values of in-medium magnetic moments of these three baryons.
In the QMC model calculations, the magnetic moments of  $\Lambda_c^{+}$,
$\Xi_c^{+}$, $\Xi_c^{0}$
are modified very slightly as a function of density of the nuclear medium. Actually, the magnetic moment of these baryons in the QMC model (due to valence quark)  is equal to the magnetic moment $\mu_c$ of charm quark, which is observed to be modified very less with density \cite{Tsushimaptep2022}.
In our calculations, the magnetic moment  of charm quark is calculated using
the expression given in Eq. (\ref{magandmas}). Since $\mu_c^*$ is proportional to the effective mass $m_u^*$ of light quark and its modification in dense matter causes the change in 
$\mu_c^*$ and hence, in the 
magnetic moments of 
$\Lambda_c^{+}$,
$\Xi_c^{+}$, $\Xi_c^{0}$
in our calculations, as can be seen from Table \ref{table:mub_eta0fs0}. For $\Lambda_c^{+}$, $\Xi_c^{+}$, $\Xi_c^{0}$ the dominant valence component decreases with the increase in density while the contribution coming from orbital and sea components is negligibly small (Table \ref{table:mub_eta0fs0}), thus resulting in the decrease of magnetic moment ratios as observed from Fig. \ref{fig_moment_Xi_c}. 

The magnitude of doubly charmed $\Xi_{cc}^{++}$ baryon increases with an increase in the density $\rho_B$ 
of the nuclear medium, whereas, in case of	$\Xi_{cc}^{+}$ and $\Omega_{cc}^{+}$, their magnetic moment decreases with $\rho_B$.
Increase of isospin asymmetry parameter from zero to $0.5$, causes decrease in the magnitude (becomes less negative) of $\mu_{\Xi_{cc}^{++}}^{*}$, whereas the strangeness fraction causes as increase (becomes more negative).
For the $\Xi_{cc}^{+}$ and $\Omega_{cc}^{+}$ baryons, which have positive values of magnetic moments,
the isospin asymmetry causes an increase and the strangeness fraction decreases these values.

\begin{figure}[ht!] 
		\includegraphics[width= 16 cm, height=18 cm]{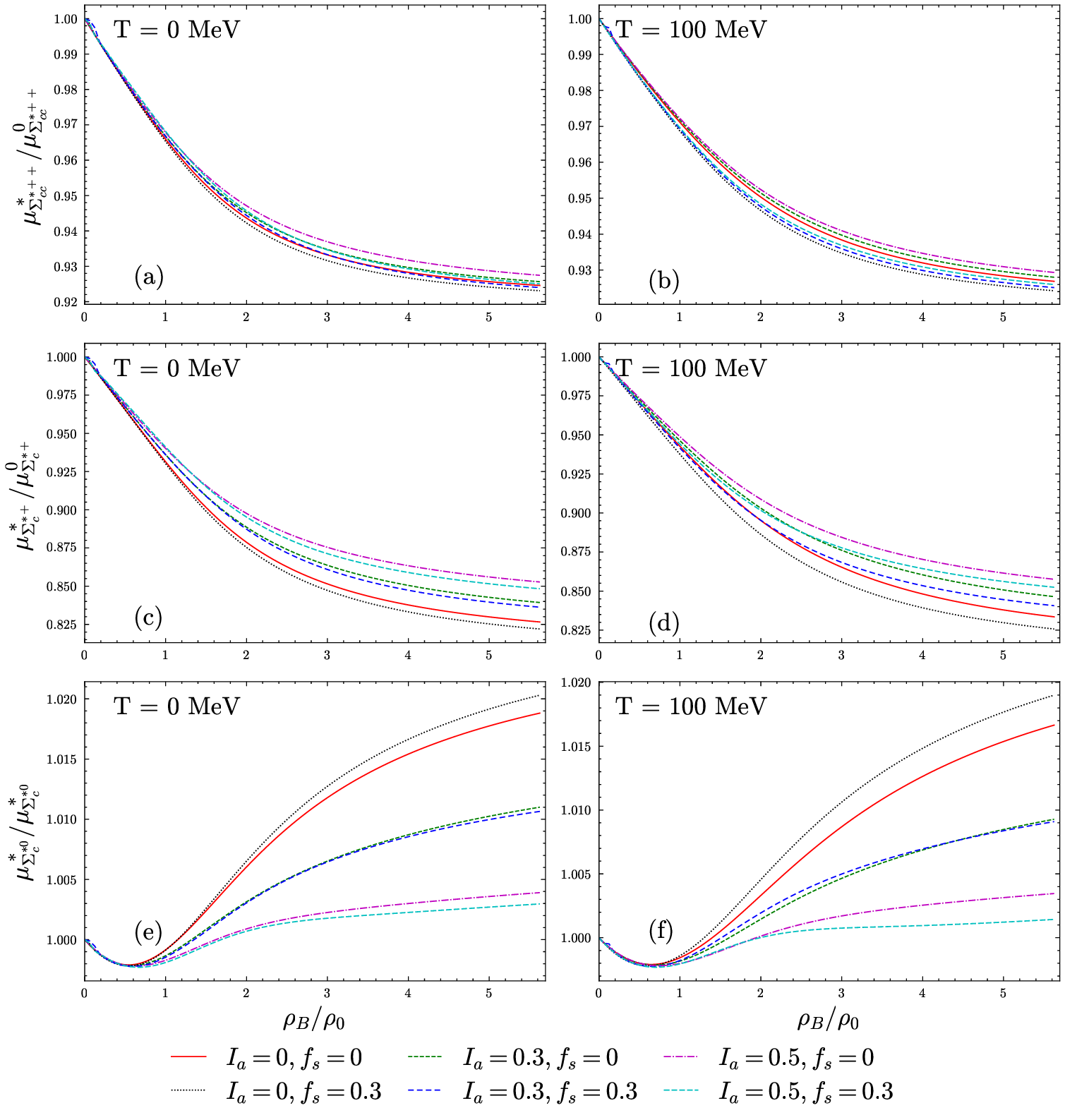}\hfill
		\caption{Effective magnetic moments for spin$-\frac{3}{2}^+$, singly charmed  $\Sigma_c^{*++},\ \Sigma_c{*+}$ and  $\Sigma^{*0}_c$ relative to their respective vacuum values are shown as a function of baryon densities $\rho_B$ (in units of nuclear saturation density $\rho_0$) at temperatures $T = 0$ [in subplots (a), (c) and (e)] and $100$ [in subplots (b), (d) and (f)] MeV. Results are shown for isospin asymmetry $I_a = 0,\ 0.3,\ 0.5$ and strangeness fractions $f_s = 0,\ 0.3$.}
\label{fig_moment_Sigma_starc}
	\end{figure}

\begin{figure}[ht!] 
	\includegraphics[width= 18 cm, height=14 cm] {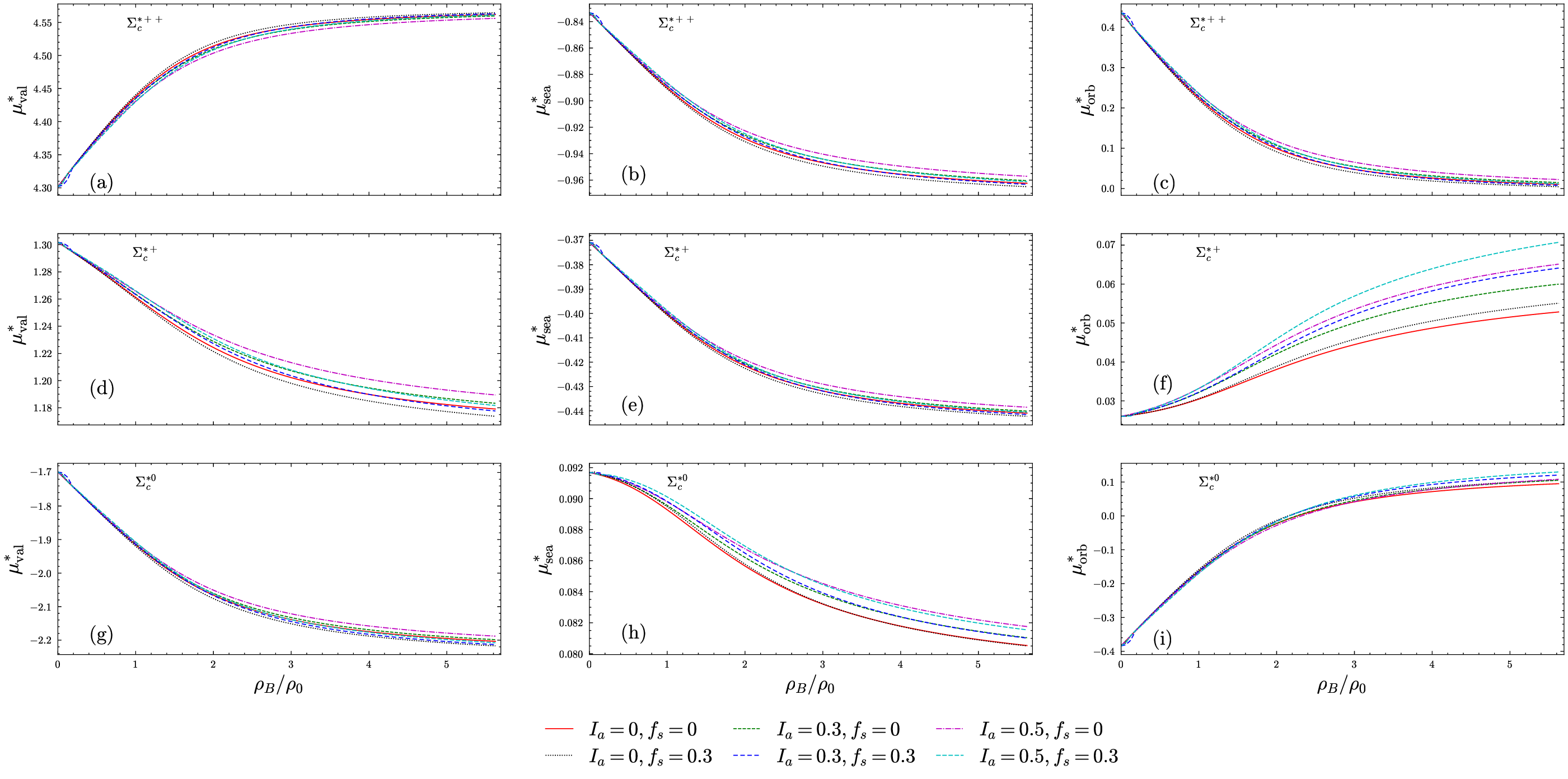}\hfill

	\caption{Contributions of valence, sea and orbital moment of sea quarks ($\mu_{\mathrm{val}}^{*}$,$\mu_{\mathrm{sea}}^{*}$, $\mu_{orbit}^{*}$) towards the total magnetic moment of \spinthhalf singly charmed $\Sigma_c^{*++}$[in subplots (a), (b) and (c)], $\Sigma_c^{*+}$[in subplots (d), (e) and (f)] and $\Sigma_c^{0}$[in subplots (g), (h) and (i)] are shown as a function of baryon density $\rho_B$ (in units of nuclear saturation density $\rho_0$) at temperatures $T = 0$. Results are shown for isospin asymmetry $I_a = 0,\ 0.3,\ 0.5$ and strangeness fractions $f_s = 0,\ 0.3$.}
	\label{fig_comp_sigmacstar}
\end{figure}

\begin{figure}[ht!]
	\includegraphics[width= 17 cm, height=18 cm] {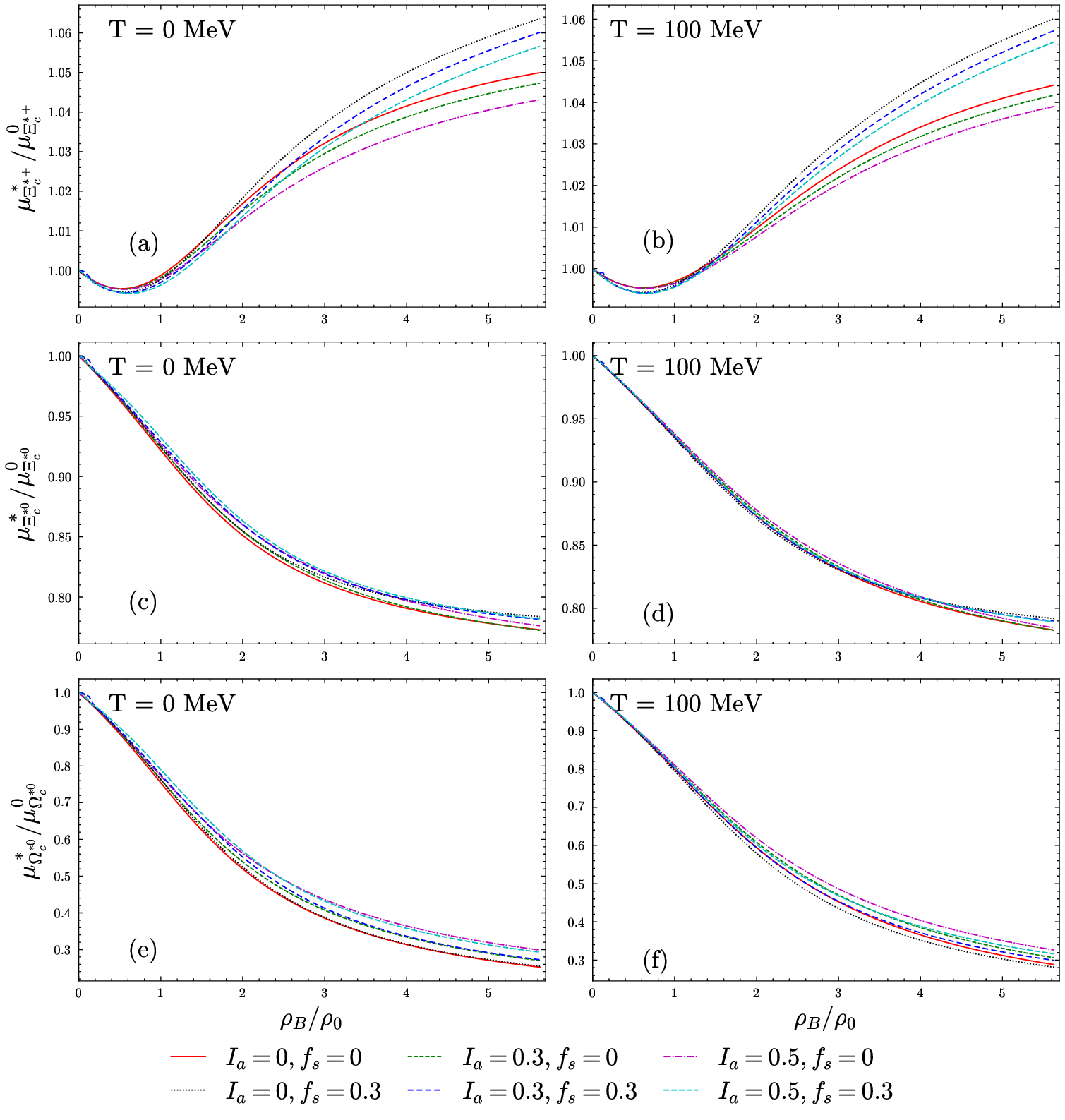}\hfill
	\caption{Effective magnetic moments for spin$-\frac{3}{2}^+$ singly charmed $\Xi_c^{*+},\ \Xi_c^{*0}$,and $\Omega_c^{*0}$ baryons relative to their respective vacuum values are shown as a function of baryon densities $\rho_B$ (in units of nuclear saturation density $\rho_0$) at temperatures $T = 0$ [in subplots (a), (c) and (e)] and $100$ [in subplots (b), (d) and (f)] MeV. Results are shown for isospin asymmetry $I_a = 0,\ 0.3,\ 0.5$ and strangeness fractions $f_s = 0,\ 0.3$.}
	\label{fig_moment_XiOmega_starc}			

	\end{figure}

\begin{figure}[ht!] 
			\includegraphics[width= 17 cm, height=18 cm]{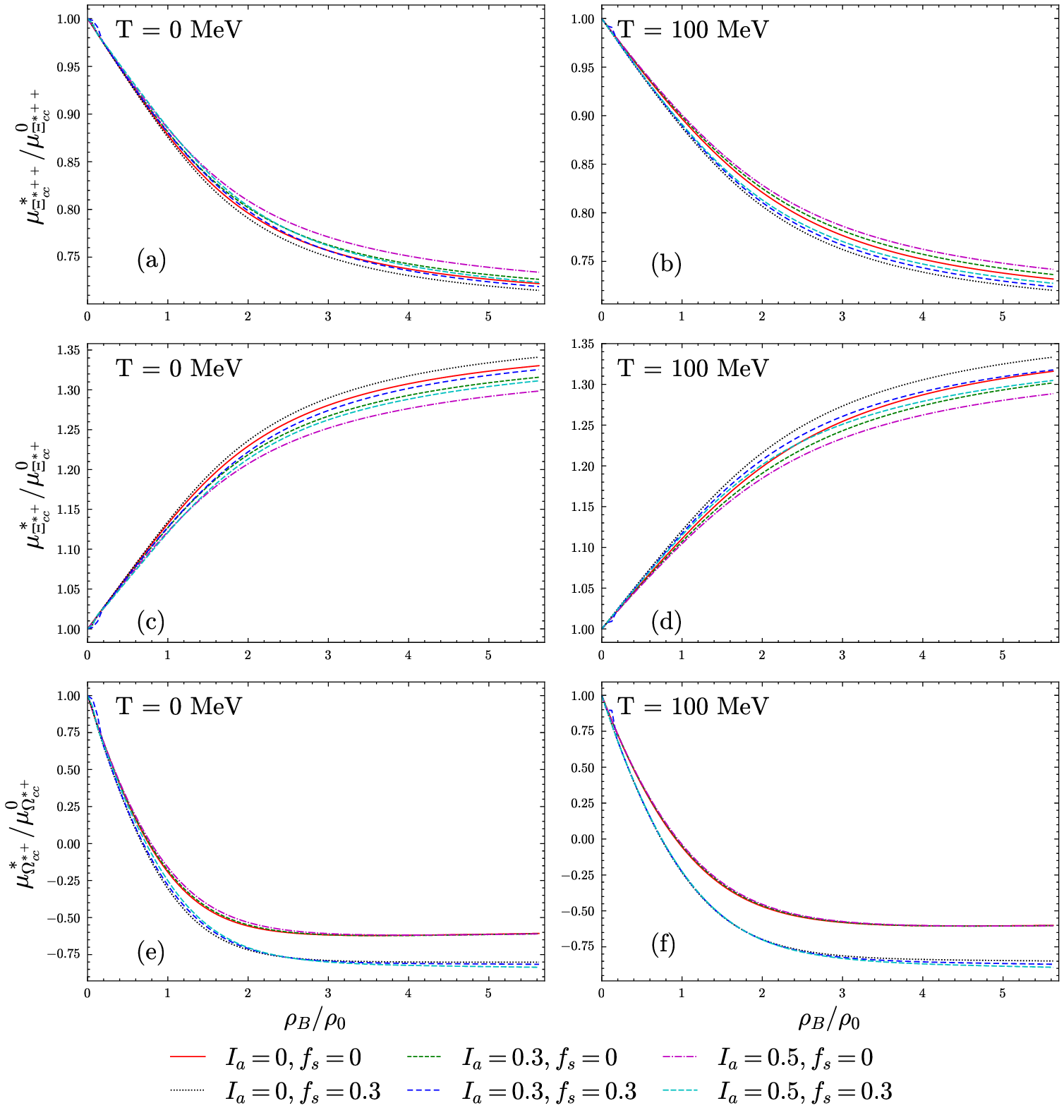}\hfill
			\caption{Effective magnetic moments for spin$-\frac{3}{2}^+$ doubly charmed  $\Xi_{cc}^{*++},\ \Xi_{cc}^{*+}$ and $\Omega_{cc}^{*+}$ baryons relative to their respective vacuum values are shown as a function of baryon densities $\rho_B$ (in units of nuclear saturation density $\rho_0$) at temperatures $T = 0$ and $100$ MeV. Results are shown for isospin asymmetry values, $I_a = 0,\ 0.3,\ 0.5$ and strangeness fractions $f_s = 0,\ 0.3$.}
			\label{fig_moment_Xi_starcc}			
		\end{figure}

In Figs. \ref{fig_moment_Sigma_starc}	
to 
\ref{fig_moment_Xi_starcc}
we have shown the magnetic moments ratio, $\mu_{B}^{*}/\mu_{B}^0$, of spin$\frac{3}{2}^+$ singly
($\Sigma_c^{*++}, \Sigma_c^{*+}, \Sigma_c^{*0},  ,  \Xi^{*\prime +}_c, \Xi^{*\prime0}_c$ and $\Omega^{*0}_c$) and doubly ($\Xi_{cc}^{*++}$, 	$\Xi_{cc}^{*+}$, $\Omega_{cc}^{*+}$) charmed 
 baryons as a function of baryon density $\rho_B$ (in units of $\rho_0$), for nuclear ($f_s = 0$) and strange  matter ($f_s = 0.3$), at temperatures $T = 0$ and $100$ MeV
 and  isospin asymmetries $I_a = 0,0.3$ and $0.5$.
 From Fig. \ref{fig_moment_Sigma_starc}, we observe that the  magnitudes of magnetic moments for
 $\Sigma_c^{*++}$ and $\Sigma_c^{*+}$ decrease with density of the dense medium, whereas the magnitude of the magnetic moment of  $\Sigma_c^{*0}$ shows a small decrease 
 initially  and then increases after certain value showing a small increase overall. In cold symmetric nuclear medium, at baryon density $3\rho_0$, the ratio of effective magnetic moment to vacuum value shows an overall decrease of the order of $15.25 \%$ and $14.83 \%$ for $\Sigma_c^{*++}$ and  $\Sigma_c^{*+}$, while for $\Sigma_c^{*0}$ a small increase of $1.1\%$ is observed. This small increase in $\Sigma_c^{*0}$ can be attributed to a large decrease in magnitude of orbital component which compensates the increase in magnitude of $\mu_{\mathrm{val}}^*$ which shows an increase of about $17.0\%$ at baryon density $3\rho_0$ for cold symmetric nuclear matter, as can be observed from Fig. \ref{fig_comp_sigmacstar}.
 For a given density of the medium, the impact of isospin asymmetry causes an increase (decrease)
 in the magnetic moment values of baryons having positive  (negative) value whereas strangeness has an opposite effect.
 In case of baryons $\Xi^{*+}_c, \Xi^{*0}_c$ and $\Omega^{*0}_c$, plotted in Fig. \ref{fig_moment_XiOmega_starc}, the
 magnitude of the magnetic moments of  $\Xi^{*0}_c$ and $\Omega^{*0}_c$
 decrease as as a function of $\rho_B$ while the 
 value of $\mu_{\Xi^{*+}_c}^{*}$ initially decreases and then increases, thus showing a small overall increase in $\rho_B$. The small change in overall magnitude of $\Xi^{*+}_c$ is attributed to orbital component whose value changes from $0.21$ at $\rho_{B} = 0$ to $0.03$ at $\rho_{B} = 3\rho_0$, thus compensating the increase in valence component of magnetic moment (Table ~\ref{table:mub_eta5fs0_3by2}).
 Since the $\Omega^{*0}_c$ baryon has no light $u$ or $d$ quark and is composed of  two strange $s$ and one charm $c$ quarks, its in-medium magnetic moments are more sensitive to the strangeness fraction as compared to the isospin asymmetry. 
The in-medium value of the magnetic moment of doubly charmed 
spin$-\frac{3}{2}^+$ $\Xi_{cc}^{*++}$ baryon
is observed to decrease as a function of $\rho_B$, as can be seen from Fig. \ref{fig_moment_Xi_starcc}.
In case of $\Xi_{cc}^{*+}$,
its magnetic moment has negative value whose magnitude increases with $\rho_B$.
For the $\mu_{\Omega_{cc}^{*+}}$, 
staring from $0.05$ in the free space, it becomes more and more negative as the baryonic density is increased.
Also being composed of one strange and two charm quarks, it is sensitive 
to the strangeness fraction as compared to the isospin asymmetry of the medium.



	
	\section{Summary}
	\label{sec:summary}
	To summarize, in the current study we first investigated the modification in masses of spin$-\frac{1}{2}^+$ and spin$-\frac{3}{2}^+$ charmed baryons in hadronic medium for different values of isospin asymmetry $I_a$ and strangeness fraction $f_s$. The impact of medium was simulated through chiral SU(3) quark mean field model and modification in masses of charmed baryons is calculated through the behavior of constituents $u$, $d$ and $s$. In the present study mass of $c$-quark, $m_c$ is considered to remain same with the medium  which is going to have an minimal impact on our calculation because of its higher mass in comparison to other three quarks. The constituent quark masses are observed to decrease with density of the hadronic medium which is also reflected in the masses of charmed baryons as well. The introduction of isospin asymmetry in the medium, is observed to cause a splitting in $u$ and $d$ quark masses which also leads to splitting in masses of charmed baryons as well. The increase in strangeness fraction causes a higher decrease in the strange quark mass than the light quark masses, thus having an higher impact on the charmed baryons consisting of valence strange quarks in them. 	
	 Isospin asymmetry effects are observed to be more significant for the baryons composed of light  $u$ or $d$ quarks, whereas 
	strangeness effects became appreciable for the baryons having strange quarks.  The variation in effective masses is found to effect the magnetic moments of the charmed baryon as the effective magnetic moments of the constituent quarks are dependent upon the effective mass of the charmed baryons under consideration. In the present study, we have calculated the ratio of effective magnetic moments to the values at free space using SU(4) chiral constituent quark model by considering valence quark contributions with configuration mixing, contribution from sea quarks and orbital contribution of sea quarks. For the charmed baryons, it has been observed that even though the contribution from valence quarks is  dominating  but the sea and orbital components also have an significant impact on the variation of effective magnetic moments with density. Only in case of $\Omega_{cc}^{0}$, $\Omega_{cc}^+$, $\Xi_c^+$, $\Xi_c^0$ and $\Lambda_c^+$ both the sea and orbital contributions are very small in comparison to the valence component as these baryons are dominated by heavy quarks in the valence structure. Also the contribution of orbital part at vacuum also has the same sign as that of the contribution from valence quarks while sea quraks are having the contribution with opposite sign to both of them (except for $\Xi_c^0$ and $\Omega_{cc}^{*+}$). In some of the cases these contributions are found to change sign at the higher densities, e.g., sea quark contribution to $\Omega_c^0$. 
The present study of in-medium masses and magnetic moments of baryons will be useful for experimental facilities where dense matter at finite baryon densities may  be produced.
	

\begin{table}
	\begin{tabular}{|c|c|c|c|c|c|c|c|}
		\hline
		Baryon & Present Study & \cite{Tsushimaptep2022} & \cite{Aliev2015} & \cite{Faessler2006} & \cite{Patel2008} & \cite{Barik1983}  & \cite{Jena1986} \\
		\hline
		$ \Sigma_c^{++} $ & 2.187  & 1.378  & 2.4 & 1.76 & 2.279 &2.44 & 2.448  \\ \hline
		$ \Sigma_c^{+} $  & 0.333  & 0.238 & 0.5 & 0.36 & 0.501 & 0.525 & 0.524  \\ \hline
		$ \Sigma_c^{0} $ & $-1.513$  & $-0.903$ & $-1.5$ & $-1.04$ & $-1.015$ & $-1.391$ & $-1.400$ \\ \hline
		$ \Xi_c^{\prime+} $ & 0.712  & $-- $& $-- $ &  $-- $ &  $-- $ &   $-- $ &  $-- $ \\ \hline
		$ \Xi_c^{\prime0} $ & $-1.138  $&  $-- $&  $-- $&  $-- $ &  $-- $ &  $-- $ &   $-- $\\ \hline
		$ \Omega_c^{0} $ & $-0.763$  &  $-- $ &  $-- $ &  $-- $ &  $-- $ &   $-- $ &  $-- $ \\ \hline
		$ \Xi_c^{+} $ & 0.324  & 0.424 & 0.8 & 0.41 & 0.711 & 0.796  & 0.779 \\ \hline
		$ \Xi_c^{0} $ & 0.209  & 0.424 & $-1.2 $& 0.39 & $-0.966  $& $-1.12 $& $-1.145 $\\ \hline
		$ \Lambda_c^{+} $ & 0.299  & 0.423 &- &0.42 &0.385 &0.341  &0.352  \\ \hline
		$ \Xi_{cc}^{++} $ & $-0.085  $& $-- $  &  $-- $&  $-- $&   $-- $&  $-- $ &  $-- $ \\ \hline
		$ \Xi_{cc}^{+} $ & 0.667  &  $-- $&  $-- $ &   $-- $&  $-- $ &  $-- $ &  $-- $ \\ \hline
		$ \Omega_{cc}^{+} $ & 0.515  &  $-- $ &  $-- $&  $-- $ &  $-- $ &   $-- $ &  $-- $\\ \hline

%

	\end{tabular}
	\caption{Comparison of vacuum values of magnetic moments of  spin$-\frac{1}{2}^{+}$ charmed baryons with various theoretical models.} 	\label{table:comparison_theor_models}
\end{table}

\begin{table} 
\begin{tabular}{|c|c|c|c|c|c|c|c|}
		\hline
		Baryon & Present Study & \cite{Choudhury:1976dn} & \cite{Aliev:2000cy} \\
		\hline
		$ \Sigma_c^{*++} $ & 3.904  & 4.39  & $4.81\pm1.22$   \\ \hline
		$ \Sigma_c^{*+} $  & 0.957  & 1.39 & $2.00
		\pm 0.46$    \\ \hline
		$ \Sigma_c^{*0} $ & $-1.991$  & $-1.61$ & $-0.81 \pm 0.20$  \\ \hline
		$ \Xi_c^{*+} $ & 1.554  & $1.74 $& $1.68 
		\pm 0.42 $   \\ \hline
		$ \Xi_c^{*0} $ & $-1.393  $&  $-1.26 $&  $-0.68 \pm 0.18 $ \\ \hline
		$ \Omega_c^{*0} $ & $-0.796$  &  $-0.91 $ &  $-0.62 \pm 0.18 $   \\ \hline
		$ \Xi_{cc}^{*++} $ & 1.813  & 2.78 & $--$   \\ \hline
		$ \Xi_{cc}^{*+} $ & -0.79  & -0.22 & $-- $ \\ \hline
		$ \Omega_{cc}^{*+} $ & 0.05  & 0.13 &$--$    \\ \hline
	\end{tabular}
	\caption{Comparison of vacuum values of magnetic moments of  spin$-\frac{3}{2}^{+}$ charmed baryons with various theoretical models.} 	\label{table:comparison_theor_models3_2}
\end{table}


		\begin{sidewaystable}
			\begin{tabular}{|c|c|c|c|c|c|c|c|c|c|c|c|c|}
				\hline Baryon
			 & \multicolumn{4}{|c|}{$\rho_B=0$} & \multicolumn{4}{|c|}{$\rho_B=\rho_0$} & \multicolumn{4}{|c|}{$\rho_B=3 \rho_0$} \\
				\hline & $\mu_{B,\mathrm {val}}^{*}$ & $\mu_{B,\text {sea}}^{*}$ & $\mu_{B,\text {orbital }}^{*}$ & $\mu_B^*$ & $\mu_{B,\mathrm{val}}^{*}$ & $\mu_{B, \text {sea }}^{*}$ & $\mu_{B,\text {orbital }}^{*}$ & $\mu_B^*$ & $\mu_{B,\mathrm{val}}^{*}$ & $\mu_{B,\text {sea }}^{*}$ & $\mu_{B,\text {orbital }}^{*}$ & $\mu_B^*$ \\
				
\hline $\mu_{ \Sigma_c^{++} }^*(\mu_N )$ & 2.434 &  $-0.52$ &  0.274 &  2.187 &  2.611 &  $-0.557$ &  0.14 &  2.194 &  2.782 &  $-0.594$ &  0.03 &  2.217 \\
\hline $\mu_{ \Sigma_c^{+} }^*(\mu_N )$ & 0.551 &  $-0.234$ &  0.016 &  0.333 &  0.614 &  $-0.254$ &  0.019 &  0.379 &  0.678 &  $-0.275$ &  0.028 &  0.432 \\
\hline $\mu_{ \Sigma_c^{0} }^*(\mu_N )$ & $-1.332 $&  0.06 &  $-0.241 $&  $-1.513 $&  $-1.383 $&  0.058 &  $-0.102 $&  $-1.427 $&  $-1.425 $&  0.054 &  0.026 &  $-1.345$\\
\hline $\mu_{ \Xi_c^{\prime+} }^*(\mu_N )$ & 0.826 &  $-0.246 $&  0.132 &  0.712 &  1.003 &  $-0.269 $&  0.067 &  0.801 &  1.215 &  $-0.294 $&  0.014 &  0.935 \\
\hline $\mu_{ \Xi_c^{\prime0} }^*(\mu_N )$ & $-1.057 $&  0.044 &  $-0.126 $&  $-1.138 $&  $-0.961 $&  0.034 &  $-0.054 $&  $-0.981 $&  $-0.817 $&  0.02 &  0.012 &  $-0.785$\\
\hline $\mu_{ \Omega_c^{0} }^*(\mu_N )$ & $-0.781 $&  0.028 &  $-0.01 $&  $-0.763 $&  $-0.553 $&  0.011 &  $-0.006 $&  $-0.548 $&  $-0.25 $&  $-0.013 $&  $-0.002 $&  $-0.265$\\
\hline $\mu_{ \Xi_c^{+} }^*(\mu_N )$ & 0.334 &  $-0.018 $&  0.008 &  0.324 &  0.25 &  $-0.019 $&  0.004 &  0.235 &  0.157 &  $-0.02 $&  0.001 &  0.137 \\
\hline $\mu_{ \Xi_c^{0} }^*(\mu_N )$ & 0.217 &  $0.0 $&  $-0.008 $&  0.209 &  0.128 &  $0.0 $&  $-0.003$ &  0.124 &  0.03&  $0.001 $&  0.001 &  0.03 \\
\hline $\mu_{ \Lambda_c^{+} }^*(\mu_N )$ & 0.317 &  $-0.019 $&  0.001 &  0.299 &  0.229 &  $-0.02 $&  0.001 &  0.21 &  0.127 &  $-0.021 $&  0.002 &  0.107 \\
\hline $\mu_{ \Xi_{cc}^{++} }^*(\mu_N )$ & $-0.132 $&  0.102 &  $-0.056 $&  $-0.085 $&  $-0.276 $&  0.106 &  $-0.029 $&  $-0.199 $&  $-0.426 $&  0.11 &  $-0.006 $&  $-0.322$\\
\hline $\mu_{ \Xi_{cc}^{+} }^*(\mu_N )$ & 0.634 &  $-0.016 $&  0.049 &  0.667 &  0.505 &  $-0.014 $&  0.021 &  0.511 &  0.368 &  $-0.013 $&  $-0.005$ &  0.35 \\
\hline $\mu_{ \Omega_{cc}^{+} }^*(\mu_N )$ & 0.522 &  $-0.009 $&  0.002 &  0.515 &  0.344 &  $-0.005 $&  0.001 &  0.34 &  0.148 &  0.0 &  0.0 &  0.148 \\
				
\hline
\end{tabular}
\caption{Values of effective magnetic moments of spin$-\frac{1}{2}^{+}$ charmed baryons in symmetric nuclear matter ($I_{a} =0$ and $f_s =0$) are tabulated above at temperature $T=0$ MeV and compared with values at $\rho_B =0$.}
\label{table:mub_eta0fs0}
\end{sidewaystable}

\begin{sidewaystable}
	\begin{tabular}{|c|c|c|c|c|c|c|c|c|c|c|c|c|}
		\hline Baryon
		& \multicolumn{4}{|c|}{$\rho_B=0$} & \multicolumn{4}{|c|}{$\rho_B=\rho_0$} & \multicolumn{4}{|c|}{$\rho_B=3 \rho_0$} \\
		\hline & $\mu_{B,\mathrm {val}}^{*}$ & $\mu_{B,\text {sea}}^{*}$ & $\mu_{B,\text {orbital }}^{*}$ & $\mu_B^*$ & $\mu_{B,\mathrm{val}}^{*}$ & $\mu_{B, \text {sea }}^{*}$ & $\mu_{B,\text {orbital }}^{*}$ & $\mu_B^*$ & $\mu_{B,\mathrm{val}}^{*}$ & $\mu_{B,\text {sea }}^{*}$ & $\mu_{B,\text {orbital }}^{*}$ & $\mu_B^*$ \\
		
		\hline $\mu_{ \Sigma_c^{++} }^*(\mu_N )$ & 2.434 &  $-0.52 $&  0.274 &  2.187 &  2.601 &  $-0.555 $&  0.147 &  2.193 &  2.762 &  $-0.59 $&  0.041 &  2.213 \\
		\hline $\mu_{ \Sigma_c^{+} }^*(\mu_N )$ & 0.551 &  $-0.234 $&  0.016 &  0.333 &  0.611 &  $-0.253 $&  0.021 &  0.379 &  0.672 &  $-0.273 $&  0.034 &  0.433 \\
		\hline $\mu_{ \Sigma_c^{0} }^*(\mu_N )$ & $-1.332 $&  0.06 &  $-0.241 $&  $-1.513 $&  $-1.385 $&  0.059 &  $-0.106 $&  $-1.431 $&  $-1.427 $&  0.055 &  0.026 &  $-1.346$\\
		\hline $\mu_{ \Xi_c^{\prime+} }^*(\mu_N )$ & 0.826 &  $-0.246 $&  0.132 &  0.712 &  0.992 &  $-0.267 $&  0.071 &  0.795 &  1.188 &  $-0.291 $&  0.019 &  0.916 \\
		\hline $\mu_{ \Xi_c^{\prime0} }^*(\mu_N )$ & $-1.057 $&  0.044 &  $-0.126 $&  $-1.138 $&  $-0.969 $&  0.035 &  $-0.056 $&  $-0.99 $&  $-0.839 $&  0.022 &  0.012 &  $-0.805$\\
		\hline $\mu_{ \Omega_c^{0} }^*(\mu_N )$ & $-0.781 $&  0.028 &  $-0.01 $&  $-0.763 $&  $-0.569 $&  0.012 &  $-0.006 $&  $-0.563 $&  $-0.29 $&  $-0.01 $&  $-0.003 $&  $-0.302$\\
		\hline $\mu_{ \Xi_c^{+} }^*(\mu_N )$ & 0.334 &  $-0.018 $&  0.008 &  0.324 &  0.255 &  $-0.019 $&  0.004 &  0.24 &  0.168 &  $-0.02 $&  0.001 &  0.149 \\
		\hline $\mu_{ \Xi_c^{0} }^*(\mu_N )$ & 0.217 &  $-0.0 $&  $-0.008 $&  0.209 &  0.133 &  $-0.0 $&  $-0.003 $&  0.129 &  0.042 &  $-0.001 $&  0.001 &  0.042 \\
		\hline $\mu_{ \Lambda_c^{+} }^*(\mu_N )$ & 0.317 &  $-0.019 $&  0.001 &  0.299 &  0.235 &  $-0.02 $&  0.001 &  0.216 &  0.14 &  $-0.021 $&  0.002 &  0.121 \\
		\hline $\mu_{ \Xi_{cc}^{++} }^*(\mu_N )$ & $-0.132 $&  0.102 &  $-0.056 $&  $-0.085 $&  $-0.268 $&  0.106 &  $-0.03 $&  $-0.192 $&  $-0.408 $&  0.109 &  $-0.008 $&  $-0.307$\\
		\hline $\mu_{ \Xi_{cc}^{+} }^*(\mu_N )$ & 0.634 &  $-0.016 $&  0.049 &  0.667 &  0.513 &  $-0.014 $&  0.021 &  0.52 &  0.386 &  $-0.013 $&  $-0.005 $&  0.368 \\
		\hline $\mu_{ \Omega_{cc}^{+} }^*(\mu_N )$ & 0.522 &  $-0.009 $&  0.002 &  0.515 &  0.355 &  $-0.005 $&  0.001 &  0.351 &  0.172 &  $-0.0 $&  0.001 &  0.172 \\
		
		\hline
	\end{tabular}
	\caption{Values of effective magnetic moments of  spin$-\frac{1}{2}^{+}$ charmed baryons in asymmetric nuclear matter ($I_{a} =0.5$ and $f_s =0$) are tabulated above at temperature $T=0$ MeV 		and compared with values at $\rho_B =0$.}
	\label{table:mub_eta5fs0}
\end{sidewaystable}
	
\begin{sidewaystable}
	\begin{tabular}{|c|c|c|c|c|c|c|c|c|c|c|c|c|}
		\hline Baryon 
		& \multicolumn{4}{|c|}{$\rho_B=0$} & \multicolumn{4}{|c|}{$\rho_B=\rho_0$} & \multicolumn{4}{|c|}{$\rho_B=3 \rho_0$} \\
		\hline & $\mu_{B,\mathrm {val}}^{*}$ & $\mu_{B,\text {sea}}^{*}$ & $\mu_{B,\text {orbital }}^{*}$ & $\mu_B^*$ & $\mu_{B,\mathrm{val}}^{*}$ & $\mu_{B, \text {sea }}^{*}$ & $\mu_{B,\text {orbital }}^{*}$ & $\mu_B^*$ & $\mu_{B,\mathrm{val}}^{*}$ & $\mu_{B,\text {sea }}^{*}$ & $\mu_{B,\text {orbital }}^{*}$ & $\mu_B^*$ \\
	\hline $\mu_{ \Sigma_c^{++} }^*(\mu_N )$ & 2.434 &  $-0.52$ &  0.274 &  2.187 &  2.615 &  $-0.558 $&  0.138 &  2.194 &  2.791 &  $-0.596 $&  0.025 &  2.22 \\
	\hline $\mu_{ \Sigma_c^{+} }^*(\mu_N )$ & 0.551 &  $-0.234 $&  0.016 &  0.333 &  0.615 &  $-0.254 $&  0.019 &  0.38 &  0.682 &  $-0.275 $&  0.029 &  0.435 \\
	\hline $\mu_{ \Sigma_c^{0} }^*(\mu_N )$ & $-1.332 $&  0.06 &  $-0.241 $&  $-1.513 $&  $-1.384 $&  0.058 &  $-0.099 $&  $-1.425 $&  $-1.427 $&  0.054 &  0.033 &  $-1.34$\\
	\hline $\mu_{ \Xi_c^{\prime+} }^*(\mu_N )$ & 0.826 &  $-0.246 $&  0.132 &  0.712 &  1.005 &  $-0.269 $&  0.066 &  0.802 &  1.231 &  $-0.297 $&  0.011 &  0.946 \\
	\hline $\mu_{ \Xi_c^{\prime0} }^*(\mu_N )$ & $-1.057 $&  0.044 &  $-0.126 $&  $-1.138 $&  $-0.964 $&  0.034 &  $-0.053 $&  $-0.982 $&  $-0.819 $&  0.019 &  0.015 &  $-0.784$\\
	\hline $\mu_{ \Omega_c^{0} }^*(\mu_N )$ & $-0.781 $&  0.028 &  $-0.01 $&  $-0.763 $&  $-0.556 $&  0.011 &  $-0.006 $&  $-0.551 $&  $-0.245 $&  $-0.013 $&  $-0.002 $&  $-0.261$\\
	\hline $\mu_{ \Xi_c^{+} }^*(\mu_N )$ & 0.334 &  $-0.018 $&  0.008 &  0.324 &  0.248 &  $-0.019 $&  0.004 &  0.233 &  0.152 &  $-0.02 $&  0.001 &  0.132 \\
	\hline $\mu_{ \Xi_c^{0} }^*(\mu_N )$ & 0.217 &  $0.0 $&  $-0.008 $&  0.209 &  0.126 &  $0.0 $&  $-0.003 $&  0.122 &  0.024 &  $-0.001 $&  0.001 &  0.024 \\
	\hline $\mu_{ \Lambda_c^{+} }^*(\mu_N )$ & 0.317 &  $-0.019 $&  0.001 &  0.299 &  0.227 &  $-0.02 $&  0.001 &  0.208 &  0.12 &  $-0.021 $&  0.002 &  0.101 \\
	\hline $\mu_{ \Xi_{cc}^{++} }^*(\mu_N )$ & $-0.132 $&  0.102 &  $-0.056 $&  $-0.085 $&  $-0.279$ &  0.106 &  $-0.028 $&  $-0.201 $&  $-0.435 $&  0.11 &  $-0.005 $&  $-0.33$\\
	\hline $\mu_{ \Xi_{cc}^{+} }^*(\mu_N )$ & 0.634 &  $-0.016 $&  0.049 &  0.667 &  0.502 &  $-0.014 $&  0.02 &  0.508 &  0.36 &  $-0.012 $&  $-0.007 $&  0.341 \\
	\hline $\mu_{ \Omega_{cc}^{+} }^*(\mu_N )$ & 0.522 &  $-0.009 $&  0.002 &  0.515 &  0.342 &  $-0.005 $&  0.001 &  0.338 &  0.139 &  0.0 &  0.0 &  0.14 \\
		\hline
	\end{tabular}
	\caption{Values of effective magnetic moments of  spin$-\frac{1}{2}^{+}$ charmed baryons in symmetric strange matter ($I_{a} =0.0$ and $f_s =0.3$) are tabulated above at temperature $T=0$ MeV
	and compared with values at $\rho_B =0$.}
	\label{table:mub_eta0fs3_1by2}
\end{sidewaystable}		
	
\begin{sidewaystable}
	\begin{tabular}{|c|c|c|c|c|c|c|c|c|c|c|c|c|}
		\hline Baryon 
		& \multicolumn{4}{|c|}{$\rho_B=0$} & \multicolumn{4}{|c|}{$\rho_B=\rho_0$} & \multicolumn{4}{|c|}{$\rho_B=3 \rho_0$} \\
		\hline & $\mu_{B,\mathrm {val}}^{*}$ & $\mu_{B,\text {sea}}^{*}$ & $\mu_{B,\text {orbital }}^{*}$ & $\mu_B^*$ & $\mu_{B,\mathrm{val}}^{*}$ & $\mu_{B, \text {sea }}^{*}$ & $\mu_{B,\text {orbital }}^{*}$ & $\mu_B^*$ & $\mu_{B,\mathrm{val}}^{*}$ & $\mu_{B,\text {sea }}^{*}$ & $\mu_{B,\text {orbital }}^{*}$ & $\mu_B^*$ \\
	\hline $\mu_{ \Sigma_c^{++} }^*(\mu_N )$ & 2.434 &  $-0.52 $&  0.274 &  2.187 &  2.6 &  $-0.555 $&  0.148 &  2.193 &  2.775 &  $-0.593 $&  0.034 &  2.216 \\
	\hline $\mu_{ \Sigma_c^{+} }^*(\mu_N )$ & 0.551 &  $-0.234 $&  0.016 &  0.333 &  0.611 &  $-0.253 $&  0.021 &  0.379 &  0.677 &  $-0.274 $&  0.036 &  0.439 \\
	\hline $\mu_{ \Sigma_c^{0} }^*(\mu_N )$ & $-1.332 $&  0.06 &  $-0.241 $&  $-1.513 $&  $-1.384 $&  0.059 &  $-0.107 $&  $-1.432 $&  $-1.43 $&  0.055 &  0.038 &  $-1.337$\\
	\hline $\mu_{ \Xi_c^{\prime+} }^*(\mu_N )$ & 0.826 &  $-0.246 $&  0.132 &  0.712 &  0.989 &  $-0.267 $&  0.071 &  0.793 &  1.207 &  $-0.294 $&  0.016 &  0.929 \\
	\hline $\mu_{ \Xi_c^{\prime0} }^*(\mu_N )$ & $-1.057 $&  0.044 &  $-0.126 $&  $-1.138 $&  $-0.975 $&  0.035 &  $-0.056 $&  $-0.996$ &  $-0.838 $&  0.021 &  0.018 &  $-0.799$\\
	\hline $\mu_{ \Omega_c^{0} }^*(\mu_N )$ & $-0.781 $&  0.028 &  $-0.01 $&  $-0.763 $&  $-0.58 $&  0.013 &  $-0.006 $&  $-0.573 $&  $-0.28 $&  $-0.011 $&  $-0.002 $&  $-0.293$\\
	\hline $\mu_{ \Xi_c^{+} }^*(\mu_N )$ & 0.334 &  $-0.018 $&  0.008 &  0.324 &  0.256 &  $-0.019 $&  0.004 &  0.241 &  0.161 &  $-0.02 $&  0.001 &  0.142 \\
	\hline $\mu_{ \Xi_c^{0} }^*(\mu_N )$ & 0.217 &  $-0.0 $&  $-0.008 $&  0.209 &  0.134 &  $-0.0 $&  $-0.004 $&  0.13 &  0.034 &  $-0.001 $&  0.001 &  0.034 \\
	\hline $\mu_{ \Lambda_c^{+} }^*(\mu_N )$ & 0.317 &  $-0.019 $&  0.001 &  0.299 &  0.236 &  $-0.02 $&  0.001 &  0.217 &  0.131 &  $-0.021 $&  0.002 &  0.113 \\
	\hline $\mu_{ \Xi_{cc}^{++} }^*(\mu_N )$ & $-0.132 $&  0.102 &  $-0.056 $&  $-0.085 $&  $-0.267 $&  0.106 &  $-0.03 $&  $-0.191$ &  $-0.42 $&  0.11 &  $-0.007 $&  $-0.317$\\
	\hline $\mu_{ \Xi_{cc}^{+} }^*(\mu_N )$ & 0.634 &  $-0.016 $&  0.049 &  0.667 &  0.514 &  $-0.015 $&  0.022 &  0.521 &  0.375 &  $-0.013 $&  $-0.008 $&  0.354 \\
	\hline $\mu_{ \Omega_{cc}^{+} }^*(\mu_N )$ & 0.522 &  $-0.009 $&  0.002 &  0.515 &  0.358 &  $-0.005 $&  0.001 &  0.354 &  0.16 &  $-0.0 $&  0.0 &  0.16 \\
		\hline
	\end{tabular}
	\caption{Values of effective magnetic moments of  spin$-\frac{1}{2}^{+}$ charmed baryons in asymmetric strange matter ($I_{a} =0.5$ and $f_s =0.3$) are tabulated above at temperature $T=0$ MeV and compared with values at $\rho_B =0$.}
	\label{table:mub_eta5fs3_1by2}
\end{sidewaystable}

\begin{sidewaystable}
	\begin{tabular}{|c|c|c|c|c|c|c|c|c|c|c|c|c|}
		\hline Baryon 
		& \multicolumn{4}{|c|}{$\rho_B=0$} & \multicolumn{4}{|c|}{$\rho_B=\rho_0$} & \multicolumn{4}{|c|}{$\rho_B=3 \rho_0$} \\
		\hline & $\mu_{B,\mathrm {val}}^{*}$ & $\mu_{B,\text {sea}}^{*}$ & $\mu_{B,\text {orbital }}^{*}$ & $\mu_B^*$ & $\mu_{B,\mathrm{val}}^{*}$ & $\mu_{B, \text {sea }}^{*}$ & $\mu_{B,\text {orbital }}^{*}$ & $\mu_B^*$ & $\mu_{B,\mathrm{val}}^{*}$ & $\mu_{B,\text {sea }}^{*}$ & $\mu_{B,\text {orbital }}^{*}$ & $\mu_B^*$ \\
		\hline $\mu_{ \Sigma_c^{*++} }^*(\mu_N )$ & 4.3 &  $-0.833$ &  0.436 &  3.904 &  4.438 &  $-0.89$ &  0.224 &  3.771 &  4.543 &  $-0.947$ &  0.048 &  3.644 \\
		\hline $\mu_{ \Sigma_c^{*+} }^*(\mu_N )$ & 1.3 &  $-0.371 $&  0.026 &  0.957 &  1.261 &  $-0.4 $&  0.03 &  0.891 &  1.202 &  $-0.432 $&  0.044 &  0.815 \\
		\hline $\mu_{ \Sigma_c^{*0} }^*(\mu_N )$ & $-1.7 $&  0.092 &  $-0.384 $&  $-1.991 $&  $-1.915 $& $ 0.089 $&  $-0.163 $&  $-1.989 $&  $-2.139 $&  0.083 &  0.041 &  $-2.014$\\
		\hline $\mu_{ \Xi_c^{*+} }^*(\mu_N )$ & 1.74 &  $-0.396$ &  0.21 &  1.554 &  1.876 &  $-0.431$ &  0.107 &  1.552 &  2.052 &  $-0.469$ &  0.022 &  1.604 \\
		\hline $\mu_{ \Xi_c^{*0} }^*(\mu_N )$ & $-1.26$ &  0.066 &  $-0.2$ &  $-1.393$ &  $-1.249$ &  0.051 &  $-0.086$ &  $-1.284$ &  $-1.178$ &  0.028 &  0.019 &  $-1.131$\\
		\hline $\mu_{ \Omega_c^{*0} }^*(\mu_N )$ & $-0.82$ &  0.04 &  $-0.016$ &  $-0.796$ &  $-0.604$ &  0.013 &  $-0.009$ &  $-0.601$ &  $-0.28$ &  $-0.023$ &  $-0.004$ &  $-0.307$\\
		\hline $\mu_{ \Xi_{cc}^{*++} }^*(\mu_N )$ & 1.9 &  $-0.308$ &  0.218 &  1.813 &  1.775 &  $-0.316$ &  0.112 &  1.571 &  1.636 &  $-0.324$ &  0.024 &  1.335 \\
		\hline $\mu_{ \Xi_{cc}^{*+} }^*(\mu_N )$ & $-0.29$ &  $-0.308$ &  $-0.192$ &  $-0.79$ &  $-0.46$ &  $-0.316$ &  $-0.081$ &  $-0.858$ &  $-0.637$ &  $-0.324$ &  0.021 &  $-0.94$\\
		\hline $\mu_{ \Omega_{cc}^{*+} }^*(\mu_N )$ & 0.04 &  0.016 &  $-0.008$ &  0.05 &  $-0.008$ &  0.003 &  $-0.005$ &  $-0.01$ &  $-0.015$ &  $-0.014$ &  $-0.002$ &  $-0.031$\\
		
		\hline
	\end{tabular}
	\caption{Values of effective magnetic moments of spin$-\frac{3}{2}^{+}$ charmed baryons in symmetric nuclear matter ($I_{a} =0$ and $f_s =0$) are tabulated above at temperature $T=0$ MeV and compared with values at $\rho_B =0$.}
	\label{table:mub_eta0fs0_3by2}
\end{sidewaystable}

\begin{sidewaystable}
	\begin{tabular}{|c|c|c|c|c|c|c|c|c|c|c|c|c|}
		\hline Baryon
		& \multicolumn{4}{|c|}{$\rho_B=0$} & \multicolumn{4}{|c|}{$\rho_B=\rho_0$} & \multicolumn{4}{|c|}{$\rho_B=3 \rho_0$} \\
		\hline & $\mu_{B,\mathrm {val}}^{*}$ & $\mu_{B,\text {sea}}^{*}$ & $\mu_{B,\text {orbital }}^{*}$ & $\mu_B^*$ & $\mu_{B,\mathrm{val}}^{*}$ & $\mu_{B, \text {sea }}^{*}$ & $\mu_{B,\text {orbital }}^{*}$ & $\mu_B^*$ & $\mu_{B,\mathrm{val}}^{*}$ & $\mu_{B,\text {sea }}^{*}$ & $\mu_{B,\text {orbital }}^{*}$ & $\mu_B^*$ \\
\hline $\mu_{ \Sigma_c^{*++} }^*(\mu_N )$ & 4.302 &  $-0.833 $&  0.436 &  3.904 &  4.431 &  $-0.887 $&  0.235 &  3.779 &  4.533 &  $-0.94 $&  0.065 &  3.658 \\
\hline $\mu_{ \Sigma_c^{*+} }^*(\mu_N )$ & 1.302 &  $-0.371 $&  0.026 &  0.957 &  1.266 &  $-0.399 $&  0.033 &  0.9 &  1.213 &  $-0.429 $&  0.053 &  0.838 \\
\hline $\mu_{ \Sigma_c^{*0} }^*(\mu_N )$ & $-1.698 $&  0.092 &  $-0.384 $&  $-1.991 $&  $-1.909 $&  0.09 &  $-0.168 $&  $-1.987 $&  $-2.121 $&  0.085 &  0.042 &  $-1.995 $\\
\hline $\mu_{ \Xi_c^{*+} }^*(\mu_N )$ & 1.741 &  $-0.396 $&  0.21 &  1.554 &  1.867 &  $-0.429 $&  0.112 &  1.551 &  2.029 &  $-0.465 $&  0.03 &  1.595 \\
\hline $\mu_{ \Xi_c^{*0} }^*(\mu_N )$ & $-1.259 $&  0.066 &  $-0.2 $&  $-1.393 $&  $-1.254 $&  0.052 &  $-0.089 $&  $-1.291 $&  $-1.193 $&  0.032 &  0.019 &  $-1.142 $\\
\hline $\mu_{ \Omega_c^{*0} }^*(\mu_N )$ & $-0.82 $&  0.04 &  $-0.016 $&  $-0.796 $&  $-0.621 $&  0.015 &  $-0.01 $&  $-0.615 $&  $-0.325 $&  $-0.018 $&  $-0.004 $&  $-0.348 $\\
\hline $\mu_{ \Xi_{cc}^{*++} }^*(\mu_N )$ & 1.902 &  $-0.308 $&  0.218 &  1.813 &  1.782 &  $-0.315 $&  0.117 &  1.584 &  1.653 &  $-0.323$ &  0.033 &  1.363 \\
\hline $\mu_{ \Xi_{cc}^{*+} }^*(\mu_N )$ & $-0.29 $&  $-0.308 $&  $-0.192 $&  $-0.79 $&  $-0.451 $&  $-0.316 $&  $-0.084 $&  $-0.851 $&  $-0.616 $&  $-0.324 $&  0.021 &  $-0.919 $\\
\hline $\mu_{ \Omega_{cc}^{*+} }^*(\mu_N )$ & 0.042 &  0.016 &  $-0.008 $&  0.05 &  $-0.007 $&  0.004 &  $-0.005 $&  $-0.008 $&  $-0.017 $&  $-0.012 $&  $-0.002 $&  $-0.03 $\\
				\hline
\end{tabular}
\caption{Values of effective magnetic moments of  spin$-\frac{3}{2}^{+}$ charmed baryons in asymmetric nuclear matter ($I_{a} =0.5$ and $f_s =0$) are tabulated above at temperature $T=0$ MeV 
	and compared with values at $\rho_B =0$.}
\label{table:mub_eta5fs0_3by2}
\end{sidewaystable}

\begin{sidewaystable}
	\begin{tabular}{|c|c|c|c|c|c|c|c|c|c|c|c|c|}
		\hline Baryon 
		& \multicolumn{4}{|c|}{$\rho_B=0$} & \multicolumn{4}{|c|}{$\rho_B=\rho_0$} & \multicolumn{4}{|c|}{$\rho_B=3 \rho_0$} \\
		\hline & $\mu_{B,\mathrm {val}}^{*}$ & $\mu_{B,\text {sea}}^{*}$ & $\mu_{B,\text {orbital }}^{*}$ & $\mu_B^*$ & $\mu_{B,\mathrm{val}}^{*}$ & $\mu_{B, \text {sea }}^{*}$ & $\mu_{B,\text {orbital }}^{*}$ & $\mu_B^*$ & $\mu_{B,\mathrm{val}}^{*}$ & $\mu_{B,\text {sea }}^{*}$ & $\mu_{B,\text {orbital }}^{*}$ & $\mu_B^*$ \\
\hline $\mu_{ \Sigma_c^{*++} }^*(\mu_N )$ & 4.302 &  $-0.833 $&  0.436 &  3.904 &  4.44 &  $-0.891 $&  0.22 &  3.769 &  4.547 &  $-0.949 $&  0.04 &  3.637 \\
\hline $\mu_{ \Sigma_c^{*+} }^*(\mu_N )$ & 1.302 &  $-0.371 $&  0.026 &  0.957 &  1.26 &  $-0.401 $&  0.031 &  0.89 &  1.198 &  $-0.433 $&  0.046 &  0.811 \\
\hline $\mu_{ \Sigma_c^{*0} }^*(\mu_N )$ & $-1.698 $&  0.092 &  $-0.384 $&  $-1.991 $&  $-1.92 $&  0.089 &  $-0.158 $&  $-1.989 $&  $-2.151 $&  0.083 &  0.052 &  $-2.016 $\\
\hline $\mu_{ \Xi_c^{*+} }^*(\mu_N )$ & 1.741 &  $-0.396 $&  0.21 &  1.554 &  1.877 &  $-0.432 $&  0.105 &  1.551 &  2.068 &  $-0.474 $&  0.018 &  1.612 \\
\hline $\mu_{ \Xi_c^{*0} }^*(\mu_N )$ & $-1.259 $&  0.066 &  $-0.2 $&  $-1.393 $&  $-1.256 $&  0.051 &  $-0.084 $&  $-1.289 $&  $-1.19 $&  0.028 &  0.024 &  $-1.138 $\\
\hline $\mu_{ \Omega_c^{*0} }^*(\mu_N )$ & $-0.82 $&  0.04 &  $-0.016 $&  $-0.796 $&  $-0.612 $&  0.014 &  $-0.009 $&  $-0.607 $&  $-0.281 $&  $-0.024 $&  $-0.003 $&  $-0.308 $\\
\hline $\mu_{ \Xi_{cc}^{*++} }^*(\mu_N )$ & 1.902 &  $-0.308 $&  0.218 &  1.813 &  1.772 &  $-0.316 $&  0.11 &  1.566 &  1.627 &  $-0.324 $&  0.02 &  1.323 \\
\hline $\mu_{ \Xi_{cc}^{*+} }^*(\mu_N )$ & $-0.29 $&  $-0.308 $&  $-0.192 $&  $-0.79 $&  $-0.464 $&  $-0.316 $&  $-0.079 $&  $-0.859 $&  $-0.647 $&  $-0.324 $&  0.026 &  $-0.946 $\\
\hline $\mu_{ \Omega_{cc}^{*+} }^*(\mu_N )$ & 0.042 &  0.016 &  $-0.008 $&  0.05 &  $-0.014 $&  0.003 &  $-0.005 $&  $-0.015 $&  $-0.024 $&  $-0.014 $&  $-0.002 $&  $-0.04 $\\

		\hline
	\end{tabular}
	\caption{Values of effective magnetic moments of spin$-\frac{3}{2}^+$ charmed baryons in symmetric strange matter ($I_{a} =0.0$ and $f_s =0.3$) are tabulated above at temperature $T=0$ MeV	and compared with values at $\rho_B =0$.}
	\label{table:mub_eta0fs3_3by2}
\end{sidewaystable}		
\begin{sidewaystable}
	
	\begin{tabular}{|c|c|c|c|c|c|c|c|c|c|c|c|c|}
		\hline Baryon 
		& \multicolumn{4}{|c|}{$\rho_B=0$} & \multicolumn{4}{|c|}{$\rho_B=\rho_0$} & \multicolumn{4}{|c|}{$\rho_B=3 \rho_0$} \\
		\hline & $\mu_{B,\mathrm {val}}^{*}$ & $\mu_{B,\text {sea}}^{*}$ & $\mu_{B,\text {orbital }}^{*}$ & $\mu_B^*$ & $\mu_{B,\mathrm{val}}^{*}$ & $\mu_{B, \text {sea }}^{*}$ & $\mu_{B,\text {orbital }}^{*}$ & $\mu_B^*$ & $\mu_{B,\mathrm{val}}^{*}$ & $\mu_{B,\text {sea }}^{*}$ & $\mu_{B,\text {orbital }}^{*}$ & $\mu_B^*$ \\
\hline $\mu_{ \Sigma_c^{*++} }^*(\mu_N )$ & 4.302 &  $-0.833 $&  0.436 &  3.904 &  4.43 &  $-0.886 $&  0.236 &  3.78 &  4.54 &  $-0.944 $&  0.053 &  3.649 \\
\hline $\mu_{ \Sigma_c^{*+} }^*(\mu_N )$ & 1.302 &  $-0.371 $&  0.026 &  0.957 &  1.266 &  $-0.399 $&  0.033 &  0.901 &  1.208 &  $-0.431 $&  0.057 &  0.834 \\
\hline $\mu_{ \Sigma_c^{*0} }^*(\mu_N )$ & $-1.698 $&  0.092 &  $-0.384 $&  $-1.991 $&  $-1.907 $&  0.09 &  $-0.17 $&  $-1.987 $&  $-2.139 $&  0.084 &  0.06 &  $-1.994 $\\
\hline $\mu_{ \Xi_c^{*+} }^*(\mu_N )$ & 1.741 &  $-0.396 $&  0.21 &  1.554 &  1.864 &  $-0.429 $&  0.113 &  1.548 &  2.047 &  $-0.47 $&  0.025 &  1.602 \\
\hline $\mu_{ \Xi_c^{*0} }^*(\mu_N )$ & $-1.259 $&  0.066 &  $-0.2 $&  $-1.393 $&  $-1.262 $&  0.053 &  $-0.09 $&  $-1.299 $&  $-1.204 $&  0.031 &  0.028 &  $-1.145 $\\
\hline $\mu_{ \Omega_c^{*0} }^*(\mu_N )$ & $-0.82 $&  0.04 &  $-0.016 $&  $-0.796 $&  $-0.636 $&  0.016 &  $-0.01 $&  $-0.629 $&  $-0.321 $&  $-0.02 $&  $-0.004 $&  $-0.344 $\\
\hline $\mu_{ \Xi_{cc}^{*++} }^*(\mu_N )$ & 1.902 &  $-0.308 $&  0.218 &  1.813 &  1.783 &  $-0.315 $&  0.118 &  1.586 &  1.642 &  $-0.324 $&  0.027 &  1.345 \\
\hline $\mu_{ \Xi_{cc}^{*+} }^*(\mu_N )$ & $-0.29 $&  $-0.308 $&  $-0.192 $&  $-0.79 $&  $-0.45 $&  $-0.316 $&  $-0.085 $&  $-0.85 $&  $-0.631 $&  $-0.324 $&  0.03 &  $-0.924 $\\
\hline $\mu_{ \Omega_{cc}^{*+} }^*(\mu_N )$ & 0.042 &  0.016 &  $-0.008 $&  0.05 &  $-0.012 $&  0.004 &  $-0.005 $&  $-0.013 $&  $-0.026$ &  $-0.012$ &  $-0.002 $&  $-0.04 $\\
\hline
\end{tabular}
\caption{Values of effective magnetic moments of spin$-\frac{3}{2}^+$ charmed baryons in asymmetric strange matter ($I_{a} =0.5$ and $f_s =0.3$) are tabulated above at temperature $T=0$ MeV
and compared with values at $\rho_B =0$.}
\label{table:mub_eta5fs3_3by2}
\end{sidewaystable}

	\clearpage
				\bibliographystyle{elsarticle-num}

	\end{document}